\documentclass{aa}  
\usepackage{graphicx}
\usepackage{txfonts}
\usepackage{subcaption}                              
\usepackage{lscape}             
\usepackage{longtable}                                       
\usepackage{placeins}                                        
\usepackage{url}                                
\usepackage{color}

\makeatletter
\newcommand{\daggerthanks}[1]{%
  \footnotemark[12]%
  \protected@xdef\@thanks{%
    \@thanks
    \protect\footnotetext[12]{#1}%
  }%
}
\makeatother

\begin{document}

    \title{Two \textit{Kepler} red-giant eclipsing binaries as lower red-giant branch benchmarks for asteroseismic scaling relations}
    \titlerunning{Lower-RGB benchmarks for asteroseismic scaling relations}
    \subtitle{Dynamical masses and radii from JKTEBOP and PHOEBE-2 modeling}

    \author{A. Hedlund\inst{1}\fnmsep\thanks{Corresponding author: ahedlund@lanl.gov}
    \and P. Gaulme\inst{2}\fnmsep\daggerthanks{Deceased.}
    \and J. Jackiewicz\inst{3}
    \and T. Appourchaux\inst{4}
    \and E. Guenther\inst{2}
        }

   \institute{Los Alamos National Laboratory, P.O. Box 1663, Los Alamos, NM 87545 USA
   \and Thüringer Landessternwarte, Sternwarte 5, 07778 Tautenburg, Germany
   \and Department of Astronomy, New Mexico State University, P.O. Box 30001, MSC 4500, Las Cruces, NM 88003-8001, USA
   \and Institut d’Astrophysique Spatiale, UMR 8617, Universit\'e Paris-Saclay, CNRS, B\^{a}timent Roger-Maurice-Bonnet, 91405 Orsay Cedex, France}

    \date{Received XX Month 20XX / Accepted XX Month 20XX}

\abstract
{Space-based photometry has detected solar-like oscillations in thousands of stars, whose global seismic parameters can be used to infer stellar properties through the asteroseismic scaling relations. As the scaling relations systematically overestimate red-giant masses and radii, empirical calibration is required. Double-lined eclipsing binary star systems containing oscillating red giants provide one of the few avenues to obtain precise dynamical masses and radii without external calibration, making them ideal benchmarks for calibrating the scaling relations.}
{We present KIC\,8129189 and KIC\,10920813, two \textit{Kepler} eclipsing binary systems containing oscillating red giants, as new benchmark systems for calibrating the asteroseismic scaling relations. Both systems contain less-evolved red giants, extending the calibration sample toward smaller radii.}
{We measured radial velocities for both components from high-resolution spectra and extracted global asteroseismic parameters from \textit{Kepler} light curves. The binaries were modeled with JKTEBOP and PHOEBE-2 using both full light curves and selected eclipse sets. Benchmark masses and radii were adopted from the envelope of tested model and data configurations. We then fitted correction factors $f_{\Delta\nu}$ and $f_{\nu_{\rm max}}$ to the underlying $\Delta\nu$--density and $\nu_{\rm max}$--surface-gravity relations for the adopted benchmark sample.}
{We obtain dynamical values of $M_{\rm RG}=1.413 \pm 0.037\,M_\odot$ and $R_{\rm RG}=4.296 \pm 0.061\,R_\odot$ for KIC\,8129189, and $M_{\rm RG}=1.383 \pm 0.022\,M_\odot$ and $R_{\rm RG}=5.914 \pm 0.043\,R_\odot$ for KIC\,10920813. The adopted 15-star calibration gives $f_{\Delta\nu}=0.9777 \pm 0.0048$ and $f_{\nu_{\rm max}}=1.024 \pm 0.008$, corresponding to mass and radius correction factors of $C_M=0.850 \pm 0.020$ and $C_R=0.9333 \pm 0.0090$. No significant dependence on radius is detected in either individual correction factor, nor a metallicity dependence in $f_{\nu_{\rm max}}$; model-based $f_{\Delta\nu}$ values are systematically lower than the dynamically derived values.}
{The two systems extend the benchmark sample toward less-evolved red giants with smaller radii and agree with the empirical correction inferred from previous benchmark systems. They demonstrate that continued expansion of the double-lined eclipsing-binary benchmark sample is essential for testing the scaling relations across evolutionary stage.}

   \keywords{Asteroseismology --
                Stars: binaries: eclipsing --
                Stars: binaries: spectroscopic --
                Stars: variables: general
               }
\maketitle
\nolinenumbers

\section{Introduction}
\label{sec:intro}

High-precision space-based photometry has enabled the detection of stellar pulsations in tens of thousands of stars, transforming the field of asteroseismology and opening a window into stellar interiors \citep[for reviews, see][]{ChaplinMiglio13,Kurtz22,Huber26}. Photometric observations from the \textit{Kepler} and TESS missions have been used to identify in excess of 20\,000 and 150\,000 oscillating red giant stars across the sky, respectively \citep{Hon19,Hon21}. Forthcoming observations from PLATO and the Nancy Grace Roman Space Telescope are expected to substantially expand these samples \citep{Rauer2025,Weiss2025}. The global seismic parameter $\nu_{\rm max}$, the frequency of maximum oscillation power, scales with surface gravity, while $\Delta\nu$, the large frequency separation, scales with mean density; together, they enable estimates of stellar masses and radii when combined with an effective temperature \citep[e.g.,][]{Brown91,Kjeldsen95,Kallinger14}. These measurements now underpin ensemble asteroseismology and Galactic archaeology through large spectroscopic-seismic samples such as APOKASC \citep{Pinsonneault14,Pinsonneault18,Pinsonneault25}, where red-giant masses and radii constrain stellar ages and distances, enabling population studies \citep{Miglio13,SilvaAguirre18}. As a result, even few-percent systematic errors in the scaling relations are important for the interpretation of large asteroseismic samples.

Detached eclipsing binaries are one of the cleanest empirical tests of these seismic inferences. In eclipsing double-lined spectroscopic binaries (SB2s), the combination of radial velocities and eclipse photometry allows the system to be dynamically characterized, providing independent measurements of masses and radii. Systems in which one component is an oscillating red giant are therefore particularly valuable: the same star has both a direct dynamical measurement and a seismic estimate. Hierarchical triples with eclipse-timing variations provide a complementary route to benchmark masses in favorable configurations, as demonstrated for KIC\,7955301 \citep{Gaulme22}. Previous \textit{Kepler} studies have demonstrated the power of this comparison and motivated the need for calibration. \citet{Gaulme16} found that uncorrected red-giant scaling relations overestimated dynamically derived masses and radii by about 15\% and 5\%, respectively. \citet{Benbakoura21} expanded the sample and confirmed the same broad discrepancy, while showing that empirical reference-value choices and model-based seismic corrections can reduce this bias. Complementary stellar model-based calibrations of large samples of \textit{Kepler} red giant branch stars reach similar conclusions. \citet{Li22} used radial-mode frequencies, global seismic parameters, and spectroscopic constraints to model 3\,642 red giants and found that the uncorrected scaling relations overestimate masses and radii by 15\% and 7\%, respectively.

The benchmark sample of double-lined spectroscopic eclipsing binaries remains small and does not cover the full extent of red-giant parameter space. Most well-characterized red-giant eclipsing-binary benchmarks are more evolved than the lower red-giant branch systems presented in this work, making additional smaller-radius systems important for testing whether the same correction factors vary with evolutionary state. This broader mapping is especially important because recent large-sample analyses indicate that the performance of the scaling relations is not uniform: the relations are relatively well behaved on the lower RGB and in the red clump, but become increasingly model dependent for luminous giants and can break down near the RGB tip \citep{Yu20,Zinn23,Pinsonneault25,Ash24}. 

This work presents a spectroscopic, asteroseismic, and dynamical analysis of KIC\,8129189 and KIC\,10920813, two detached eclipsing binaries containing solar-like oscillating red giants with main-sequence companions. We confirmed their SB2 nature, measured radial velocities for both components, extracted the global seismic properties of the red giants, and determined their atmospheric parameters from high-resolution spectra. We then modeled both systems with the binary-modeling tools JKTEBOP and PHOEBE-2. Because the \textit{Kepler} eclipse morphology was affected by time-variable photometric structure, we compared models based on the full stitched light curves with models based on selected eclipse sets. This comparison is essential for defining robust benchmark uncertainties because, for systems intended as calibration benchmarks, the relevant uncertainty should reflect not only the formal precision of one preferred model but also the stability of the inferred red-giant mass and radius under reasonable choices of model prescription and eclipse selection.

We used the resulting dynamical masses and radii to update the empirical red-giant scaling-relation calibration. Rather than fitting independent offsets to mass and radius, we fitted shared correction factors $f_{\Delta\nu}$ and $f_{\nu_{\rm max}}$ at the level of the underlying $\Delta\nu$--density and $\nu_{\rm max}$--surface-gravity relations. Alongside concerns about the applicability of the scaling relations across red-giant parameter space, studies have also suggested that the correction factors themselves may depend on stellar properties. Model-based prescriptions have predicted that $f_{\Delta\nu}$ may depend on stellar properties and evolutionary state \citep{Li23,Schimak26,Valle25}, while studies of the $\nu_{\rm max}$ relation have suggested a possible composition dependence \citep{Viani17,Li24}. The sample-level calibration is intended to quantify systematic corrections to seismic masses and radii relevant to ensemble applications; the star-by-star behavior of the underlying seismic relations is examined separately for potential trends with stellar properties.

The paper is organized as follows. Sect.~\ref{sec:data} describes the \textit{Kepler} photometry and spectroscopic observations. Sects.~\ref{sec:spec} and \ref{sec:phot} present the radial-velocity, atmospheric, and asteroseismic analyses. Sect.~\ref{sec:dyn} describes the JKTEBOP and PHOEBE-2 dynamical modeling, compares the model and data configurations, and defines the adopted benchmark masses and radii. Sect.~\ref{sec:grid} presents the grid-based stellar modeling. Sect.~\ref{sec:discussion} compares the resulting benchmark values with seismic and stellar-model estimates, performs the scaling-relation calibration, and explores potential trends with stellar properties.

\section{Data}\label{sec:data}

\subsection{The Kepler time series}\label{sec:lcs}
\begin{figure}[t]
\centering
\includegraphics[width=0.38\textwidth]{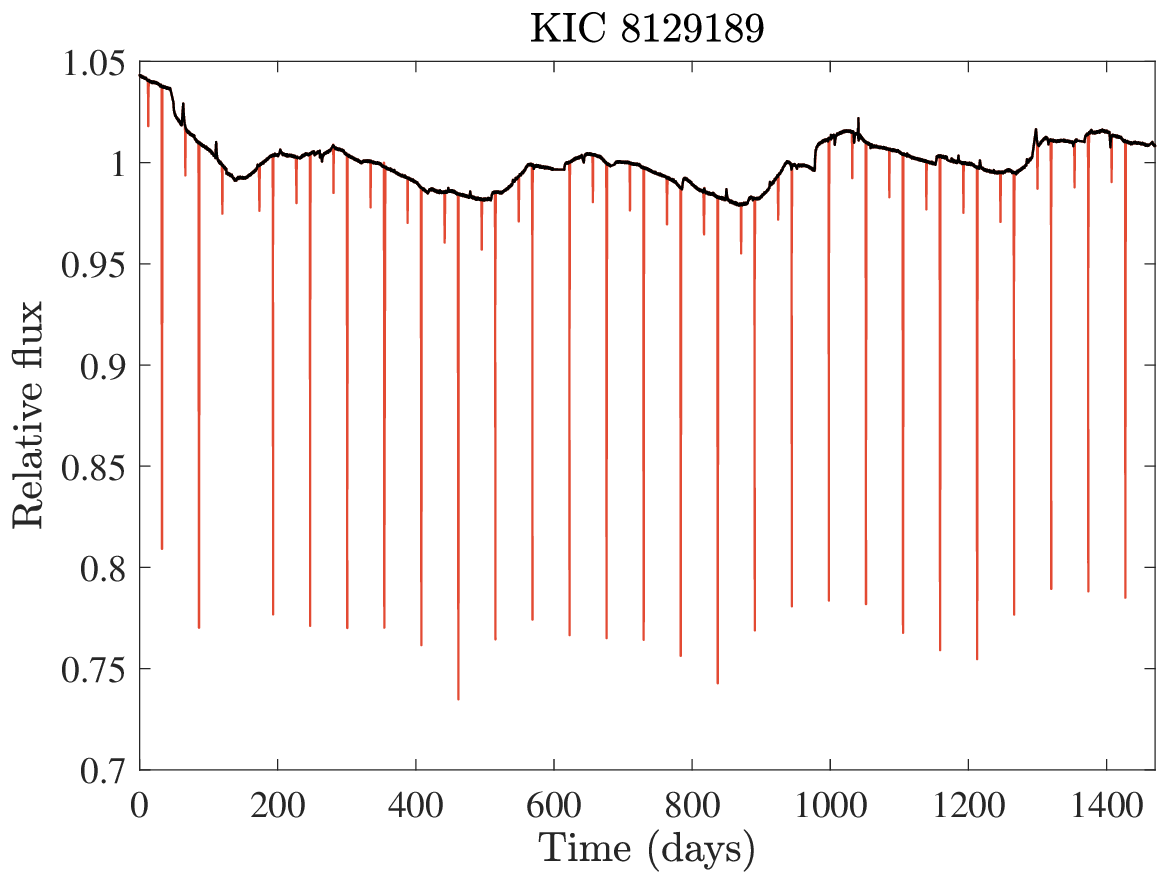}\\
\includegraphics[width=0.38\textwidth]{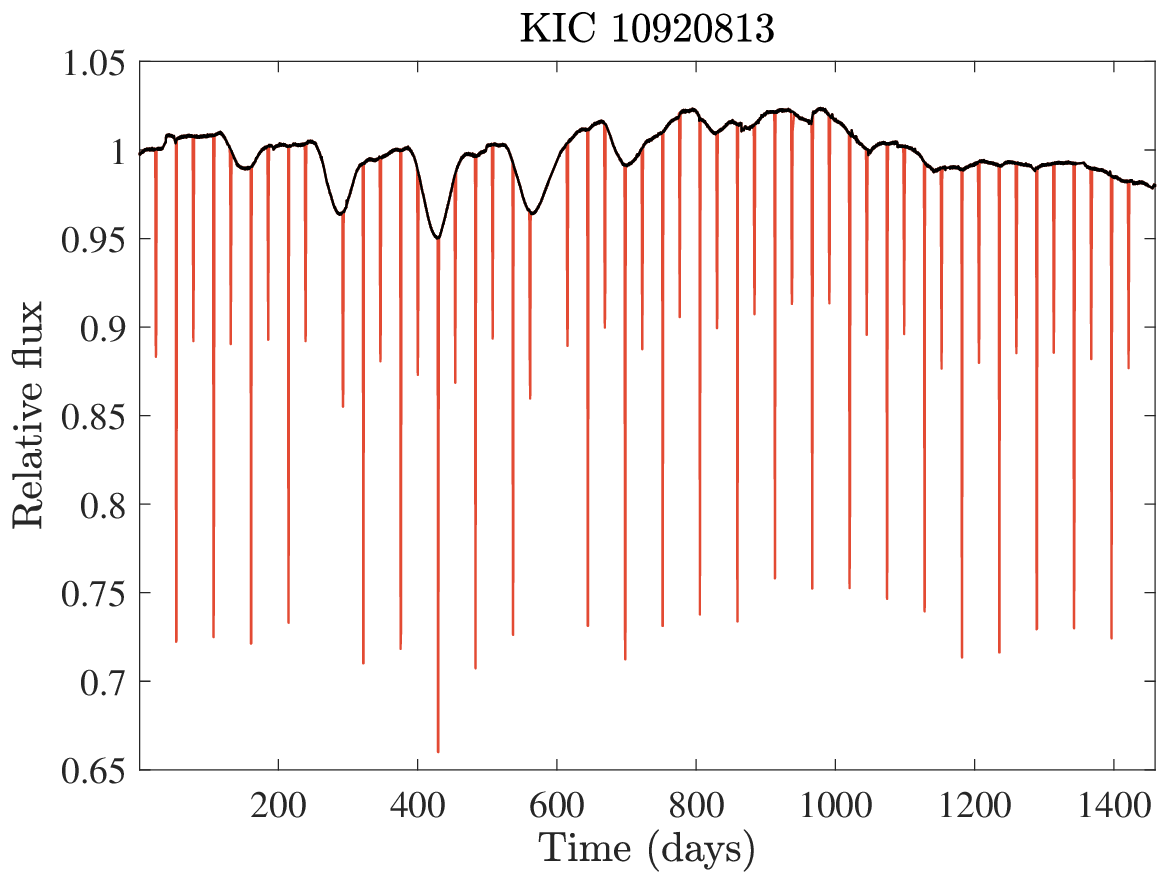}
\caption{Stitched and cleaned SAP \textit{Kepler} light curves of KIC\,8129189 (top) and KIC\,10920813 (bottom).
\label{fig:stitched_lc}}
\end{figure}

We used the public \textit{Kepler} long-cadence (29.4 min) light curves available from the Mikulski Archive for Space Telescopes (MAST). For each target, both Simple Aperture Photometry (SAP) and Pre-search Data Conditioning Simple Aperture Photometry (PDC-SAP) light curves are available. The PDC-SAP light curves are corrected for discontinuities, systematic errors, and excess flux due to aperture crowding \citep{Twicken10}, but this processing can also alter astrophysical signals such as rotational modulation and eclipse-depth variations \citep[e.g.,][]{Garcia14,Gaulme14}. We therefore used the SAP data and performed a custom detrending and stitching procedure designed to preserve long-term photometric variability while treating quarter boundaries and data gaps consistently. The stitched and cleaned SAP light curve of KIC\,8129189 shows a modulation near 372 days, consistent with the \textit{Kepler} spacecraft orbital period around the Sun, while KIC\,10920813 shows a 138-day modulation that is also visible in the PDC-SAP light curve and is indicative of rotationally modulated stellar activity. The stitched and cleaned light curves are shown in Fig.~\ref{fig:stitched_lc}.

For the asteroseismic analysis, eclipses must be removed because they introduce power into the Fourier domain and can contaminate the power density spectrum. We considered two approaches. The first was to model the eclipses, for example using the JKTEBOP code \citep{Southworth13} described in Sect.~\ref{sec:jktebop}, and subtract the model from the light curve. As noted by \citet{Gaulme16}, however, small imperfections in an eclipse model can produce artifacts when repeated over the full \textit{Kepler} time series. We therefore adopted the second approach: masking the eclipses and filling the resulting gaps with a local second-order polynomial before computing the power density spectrum.

\subsection{Spectroscopic observations}\label{sec:spectra}
Spectroscopic follow-up observations of KIC\,8129189 and KIC\,10920813 were obtained primarily with the Astrophysical Research Consortium (ARC) 3.5~m telescope at Apache Point Observatory (APO) between 2019 and 2023. We used the ARC \'echelle spectrograph (ARCES) to confirm the double-lined nature of both systems, obtain radial-velocity curves, and determine atmospheric parameters. ARCES is a high-resolution visible-light spectrograph with $R\approx31\,500$ and wavelength coverage spanning 3\,200--10\,000\,\AA\ \citep{Wang03}. Previous studies of red-giant binary systems have shown that spectra with S/N $\approx10$--$20$ provide radial-velocity uncertainties of $\sigma_{\rm RV}\approx0.5\,{\rm km\,s^{-1}}$ \citep[e.g.,][]{Rawls16,Gaulme16,Benbakoura21,Gaulme22}. We obtained 29 ARCES spectra of KIC\,8129189 and 26 ARCES spectra of KIC\,10920813, with typical S/N $\approx10$--$25$. These spectra were processed and reduced following the procedures outlined in \citet{Rawls16} and \citet{Gaulme16}.

To improve the orbital phase coverage of KIC\,8129189, we obtained two additional spectra using the coud\'e \'echelle spectrograph on the 2~m Alfred Jensch telescope at the Th\"uringer Landessternwarte Tautenburg (TLS). Using the VIS-Gism, the spectra cover the wavelength range 452--765\,nm. With the 2 arcsec slit, the resolving power is $R=\lambda/\Delta\lambda\approx35\,000$. These spectra were processed in a manner similar to the ARCES spectra and were used only to improve the radial-velocity phase coverage.

\section{Spectroscopic analysis}\label{sec:spec}

\subsection{Radial-velocity derivation}\label{sec:rvs}
Initial inspection of the one-dimensional continuum-normalized ARCES spectra confirmed that both KIC\,8129189 and KIC\,10920813 are double-lined spectroscopic binaries. Both sets of spectral lines were detected in 27 of the 29 ARCES spectra for KIC\,8129189 and in 25 of the 26 ARCES spectra for KIC\,10920813. Both components were also detected in each of the KIC\,8129189 spectra obtained at the Th\"uringer Landessternwarte Tautenburg (TLS).

Radial-velocity curves for both components are required to model the binary systems dynamically. We derived the radial velocities using the broadening-function (BF) technique \citep{Rucinski92,Rucinski02}, which uses linear deconvolution to obtain a broadening kernel from a sharp-lined spectral template. Following \citet{Benbakoura21}, we used a synthetic spectrum representative of the companion, rather than the red giant, to increase the detectability of the less luminous component in the composite spectra. The broadening functions were calculated over shorter-wavelength spectral regions to increase the companion-line contribution and avoid regions with significant telluric contamination.

\begin{figure}[t]
\centering
\includegraphics[width=0.38\textwidth]{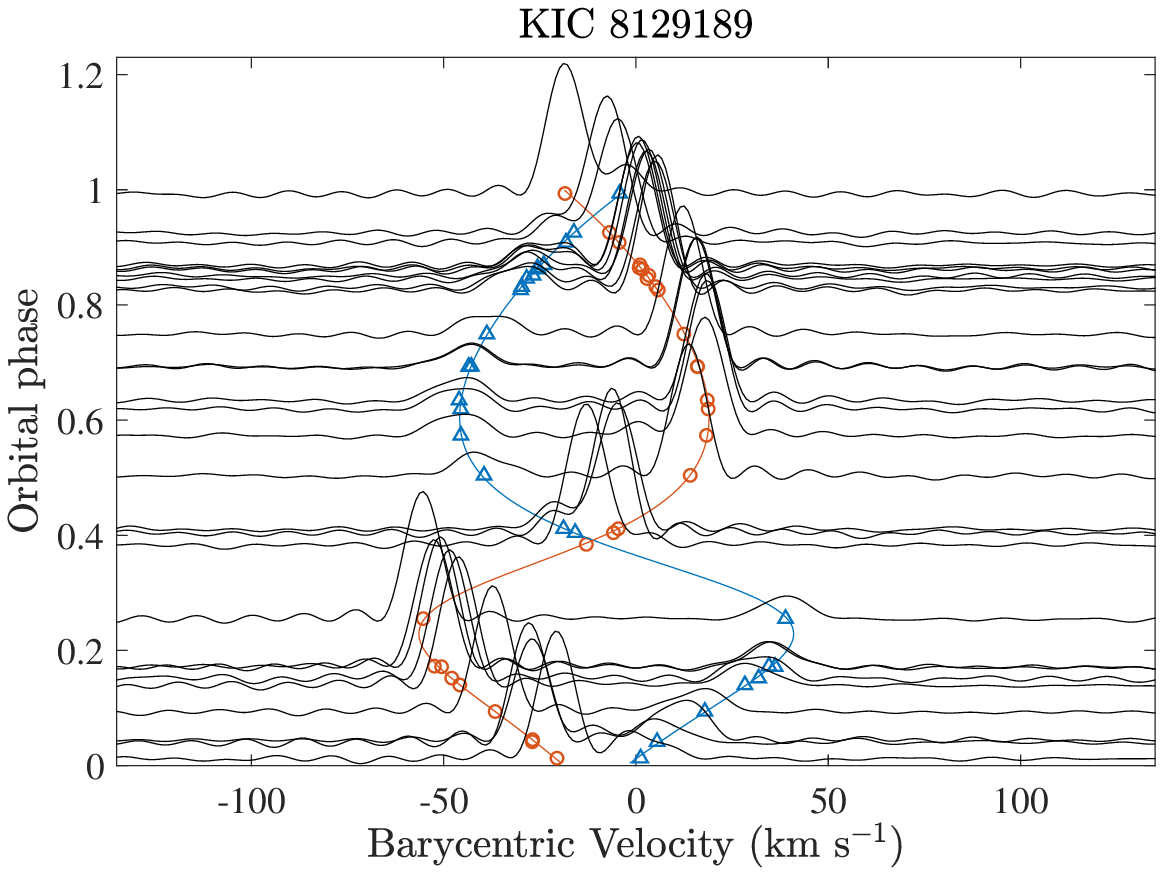}\\
\includegraphics[width=0.38\textwidth]{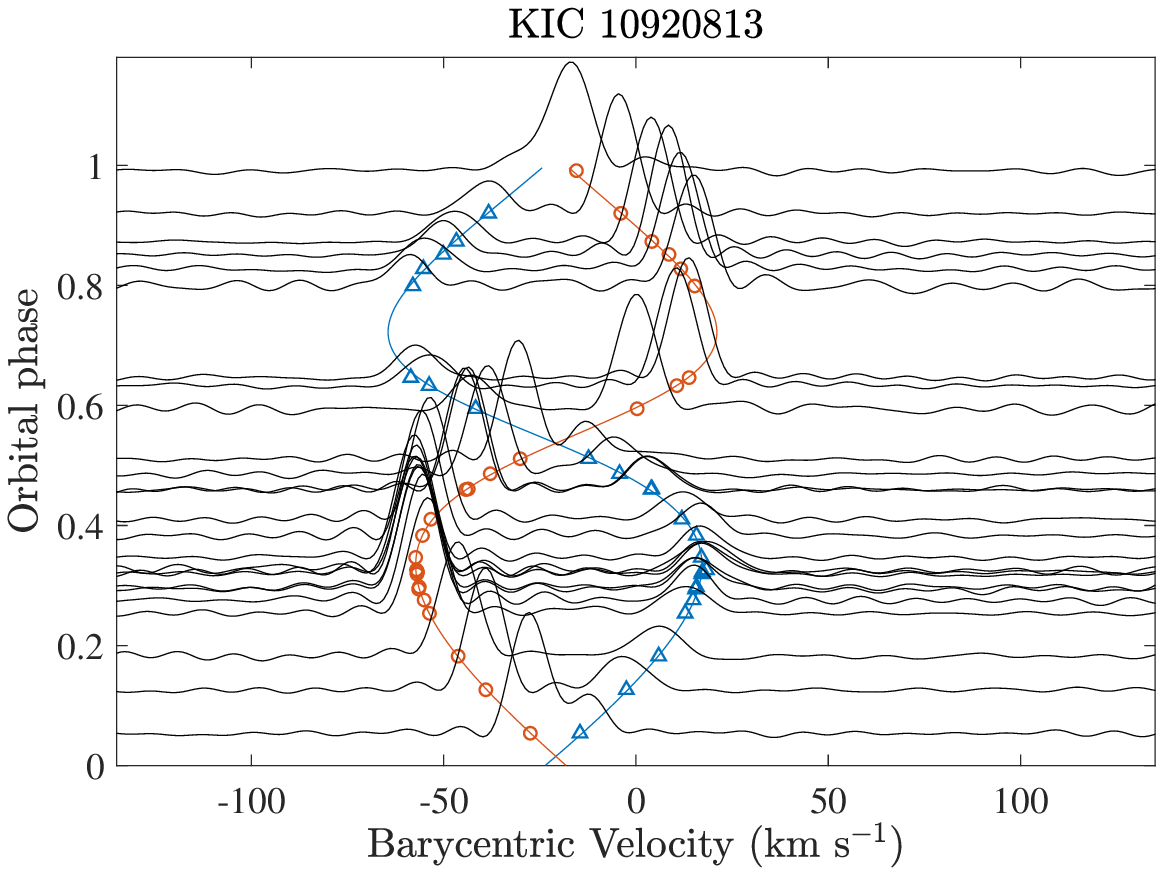}
\caption{Broadening functions of KIC\,8129189 (top) and KIC\,10920813 (bottom), sorted by orbital phase. The reference time is the epoch of secondary eclipse, when the hotter companion transits the red giant. Each line represents the BF computed from an individual spectrum. Blue triangles and red circles mark the radial-velocity measurements of the main-sequence companion and red-giant components, respectively, derived from the BF profiles.
\label{fig:BF}}
\end{figure}

We implemented this method using a Python adaptation of Rucinski's IDL routines\footnote{\url{https://www.astro.utoronto.ca/rucinski/SVDcookbook.html}} and main-sequence spectral templates with $T_{\rm eff}=5500$ and $5800\,{\rm K}$ generated with the \texttt{PHOENIX BT-Settl} code assuming solar abundances \citep{Allard03,Asplund09}. After the broadening kernels were calculated, they were smoothed with a Gaussian to suppress uncorrelated noise. We then fitted a two-component Gaussian model to the BF peaks to measure the radial velocities of both components. The BF profiles plotted as a function of orbital phase are shown in Fig.~\ref{fig:BF}.

\subsection{Spectral disentangling}

After deriving radial velocities for both components, we separated the time-series composite spectra into their individual component spectra using the spectral disentangling tool \texttt{fd3}. The \texttt{fd3} code performs Fourier-space spectral disentangling, reconstructing the component spectra from composite spectra obtained at different orbital phases \citep{Ilijic04}.

To achieve sufficient S/N in the reconstructed component spectra for the subsequent stellar-line analysis, we selected approximately a dozen spectra for each binary system. We verified that the selected spectra provided adequate orbital phase coverage, then applied barycentric corrections and subtracted the systemic velocity of each binary. The spectra were divided into wavelength chunks and resampled in logarithmic wavelength space to reduce continuum-level variations in the disentangled components \citep{Ilijic04,Beck14}. Using initial estimates of the orbital parameters ($P_{\rm orb}$, $t_{\rm peri}$, $e$, $\omega$, $K_{1}$, and $K_{2}$) and the chunked composite spectra, \texttt{fd3} reconstructed the component spectra. We considered the reconstruction successful when the radial velocities returned by \texttt{fd3} were consistent with the radial-velocity curves derived in Sect.~\ref{sec:rvs}.

\subsection{Atmospheric parameters}

We estimated spectroscopic parameters from the disentangled red-giant component spectra of KIC\,8129189 and KIC\,10920813. We used the LTE spectral-analysis code MOOG \citep{Sneden73} with the \texttt{synth} driver. The disentangled spectra have S/N $\approx40$, with higher S/N at redder wavelengths. Because the surface gravities are well constrained by the seismic and dynamical analyses, we fixed $\log g$ in the spectral analysis and solved for $T_{\rm eff}$ and $[{\rm Fe/H}]$. The resulting spectroscopic constraints are listed in Table~\ref{tab:atm_param}. Both red-giant components are slightly metal-poor relative to the Sun.

Previous studies of red giants in binary systems have found evidence of surface magnetic activity on the red-giant component, particularly in systems with orbital periods shorter than about 120 days, in spin-orbit resonant configurations, or in tidally locked systems \citep{Gaulme14,Gaulme20,Gehan22,Gehan24}. The two systems studied here have nearly identical orbital periods, $P_{\rm orb}=53.64$ days for KIC\,8129189 and $53.74$ days for KIC\,10920813, but only KIC\,10920813 displays photometric modulation characteristic of rotationally modulated surface activity.

Because KIC\,10920813 appears magnetically active, we searched for activity indicators in the individual spectra. We focused on Ca\,II H \& K emission, a long-used tracer of surface magnetic activity \citep{Babcock61,Petit08,Auriere15}, and H$\alpha$ core emission, which is also used as an activity indicator in late-type stars \citep{Cincunegui07,Newton17}. Evidence of H$\alpha$ core emission is visible in several observations of KIC\,10920813. Ca\,II H \& K line-core emission is also marginally detected, although the spectra are noisy below 4000\,\AA. Finally, we searched for the Li\,I line at 6707.8\,\AA, since lithium absorption in giants has been linked to tidal interactions in close binary systems \citep{Casey19}. We found no spectroscopic evidence of activity in KIC\,8129189, and neither star shows lithium enhancement.

\section{Photometric analysis}\label{sec:phot}

\subsection{Global asteroseismic parameters}

We derived global asteroseismic parameters from the eclipse-masked \textit{Kepler} light curves described in Sect.~\ref{sec:lcs}. Oscillations were detected using the envelope autocorrelation function (EACF) method developed by \citet{Mosser09}. We first applied the EACF to the Fourier transform of each light curve to obtain initial estimates of $\nu_{\rm max}$ and $\Delta\nu$ that are independent of the background model. Following \citet{Kallinger14}, we then modeled the background signal in the power density spectrum (PSD) as

\begin{equation}
\label{eq:background}
    S(\nu) = N(\nu)+\eta(\nu)[B(\nu)+G(\nu)],
\end{equation}
where the model consists of a noise function $N(\nu)$, a damping factor $\eta(\nu)$, the sum of three super-Lorentzian background components $B(\nu)$, and a Gaussian power excess $G(\nu)$ centered on $\nu_{\rm max}$ with height $H_{\rm max}$, given by

\begin{equation}
\label{eq:gaussian}
    G(\nu) = H_{\rm max}\exp\left[- \frac{(\nu-\nu_{\rm max})^{2}}{2 \sigma^{2}} \right].
\end{equation}

\begin{figure}[t]
\centering
\includegraphics[width=0.38\textwidth]{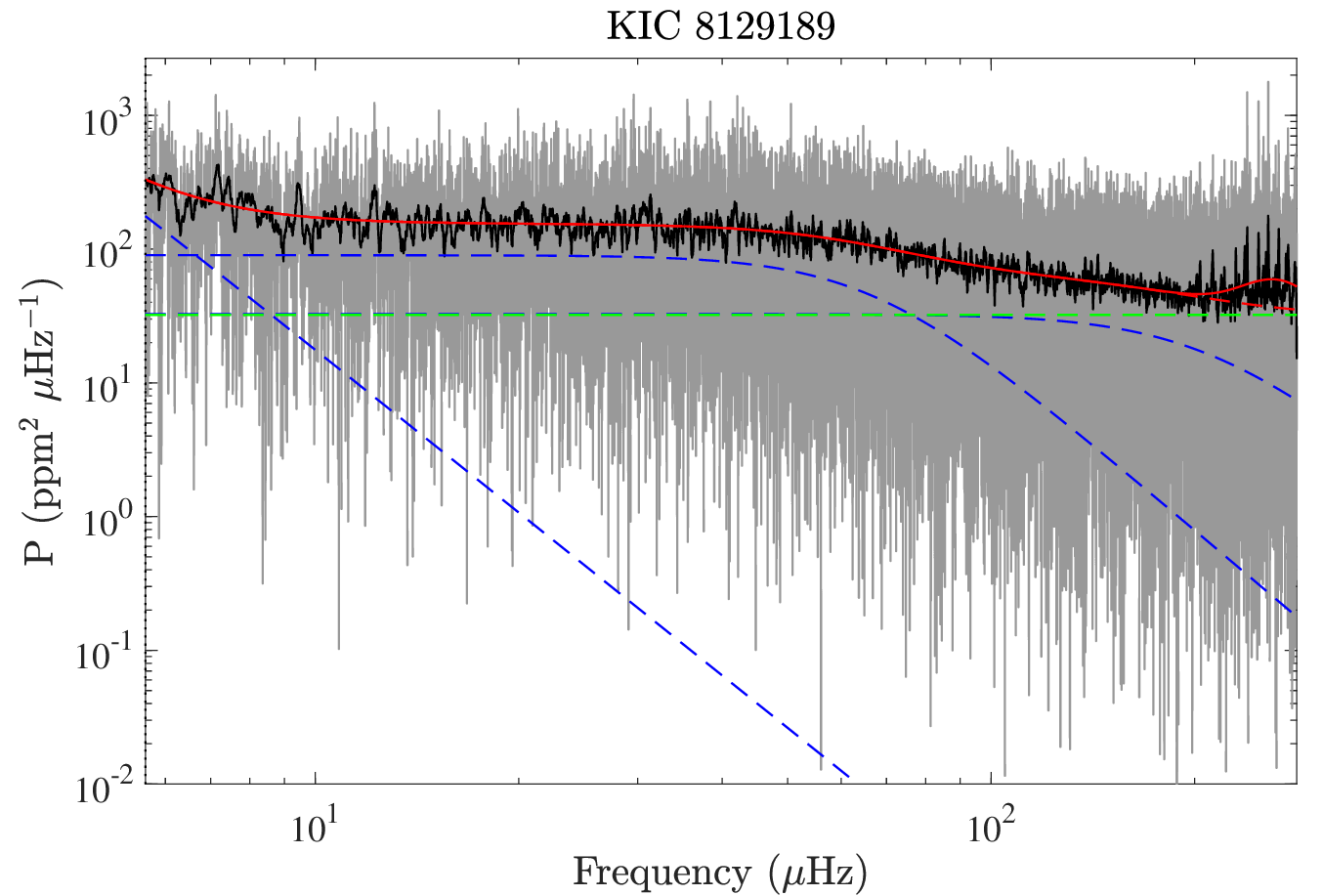}\\
\includegraphics[width=0.38\textwidth]{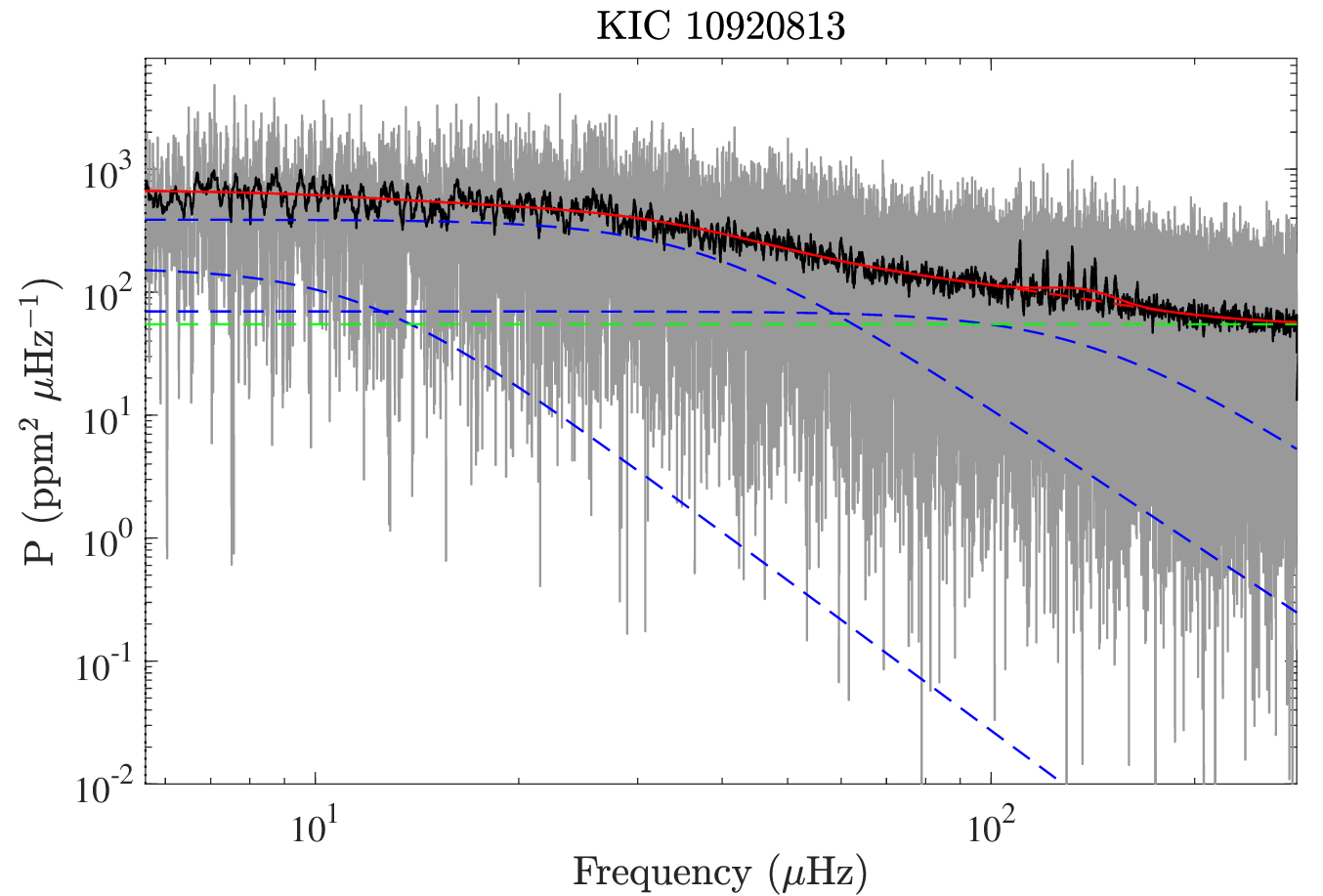}
\caption{Background and oscillation-envelope fits to the power density spectra of KIC\,8129189 (top) and KIC\,10920813 (bottom). The noise component $N(\nu)$ is shown by the green dashed line, and the three super-Lorentzian components are shown by the blue dashed lines. The combined background model is shown by the red dashed line, and the full model including the Gaussian oscillation-power excess is shown by the red solid line.}
\label{fig:background_fit}
\end{figure}

The components of the PSD fit are shown in Fig.~\ref{fig:background_fit}. We divided the PSD by the fitted background function to obtain a whitened PSD and then reapplied the EACF to refine the global seismic parameters. Oscillations were considered detected when a clear power excess was visible in the spectrum and the maximum EACF value was greater than 8 \citep{Mosser09}. The final global seismic parameters for the red-giant component of each system are listed in Table~\ref{tab:seis_param}.

\begin{figure}[t]
\centering
\includegraphics[width=0.45\textwidth]{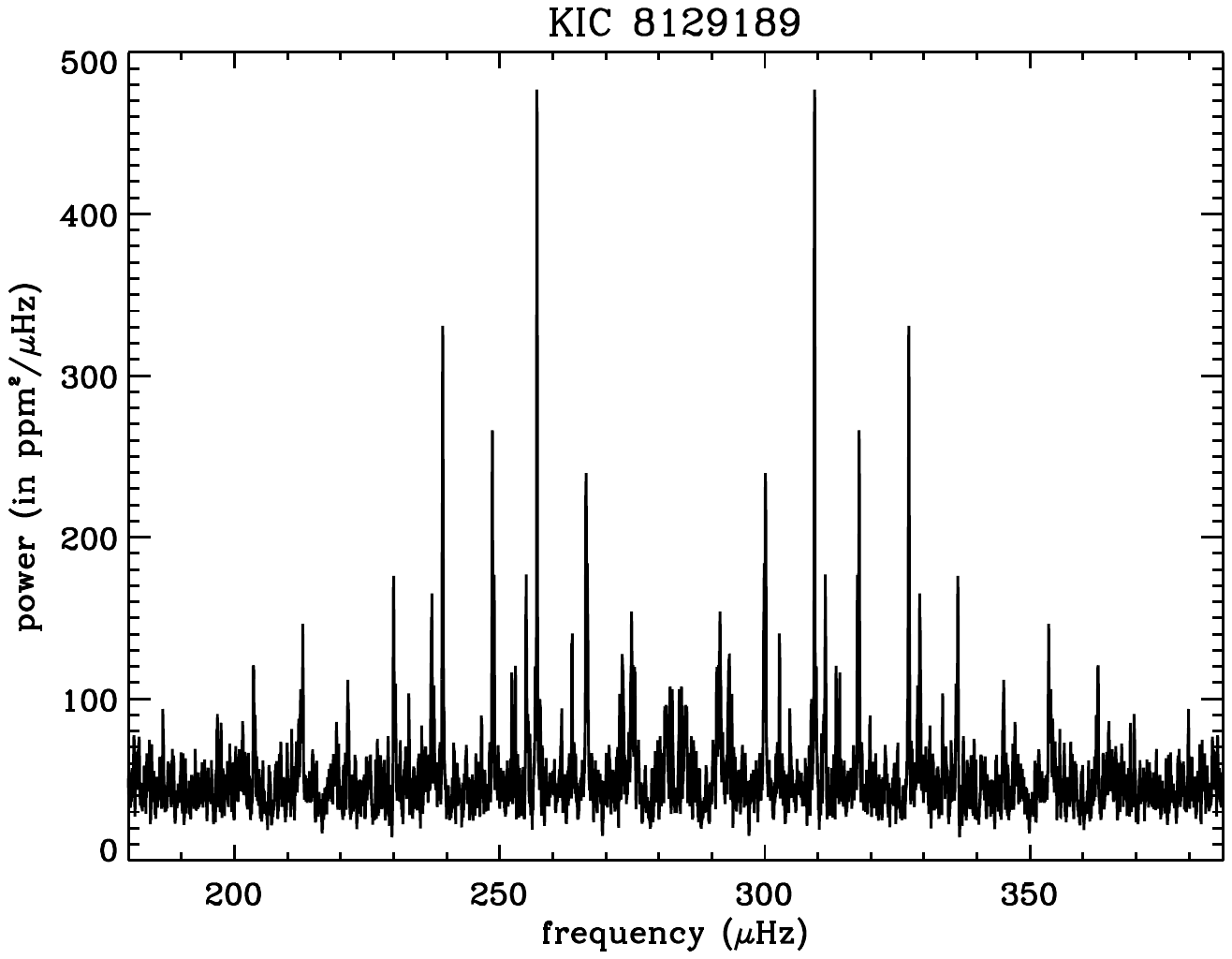}\\
\includegraphics[width=0.45\textwidth]{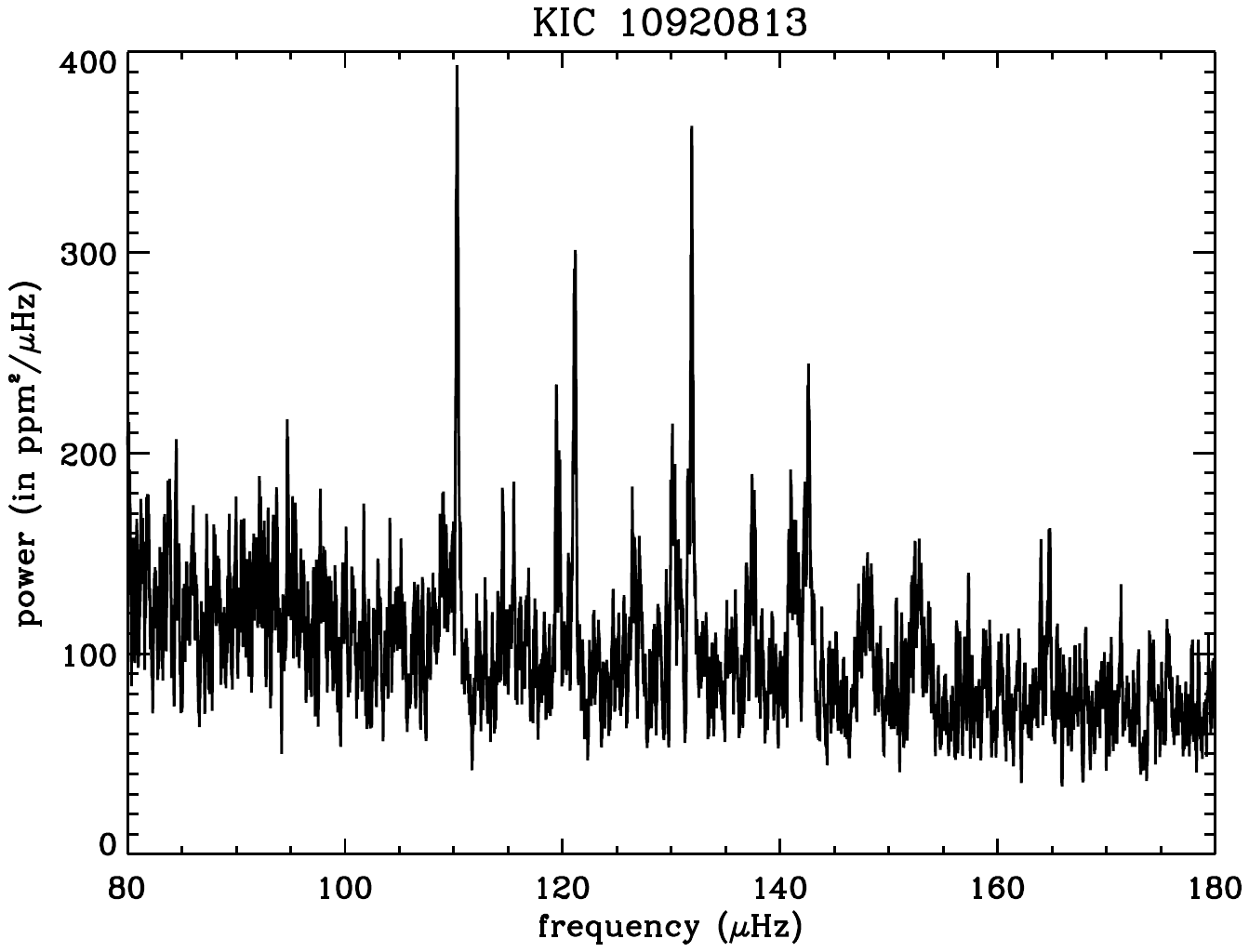}
\caption{Power density spectra of KIC\,8129189 (top) and KIC\,10920813 (bottom) around the oscillation power.  Spectra are smoothed to $0.15\,\mu{\rm Hz}$.}
\label{fig:power_spectra}
\end{figure}

\subsection{Individual oscillation modes and mixed modes}

\begin{figure}[t]
\centering
\includegraphics[width=0.34\textwidth]{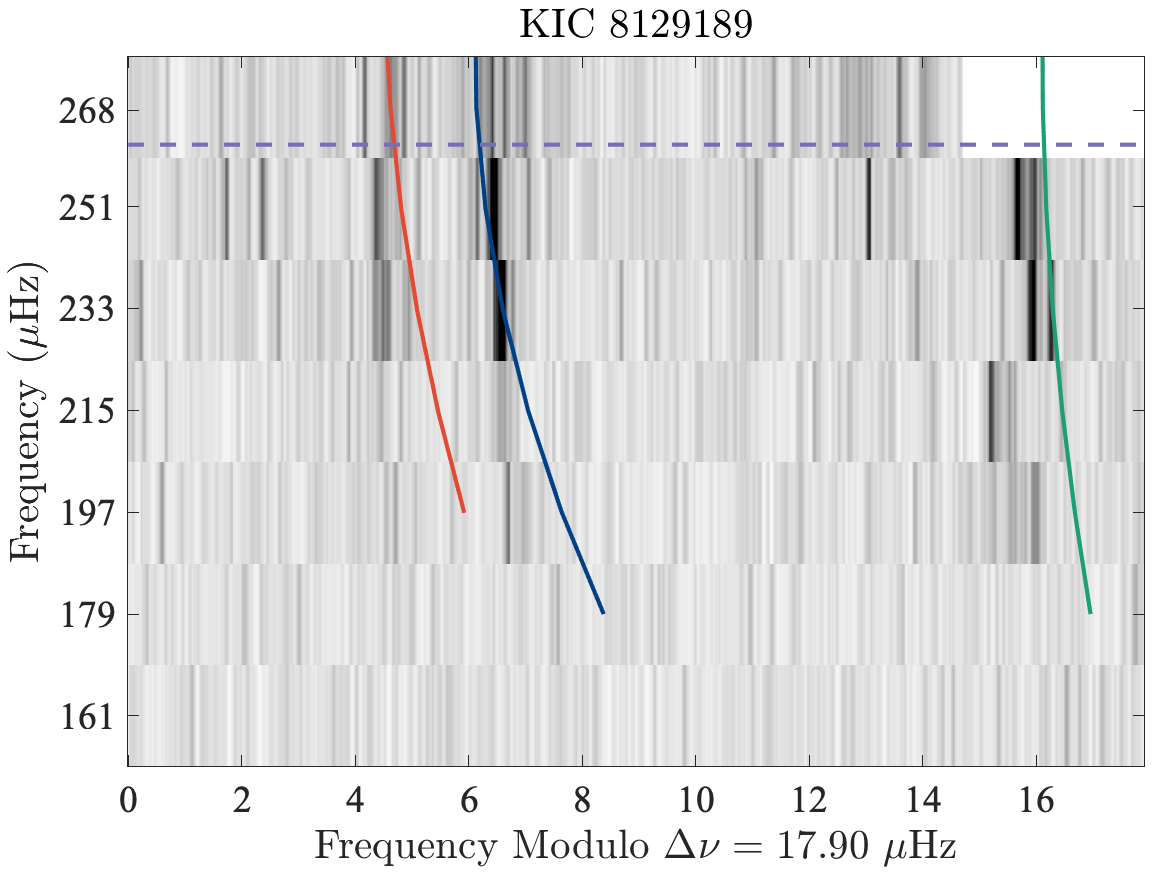}\\
\includegraphics[width=0.34\textwidth]{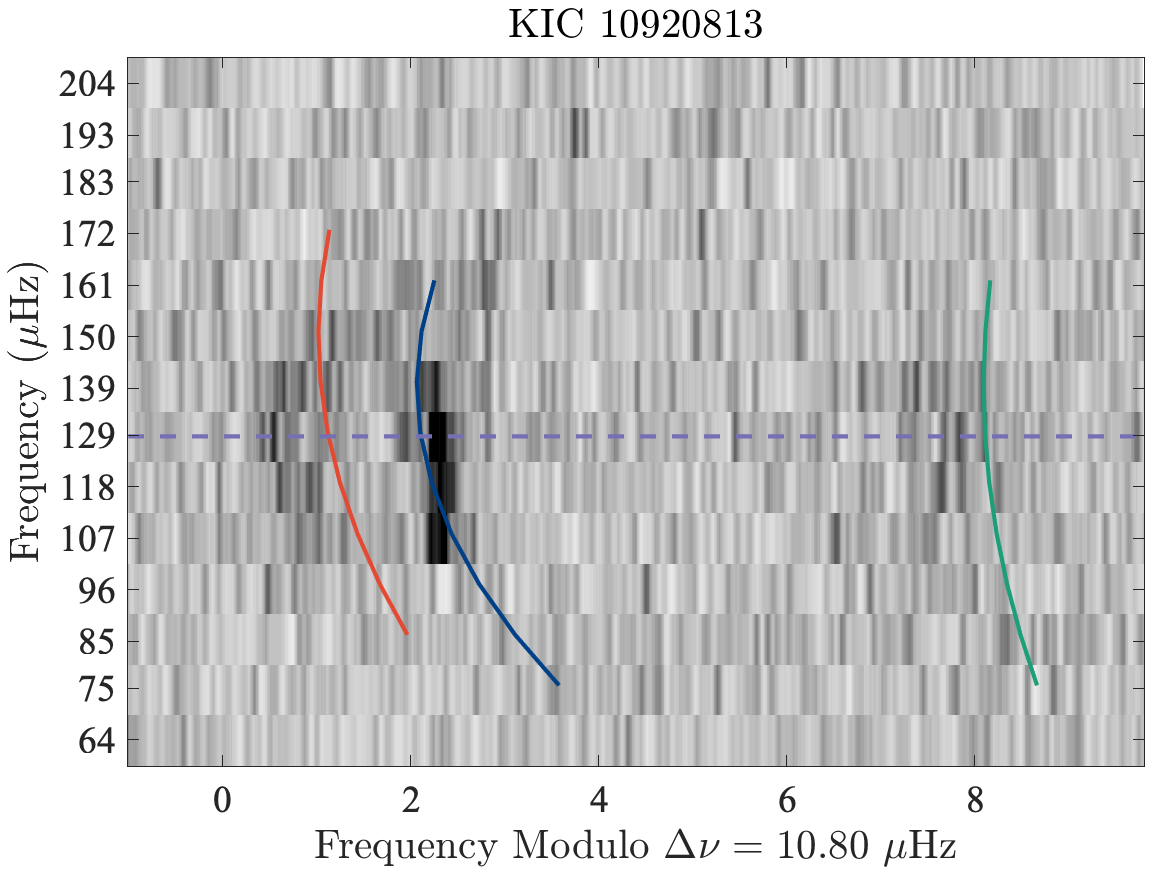}
\caption{Oscillation spectra of KIC\,8129189 (top) and KIC\,10920813 (bottom) shown as \'echelle diagrams. The power density spectra are folded over $\Delta\nu = 17.90\,\mu{\rm Hz}$ for KIC\,8129189 and $\Delta\nu = 10.80\,\mu{\rm Hz}$ for KIC\,10920813. Colored lines indicate the oscillation universal pattern and are used to identify the mode ridges: blue for $l=0$, green for $l=1$, and red for $l=2$. Power associated with modes of degree $l=3$ is visible approximately halfway between the $l=0$ and $l=1$ ridges. The horizontal dashed line marks the location of $\nu_{\rm max}$.}
\label{fig:echelle_diagram}
\end{figure}

We identified individual oscillation modes using \'echelle diagrams constructed by dividing each power density spectrum into segments of length $\Delta\nu$ and stacking them in frequency. In the universal pattern of red-giant oscillations \citep{Mosser13}, modes of a given angular degree $l$ appear as nearly vertical ridges in the \'echelle diagram. The resulting diagrams are shown in Fig.~\ref{fig:echelle_diagram}, with radial modes outlined in blue, dipole modes in green, and quadrupole modes in red.

Mixed dipole modes were identified in KIC\,8129189 following
\citet{Appourchaux20}. In this approach, the $\ell=1$ mixed-mode
pattern is described by asymptotic frequencies parameterized by the mean dipole-mode period spacing $\Delta\Pi_1$, coupling strength $q$, and g-mode phase offset $\epsilon_g$. The resulting mixed-mode parameters are listed in Table~\ref{tab:seis_param}. Based on the dipole-mode period spacing and large frequency separation, KIC\,8129189 is classified as a red-giant-branch star that has not yet ignited helium in its core \citep{Bedding11,Mosser12,Mosser14}. No mixed modes could be identified in KIC\,10920813.

The power spectra were globally fit using a procedure similar to that of \citet{Appourchaux12}, with several differences. The $l=2$ to $l=0$ mode-height ratio was fixed to 0.67, while the mode amplitudes of the $l=1$ modes were fitted independently. The background, consisting of three Lorentzian components and a white-noise component, was fit with a fixed exponent of 4. Two rotational splittings were fit: one for $l=1$ modes and one for $l=2$ modes. The inclination angle was common to all splittings, and the number of fitted $l=1$ modes depended on the detection of mixed modes. The fitting software is written in C++ and is publicly available on GitHub\footnote{\url{https://git.ias.u-psud.fr/plato_pdc/plato_wp37/wp372_msap3/msap3_04/mixfit}}. The p-mode parameter tables are given in Appendix~\ref{sec:appendosc}.  For KIC\,8129189, the peak of the oscillation power is close to the Nyquist frequency of 283.2 $\mu$Hz; only one peak was identified as an $l=1$ mode aliased from the Nyquist frequency \citep[e.g.,][]{Yu16,Liagre25}.

\begin{table}
\caption{Global seismic parameters of the red-giant components of KIC\,8129189 and KIC\,10920813.}
\label{tab:seis_param}
\footnotesize
\centering
\begin{tabular}{l l l l}
\hline
Parameter & Unit & KIC\,8129189 & KIC\,10920813\\
\hline
$\nu_{\rm max}$ & $\mu{\rm Hz}$ & $256.0 \pm 3.7$ & $130.0 \pm 2.3$ \\  
$\Delta\nu$ & $\mu{\rm Hz}$ & $17.90 \pm 0.05$ & $10.79 \pm 0.06$ \\  
$\Delta\Pi_1$ & s & 85 &  \\
$q$ & & 0.12 & \\
$d_{01}$ & & 0.026 & \\ 
$\epsilon_p$ & & 1.38 & \\
$\epsilon_g$ & & 0.36 & \\
$\delta\nu_{\mathrm{split},l=1}$ & nHz & $299.0 \pm 0.1$ & $424 \pm 7$\\
$\delta\nu_{\mathrm{split},l=2}$ & nHz & $340 \pm 7$ & $250 \pm 35$\\
$i$ & deg & $30 \pm 2$ & $35 \pm 5$\\
\hline
\end{tabular}
\tablefoot{Listed parameters are the frequency of maximum oscillation power $\nu_{\rm max}$, mean large frequency separation $\Delta\nu$, mean dipole-mode period spacing $\Delta\Pi_1$, coupling factor $q$, small frequency spacing $d_{01}$ relative to the large separation, p- and g-mode phase offsets $\epsilon_p$ and $\epsilon_g$, rotational splittings, and seismic inclination.}
\end{table}

\subsection{Photometric modulation, activity, and rotation}

As discussed in Sect.~\ref{sec:lcs}, KIC\,8129189 shows photometric variability dominated by a signal near the \textit{Kepler} spacecraft orbital period, whereas KIC\,10920813 shows significant periodic modulation with $P\approx138$ days, consistent with rotational modulation. Because no periodic photometric modulation attributable to stellar rotation was identified in KIC\,8129189, we are unable to provide an estimate of its surface rotation rate. If the $P\approx138\,{\rm d}$ modulation observed in KIC\,10920813 is attributed to rotation of the red-giant component, it corresponds to a surface rotation frequency of approximately $0.084\,\mu{\rm Hz}$. This is substantially lower than the fitted seismic rotational splittings of $0.424\,\mu{\rm Hz}$ for $\ell=1$ and $0.250\,\mu{\rm Hz}$ for $\ell=2$ (Table~\ref{tab:seis_param}). The origin of this discrepancy is unclear. Previous studies have demonstrated the complexities of inferring surface rotation from photometric variability \citep{Aigrain15}, while fitted rotational splitting and seismic inclination can also be correlated when the individual azimuthal components are not well resolved \citep{Ballot08}. The photometric variability index $S_{\rm ph}$, defined as the mean standard deviation of light-curve segments spanning five rotation periods following \citet{Mathur14}, is of order 1\% for KIC\,10920813. This places KIC\,10920813 at the high-activity end of the oscillating red-giant population in the $S_{\rm ph}$--rotation-period diagram of \citet{Gaulme20}, near the lower range of the non-oscillating active red giants. Unlike those non-oscillating systems, however, KIC\,10920813 retains clear solar-like oscillations with no evidence of strong suppression.

\section{Dynamical modeling}\label{sec:dyn}

Dynamical models provide the independent masses and radii that make these systems useful as asteroseismic benchmarks. For calibration purposes, the robustness of the benchmark values depends not only on the formal uncertainty of a single best-fitting binary model, but also on the stability of the inferred red-giant mass and radius under reasonable choices of model prescription and light-curve selection. This is especially important in the era of space-based photometry, where effects that were previously buried in the noise can appear routinely in high-precision eclipsing-binary light curves. This has motivated the development of models with improved numerical fidelity and more complete treatments of radiative and dynamical effects. We therefore used the dynamical analysis to test two sources of systematic uncertainty: the choice of binary-star modeling code and the choice of photometric data used in the fit.

To test these systematics, we modeled the light curves and radial-velocity curves of both systems with JKTEBOP \citep{Southworth13} and PHOEBE-2 \citep{Prsa05,Prsa16}. These codes represent two complementary approaches to detached eclipsing-binary modeling. JKTEBOP is fast and robust for detached systems whose components are close to spherical, while PHOEBE-2 was developed to model the richer light-curve physics made accessible by high-precision photometry, including geometry-dependent surface intensities, reflection and irradiation, gravity darkening, and Doppler boosting. To test the effect of light-curve selection, we compared solutions obtained from the full stitched light curves with solutions based on selected eclipse sets. The selected-eclipse approach was motivated by the time-variable photometric structure seen in the \textit{Kepler} light curves, particularly for KIC\,10920813, where activity-related variability can alter the local eclipse baseline and morphology. The detailed rationale for the eclipse selection is given in Sect.~\ref{sec:lc_selection}.

For clarity, we distinguish between the observational eclipse convention and the model labeling convention used in this work. In this work, we refer to the primary eclipse as the deeper minimum, when the hotter companion is eclipsed by the red giant. In the model setup, however, the red giant is labeled as star~A and the reference epoch is defined such that the shallower eclipse occurs at phase~0. Thus, in our ephemeris, the primary eclipse does not occur at phase~0.

JKTEBOP is a modern implementation of the Eclipsing Binary Orbit Program (EBOP) developed by \citet{Etzel81}, based on the Nelson-Davis-Etzel model \citep{Nelson72}. The stars are represented as biaxial spheroids for proximity effects, while eclipse shapes are computed by numerical integration over concentric annuli projected on the stellar disks. JKTEBOP provides several analytic limb-darkening laws and uses a Levenberg-Marquardt algorithm to determine the best-fitting solution \citep{Southworth13}. Because it is computationally efficient and well suited to detached binaries, JKTEBOP provides a useful baseline for exploring how the inferred dynamical parameters depend on light-curve selection.

PHOEBE-2 is a modern implementation of PHOEBE, whose original version was built on the Wilson-Devinney framework \citep{Wilson71,Prsa05,Prsa16}. PHOEBE-2 was motivated in part by the precision of modern space-based light curves, which revealed physical effects and subtle light-curve structure that could no longer be treated as negligible residuals. It represents stellar surfaces as equipotentials discretized by a triangular mesh, allowing geometry-dependent intensity variations to be computed across the visible stellar surfaces. In this framework, effects such as gravity darkening, reflection and irradiation, and Doppler boosting can be treated consistently when they become significant at modern photometric precision \citep{Prsa16}. PHOEBE-2 can also interpolate model-atmosphere intensities directly across the mesh rather than relying only on analytic limb-darkening prescriptions \citep{Prsa16}. This increased physical realism comes at substantially higher computational cost than JKTEBOP, but recent versions include a general inverse-problem framework with estimators, optimizers, and samplers \citep{Conroy20}. Comparing the JKTEBOP and PHOEBE-2 solutions therefore allows us to test whether the more physically flexible model yields materially different benchmark masses and radii for these well-detached systems.

\subsection{JKTEBOP models}\label{sec:jktebop}

We first modeled both systems with JKTEBOP to establish baseline detached-binary solutions and to test the sensitivity of the inferred parameters to the treatment of the photometric data. JKTEBOP is well suited to this comparison because both systems are well detached and the code is computationally efficient, allowing parallel analyses of the full stitched light curves and selected eclipse sets. For each system, the fitted data consisted of the \textit{Kepler} light curve, expressed in magnitudes, together with the radial-velocity curves of both components. Separate solutions were obtained for the full stitched light curve and for the selected eclipse set, while the same radial-velocity data were retained in both cases.

The JKTEBOP models adopted spherical stellar shapes, quadratic limb darkening for both stars, and fixed gravity-darkening and reflection coefficients. The photometric mass ratio used for the light-curve calculation was set to a negative value in the input files so that the stellar shapes were forced to remain spherical, appropriate for these detached systems. The gravity-darkening coefficients were fixed to 1.0 for both stars, and the reflected-light coefficients were fixed to zero. For the quadratic limb-darkening law, we fitted the first limb-darkening coefficient for both stars, labeled $A_{1}$ and $B_{1}$ in the JKTEBOP output, and fixed the second coefficient for both stars, labeled $A_{2}$ and $B_{2}$. We fixed $A_{2}=0.19$ and $B_{2}=0.32$ for KIC\,8129189, and $A_{2}=0.18$ and $B_{2}=0.32$ for KIC\,10920813. For the JKTEBOP fits, third light was fixed to the \textit{Kepler} contamination estimates, with $L_{3}=0.001$ for KIC\,8129189 and $L_{3}=0.020$ for KIC\,10920813 in both the full-light-curve and selected-eclipse solutions. Initial testing with third light left free showed a negligible effect on the orbital solution and inferred red-giant masses and radii.

Because both systems are eccentric, JKTEBOP was parameterized in terms of $e\cos\omega$ and $e\sin\omega$, both of which were adjusted. In total, 13 parameters were fitted in each solution: the surface-brightness ratio $J$, sum of fractional radii $r_{\rm A}+r_{\rm B}$, ratio of radii $k=r_{\rm B}/r_{\rm A}$, orbital inclination $i$, $e\cos\omega$, $e\sin\omega$, orbital period $P$, ephemeris timebase $t_{0}$, the two radial-velocity semi-amplitudes $K_{\rm A}$ and $K_{\rm B}$, the systemic velocity $\gamma$, and the two first limb-darkening coefficients $A_{1}$ and $B_{1}$. In both the JKTEBOP and PHOEBE-2 analyses, we fitted a single systemic velocity for each binary, shared by both components. Differential gravitational redshift and convective blueshift can introduce approximately constant component-dependent offsets in the measured radial velocities \citep{Einstein1952,Gray2009,Brogaard22}. Because these effects primarily introduce approximately constant velocity offsets rather than altering the phase-dependent radial-velocity variation, their impact on the inferred orbital parameters is expected to be small. Given the precision and phase coverage of our radial-velocity measurements, any resulting changes in the derived masses and radii are expected to be smaller than the benchmark uncertainties adopted in Sect.~\ref{sec:adopted_dyn}.

\begin{figure}[t]
\centering
\includegraphics[width=0.49\textwidth]{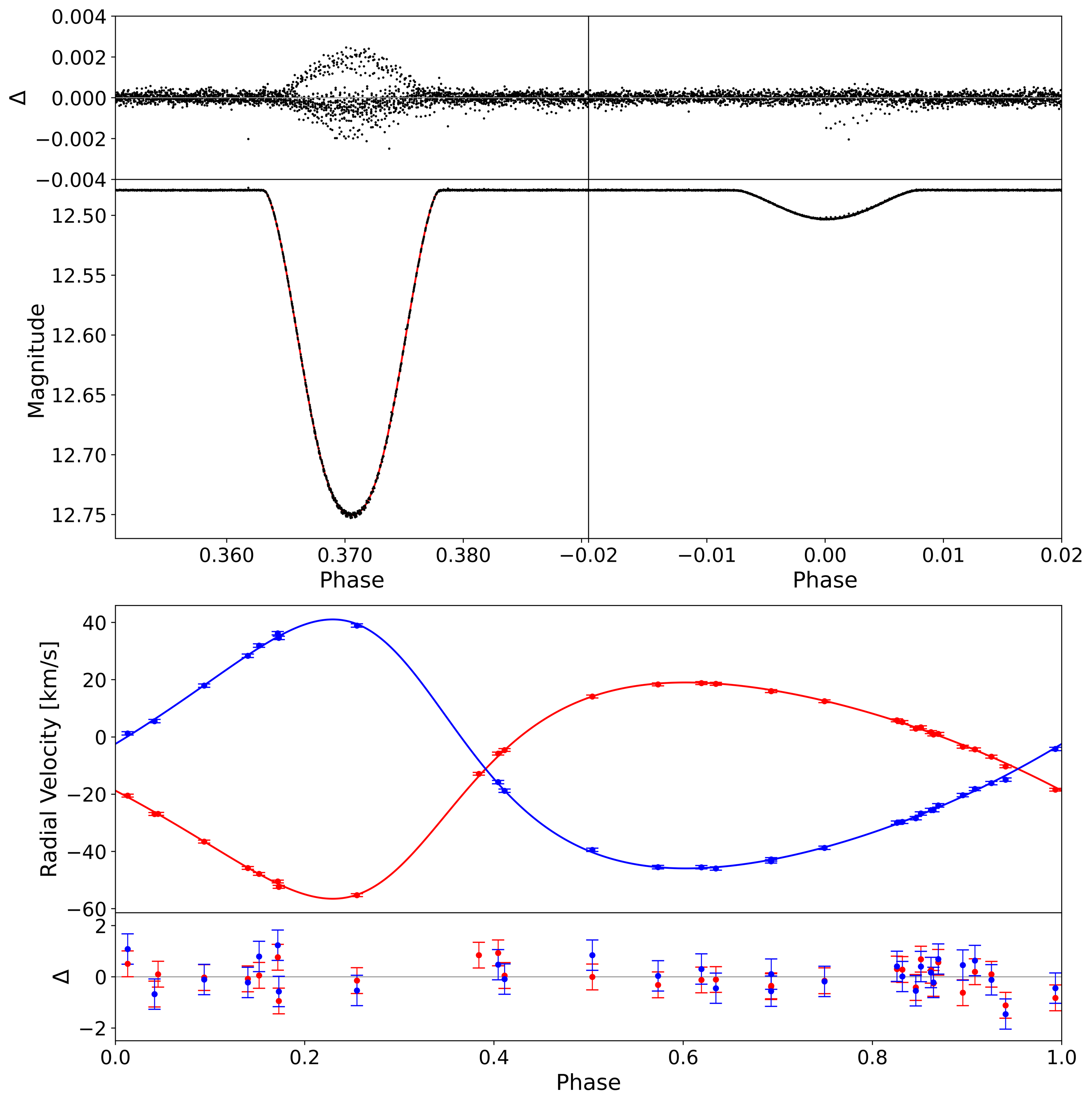}
\caption{Full-light-curve JKTEBOP solution for KIC\,8129189. Top panel: detrended \textit{Kepler} eclipse photometry, expressed in magnitudes, folded over the orbital period and centered on the primary and secondary eclipses. Black points show the detrended \textit{Kepler} data, and the red curve shows the best-fitting JKTEBOP model. Residuals, defined as light curve minus model, are shown in the upper residual panel and are expressed in mmag. Bottom panel: radial velocities of the red giant (red squares) and companion star (blue squares), folded over the orbital period. Radial-velocity residuals are shown in the lower residual panel. The adopted radial-velocity uncertainties are 0.5\,km\,s$^{-1}$ for the red giant and 1.0\,km\,s$^{-1}$ for the companion.}
\label{fig:jkte_8129}
\end{figure}

\begin{figure}[t]
\centering
\includegraphics[width=0.49\textwidth]{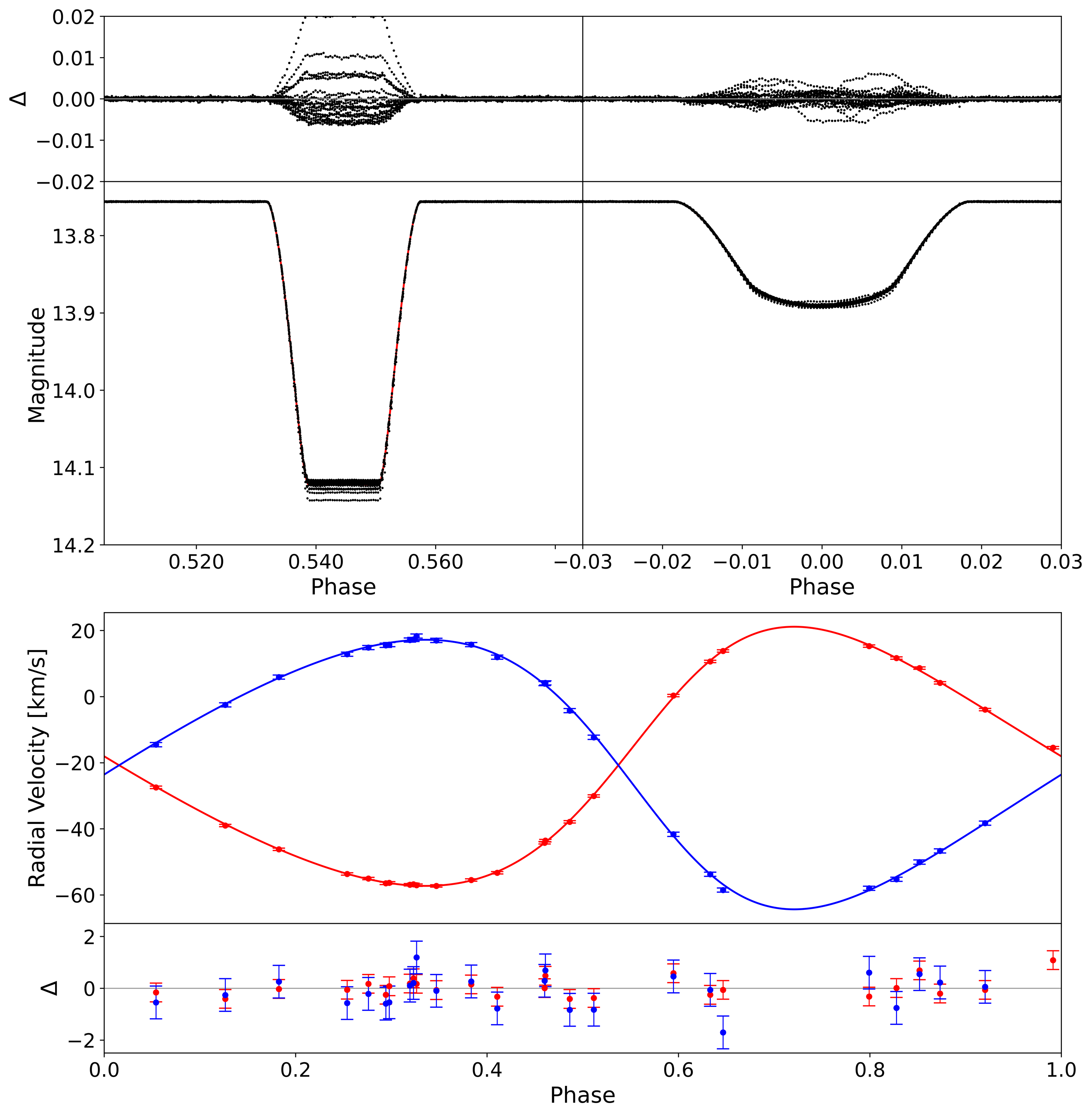}
\caption{Same as Fig.~\ref{fig:jkte_8129}, but for KIC\,10920813. The adopted radial-velocity uncertainties are 0.5\,km\,s$^{-1}$ for the red giant and 1.0\,km\,s$^{-1}$ for the companion.}
\label{fig:jkte_109}
\end{figure}

Parameter uncertainties were estimated with the JKTEBOP bootstrap routine, TASK~7, using 1000 bootstrap realizations for each fitted configuration. In each realization, the light curve and radial-velocity measurements were resampled and refitted, and the resulting parameter distributions were adopted as the basis for the quoted uncertainties. The full-light-curve JKTEBOP solutions are shown in Figs.~\ref{fig:jkte_8129} and \ref{fig:jkte_109}. The corresponding fitted and derived parameters are summarized in Tables~\ref{tab:jktebop_fit_full} and \ref{tab:jktebop_deduced_full}. The selected-eclipse JKTEBOP solutions are presented in Appendices~\ref{sec:append_812} and \ref{sec:append_109}.

\begin{table*}[t]
\caption{JKTEBOP fitted parameters for the full stitched light-curve solutions.}
\label{tab:jktebop_fit_full}
\centering
\footnotesize
\begin{tabular}{lcc}
\hline
Parameter & KIC\,8129189 & KIC\,10920813\\
\hline
$J$ & $2.550 \pm 0.094$ & $2.986 \pm 0.067$\\
$r_{\rm A}+r_{\rm B}$ & $0.06939 \pm 0.00014$ & $0.09687 \pm 0.00028$\\
$k=r_{\rm B}/r_{\rm A}$ & $0.3400 \pm 0.0018$ & $0.3709 \pm 0.0013$\\
$i$ [deg] & $87.191 \pm 0.017$ & $89.99 \pm 0.21$\\
$e\cos\omega$ & $-0.20070 \pm 0.00013$ & $0.0690166 \pm 0.0000093$\\
$e\sin\omega$ & $-0.1963 \pm 0.0031$ & $-0.18113 \pm 0.00051$\\
$P$ [days] & $53.6470701 \pm 0.0000041$ & $53.740819 \pm 0.000014$\\
$t_{0}$ [days] & $133.09234 \pm 0.00023$ & $155.55173 \pm 0.00026$\\
$K_{\rm A}$ [km\,s$^{-1}$] & $37.783 \pm 0.078$ & $39.2111 \pm 0.0022$\\
$K_{\rm B}$ [km\,s$^{-1}$] & $43.492 \pm 0.088$ & $40.7728 \pm 0.0011$\\
$\gamma$ [km\,s$^{-1}$] & $-11.20568 \pm 0.00081$ & $-20.7705 \pm 0.0014$\\
Limb darkening $A_{1}$ & $0.495 \pm 0.045$ & $0.500 \pm 0.014$\\
Limb darkening $B_{1}$ & $0.361 \pm 0.028$ & $0.419 \pm 0.055$\\
Limb darkening $A_{2}$ & 0.19 fixed & 0.18 fixed\\
Limb darkening $B_{2}$ & 0.32 fixed & 0.32 fixed\\
Third light $L_{3}$ & 0.001 fixed & 0.020 fixed\\
Reduced $\chi^2$ & 2.97 & 39.01\\
\hline
\end{tabular}
\tablefoot{Star~A denotes the red-giant component and star~B denotes the companion. $J$ is the surface-brightness ratio, $r_{\rm A}+r_{\rm B}$ is the sum of fractional radii, $k$ is the radius ratio, $A_{1}$ and $B_{1}$ are the fitted first limb-darkening coefficients, $A_{2}$ and $B_{2}$ are the fixed second limb-darkening coefficients, and $L_{3}$ is the fixed third-light contribution.}
\end{table*}

\begin{table}[t]
\caption{Derived JKTEBOP parameters for the full stitched light-curve solutions.}
\label{tab:jktebop_deduced_full}
\centering
\footnotesize
\begin{tabular}{lcc}
\hline
Parameter & KIC\,8129189 & KIC\,10920813\\
\hline
$e$ & $0.28075$ & $0.19384$\\
$\omega$ [deg] & $224.37$ & $290.86$\\
$M_{\rm A}$ [$M_\odot$] & $1.4170 \pm 0.0058$ & $1.37137 \pm 0.00029$\\
$M_{\rm B}$ [$M_\odot$] & $1.2310 \pm 0.0051$ & $1.31884 \pm 0.00024$\\
$R_{\rm A}$ [$R_\odot$] & $4.2886 \pm 0.0098$ & $5.889 \pm 0.012$\\
$R_{\rm B}$ [$R_\odot$] & $1.4581 \pm 0.0080$ & $2.184 \pm 0.011$\\
$\log g_{\rm A}$ [cgs] & 3.3248 & 3.0351\\
$\log g_{\rm B}$ [cgs] & 4.2008 & 3.8796\\
$\rho_{\rm A}$ [$\rho_\odot$] & 0.01797 & 0.00671\\
$\rho_{\rm B}$ [$\rho_\odot$] & 0.39710 & 0.12653\\
\hline
\end{tabular}
\tablefoot{Star~A denotes the red-giant component and star~B denotes the companion. Uncertainties for the masses and radii are the one-sigma TASK~7 bootstrap values; the corresponding central 68.27\% intervals are given in Appendix~\ref{sec:append_dyn}.}
\end{table}

\subsection{PHOEBE-2 models}

We next modeled both binaries with PHOEBE-2 in two configurations. The first used the same full stitched \textit{Kepler} light curves as in the JKTEBOP full-light-curve analysis, but converted from magnitudes to relative flux, together with the radial-velocity curves of both components. The second used the same selected eclipse photometry as the selected-eclipse JKTEBOP analysis, described in Sect.~\ref{sec:lc_selection}, and the same radial velocities as the full-light-curve fits. Together with the JKTEBOP models, these PHOEBE calculations allow us to separate two effects that would otherwise be entangled: the choice of binary-star model and the choice of photometric data set.

The PHOEBE models used Roche geometry with a marching triangular mesh of 1500 surface elements per star. Passband intensities and limb darkening were computed by interpolation from the CK2004 atmosphere tables for both components \citep{Castelli04}, with the light curves evaluated using photon-counting intensity weighting in the \textit{Kepler} passband. Irradiation was treated with the \citet{Horvat19} prescription, eclipses were computed with the native PHOEBE eclipse algorithm, and the radial velocities were computed with the flux-weighted option. The PHOEBE passband-luminosity mode, \texttt{pblum\_mode}, was set to \texttt{component-coupled}, and no explicit starspot model was included. Light-travel-time effects were disabled. As in the JKTEBOP analysis, star~A denotes the red-giant component. In contrast to the JKTEBOP fits, for which third light was fixed to the \textit{Kepler} contamination estimates, third light was sampled as a fractional light contribution in all PHOEBE-2 fits, though in practice it primarily serves as a nuisance parameter at the low contamination levels relevant to these systems.

Initial parameter values were taken from the corresponding JKTEBOP solutions so that the PHOEBE optimization began in the same region of parameter space. The sampled parameterization was chosen to parallel the JKTEBOP description as closely as possible while using PHOEBE's internal constraint system. We sampled the sum of fractional radii, radius ratio, inclination, eccentricity $e$, argument of periastron $\omega$, mass ratio $q=M_{\rm B}/M_{\rm A}$, projected semimajor axis $a\sin i$, orbital period, time of superior conjunction $t_{0,\mathrm{supconj}}$, effective-temperature ratio $T_{\mathrm{eff,B}}/T_{\mathrm{eff,A}}$, systemic velocity, third-light fraction $l_{3,\mathrm{frac}}$, and the light-curve jitter term $\sigma_{\ln f}$. The PHOEBE constraints were flipped so that $a\sin i$ rather than $a$ was sampled directly, and the secondary effective temperature was solved from the sampled temperature ratio. Walkers were initialized from narrow Gaussian perturbations around the seed solution, with the jitter term initialized from a uniform distribution on $-10 < \sigma_{\ln f} < 0$. \texttt{emcee} was used to perform Markov chain Monte Carlo sampling until convergence was identified from autocorrelation and trace plots, after approximately 10\,000 iterations \citep{ForemanMackey13}. Posterior samples were extracted after discarding the first 1000 steps as burn-in.

\begin{figure}[t]
\centering
\includegraphics[width=0.49\textwidth]{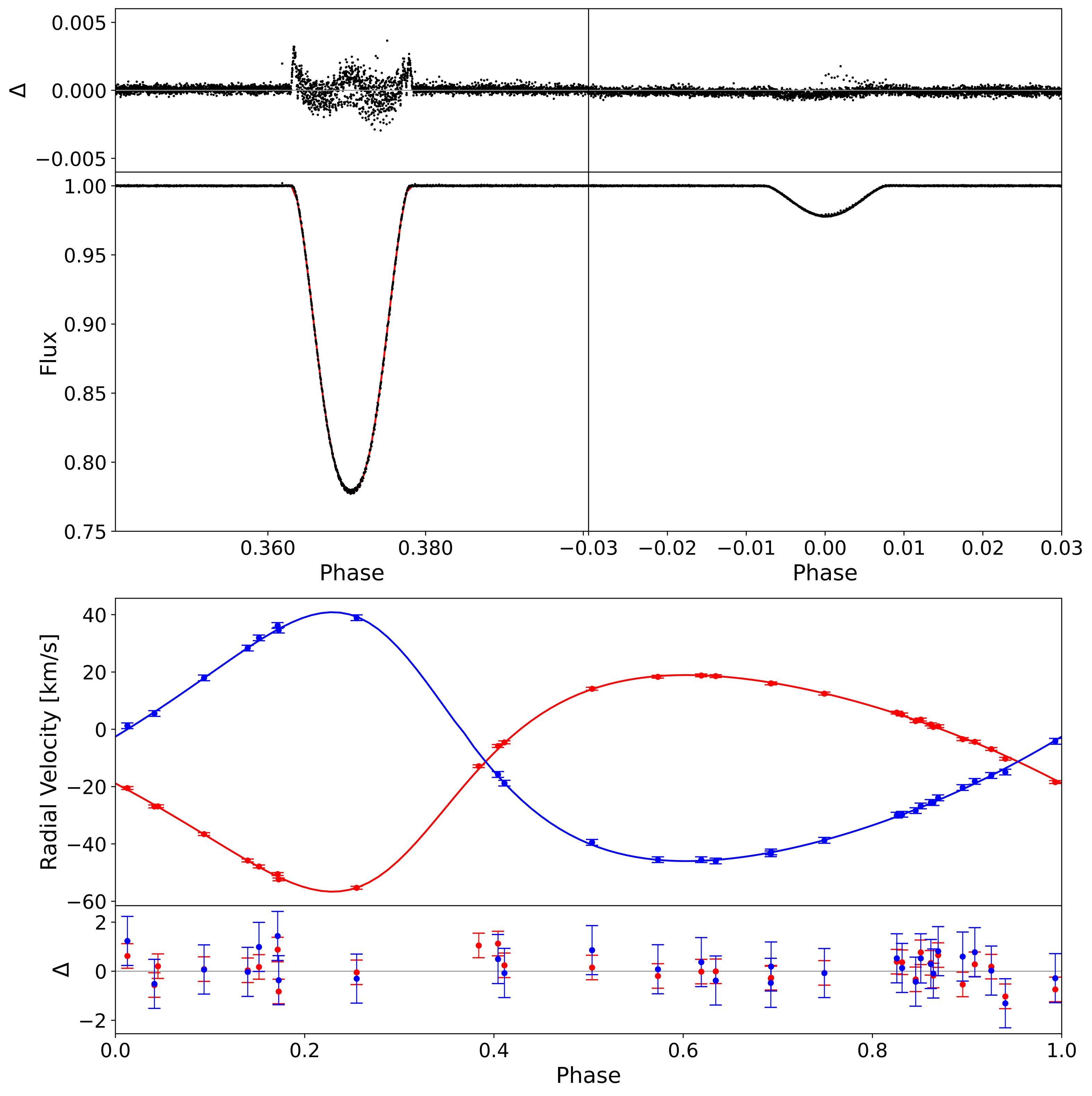}
\caption{Full-light-curve PHOEBE-2 light-curve and radial-velocity fit for KIC\,8129189. The figure follows the same layout as the JKTEBOP fit figures: the upper panels show the eclipse data, model, and residuals, while the lower panels show the radial-velocity curves and residuals for the red giant and companion. This model uses the full stitched \textit{Kepler} light curve and the same radial-velocity data as the full-light-curve JKTEBOP solution.}
\label{fig:phoebe_full_8129}
\end{figure}

\begin{figure}[t]
\centering
\includegraphics[width=0.49\textwidth]{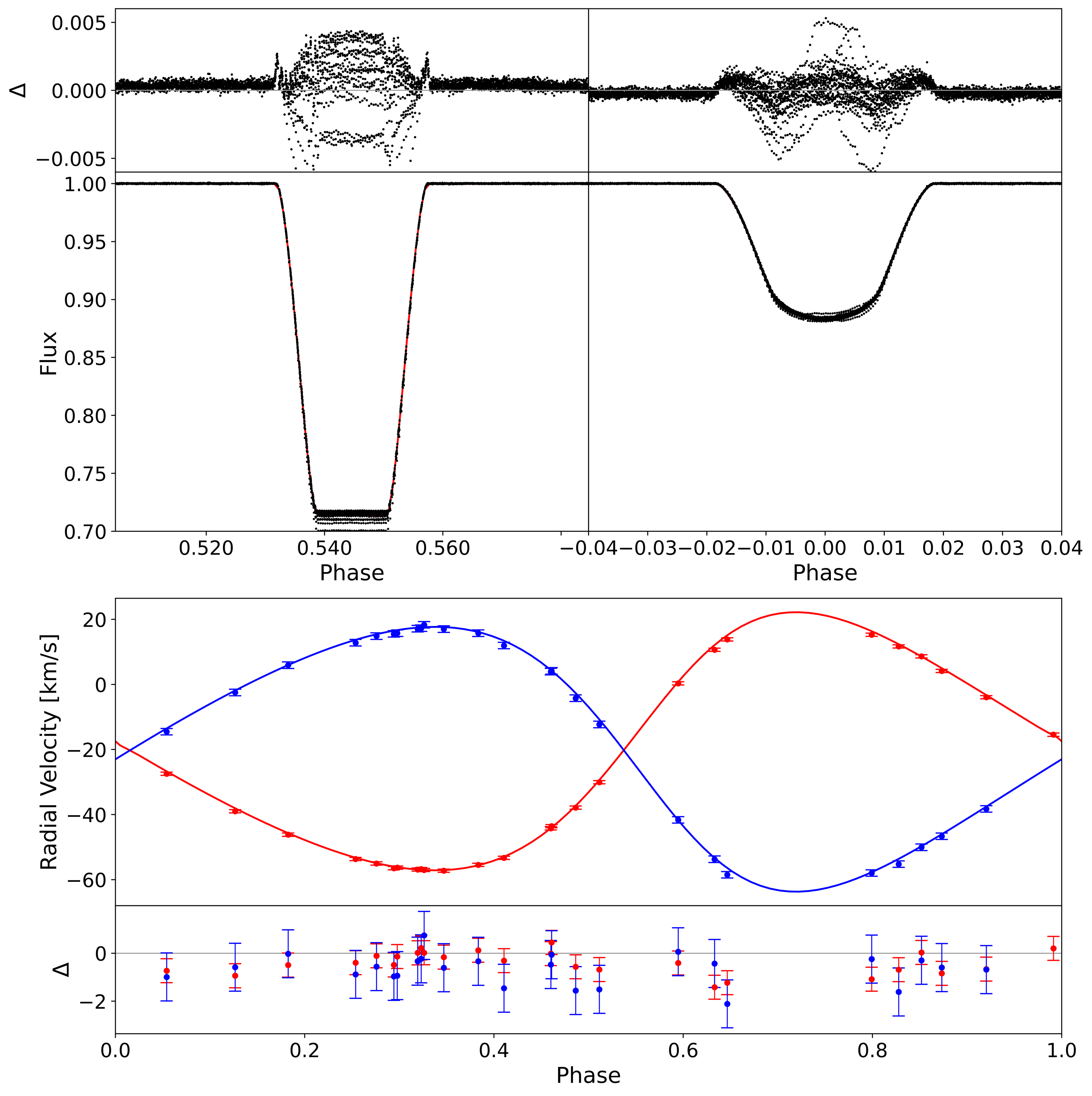}
\caption{Same as Fig.~\ref{fig:phoebe_full_8129}, but for KIC\,10920813. The comparison with Fig.~\ref{fig:jkte_109} shows the full-light-curve PHOEBE-2 and JKTEBOP solutions for the same system, while the selected-eclipse diagnostics in Appendix~\ref{sec:append_dyn} show the corresponding restricted-eclipse fits.}
\label{fig:phoebe_full_109}
\end{figure}

The full-light-curve PHOEBE-2 fits are shown in Figs.~\ref{fig:phoebe_full_8129} and \ref{fig:phoebe_full_109}.  Posterior means and standard deviations for the full-light-curve PHOEBE-2 sampled and propagated quantities are summarized in Tables~\ref{tab:phoebe_fit_full} and \ref{tab:phoebe_deduced_full}. The selected-eclipse PHOEBE-2 solutions are in  Appendices~\ref{sec:append_812} and \ref{sec:append_109}. The selected-eclipse PHOEBE and selected-eclipse JKTEBOP diagnostic figures are provided in Appendix~\ref{sec:append_dyn}; together, these figures document the four model and data configurations used in the conservative benchmark envelope.

\begin{table*}[t]
\caption{Full-light-curve PHOEBE-2 sampled parameters.}
\label{tab:phoebe_fit_full}
\centering
\footnotesize
\begin{tabular}{lcc}
\hline
Parameter & KIC\,8129189 & KIC\,10920813\\
\hline
$r_{\rm A}+r_{\rm B}$ & $0.06926 \pm 0.00009$ & $0.09698 \pm 0.00004$\\
$k=r_{\rm B}/r_{\rm A}$ & $0.342 \pm 0.002$ & $0.3653 \pm 0.0001$\\
$i$ [deg] & $87.194 \pm 0.004$ & $89.93 \pm 0.04$\\
$e$ & $0.2805 \pm 0.0004$ & $0.1993 \pm 0.0003$\\
Argument of periastron $\omega$ [deg] & $224.3 \pm 0.1$ & $290.24 \pm 0.03$\\
Mass ratio $q=M_{\rm B}/M_{\rm A}$ & $0.870 \pm 0.003$ & $0.97 \pm 0.02$\\
$a\sin i$ [$R_\odot$] & $82.68 \pm 0.03$ & $83.67 \pm 0.04$\\
$P$ [days] & $53.64716 \pm 0.00002$ & $53.7408177 \pm 0.0000001$\\
$t_{0,\mathrm{supconj}}$ [days] & $133.092652 \pm 0.000001$ & $155.551747 \pm 0.000003$\\
$T_{\mathrm{eff,B}}/T_{\mathrm{eff,A}}$ & $1.268 \pm 0.002$ & $1.2887 \pm 0.0002$\\
$\gamma$ [km\,s$^{-1}$] & $-11.31 \pm 0.01$ & $-20.23 \pm 0.26$\\
Third-light fraction $l_{3,\mathrm{frac}}$ & $0.0024 \pm 0.0006$ & $0.01225 \pm 0.00005$\\
$\sigma_{\ln f}$ & $-7.82 \pm 0.02$ & $-6.614 \pm 0.007$\\
\hline
\end{tabular}
\tablefoot{Values are posterior means and standard deviations. Star~A denotes the red-giant component. $l_{3,\mathrm{frac}}$ is the sampled fractional third-light contribution, and $\sigma_{\ln f}$ is the light-curve jitter term.}
\end{table*}

\begin{table}[t]
\caption{Full-light-curve PHOEBE-2 propagated parameters.}
\label{tab:phoebe_deduced_full}
\centering
\footnotesize
\begin{tabular}{lcc}
\hline
Parameter & KIC\,8129189 & KIC\,10920813\\
\hline
$M_{\rm A}+M_{\rm B}$ [$M_\odot$] & $2.645 \pm 0.003$ & $2.721 \pm 0.004$\\
$M_{\rm A}$ [$M_\odot$] & $1.4140 \pm 0.0003$ & $1.38 \pm 0.01$\\
$M_{\rm B}$ [$M_\odot$] & $1.231 \pm 0.003$ & $1.34 \pm 0.02$\\
$R_{\rm A}$ [$R_\odot$] & $4.272 \pm 0.002$ & $5.943 \pm 0.003$\\
$R_{\rm B}$ [$R_\odot$] & $1.462 \pm 0.006$ & $2.171 \pm 0.001$\\
$a$ [$R_\odot$] & $82.78 \pm 0.03$ & $83.67 \pm 0.04$\\
$r_{\rm A}$ & $0.05160 \pm 0.00002$ & $0.07103 \pm 0.00003$\\
$r_{\rm B}$ & $0.01766 \pm 0.00008$ & $0.02595 \pm 0.00001$\\
\hline
\end{tabular}
\tablefoot{Values are posterior means and standard deviations. Star~A denotes the red-giant component.}
\end{table}

\subsection{Comparison of dynamical solutions}

The JKTEBOP and PHOEBE-2 analyses provide four model and data configurations for each system: full-light-curve JKTEBOP, selected-eclipse JKTEBOP, full-light-curve PHOEBE-2, and selected-eclipse PHOEBE-2. These solutions allow us to assess two sources of modeling uncertainty: the treatment of the photometric time series and the choice of binary-star model. Their central mass and radius values are listed in Table~\ref{tab:dynamical_model_comparison}, and their offsets from the adopted conservative benchmark values are shown in Appendix~\ref{sec:append_offsets}, Fig.~\ref{fig:dynamical_model_comparison}.

\begin{table*}[t]
\caption{Central dynamical masses and radii from the four model and data configurations.}
\label{tab:dynamical_model_comparison}
\centering
\footnotesize
\begin{tabular}{llcccc}
\hline
System & Parameter & JKTEBOP full & JKTEBOP selected & PHOEBE full & PHOEBE selected\\
\hline
KIC\,8129189 & $M_{\rm A}$ [$M_\odot$] & $1.417$ & $1.421$ & $1.414$ & $1.410$\\
KIC\,8129189 & $M_{\rm B}$ [$M_\odot$] & $1.231$ & $1.234$ & $1.231$ & $1.226$\\
KIC\,8129189 & $R_{\rm A}$ [$R_\odot$] & $4.289$ & $4.242$ & $4.272$ & $4.255$\\
KIC\,8129189 & $R_{\rm B}$ [$R_\odot$] & $1.458$ & $1.488$ & $1.462$ & $1.448$\\
\hline
KIC\,10920813 & $M_{\rm A}$ [$M_\odot$] & $1.371$ & $1.372$ & $1.378$ & $1.383$\\
KIC\,10920813 & $M_{\rm B}$ [$M_\odot$] & $1.319$ & $1.319$ & $1.343$ & $1.320$\\
KIC\,10920813 & $R_{\rm A}$ [$R_\odot$] & $5.889$ & $5.875$ & $5.943$ & $5.931$\\
KIC\,10920813 & $R_{\rm B}$ [$R_\odot$] & $2.184$ & $2.219$ & $2.171$ & $2.180$\\
\hline
\end{tabular}
\tablefoot{The four configurations used to define the conservative benchmark uncertainties are the full stitched light curve and selected-eclipse solutions for each modeling code. Star~A denotes the red-giant component.}
\end{table*}

\subsubsection{Effect of light-curve selection}\label{sec:lc_selection}

One of the main goals of this work is to determine whether the handling of the photometric time series materially affects the inferred dynamical parameters. We therefore compared full-light-curve and selected-eclipse solutions for both JKTEBOP and PHOEBE-2. The full-light-curve solutions use the stitched \textit{Kepler} time series, while the selected-eclipse solutions emphasize local eclipse morphology at epochs chosen to minimize apparent spot-related effects.

This comparison is especially relevant for KIC\,10920813, where the secondary-eclipse depth varies measurably from epoch to epoch (Appendix~\ref{sec:append_dyn}, Fig.~\ref{fig:var_depth}). These variations are likely caused by changes in the visible surface-brightness distribution of the red giant. In principle, such variations can bias the inferred eclipse geometry if all eclipses are modeled simultaneously with a single spot-free prescription. KIC\,8129189 does not show comparably strong secondary-eclipse-depth variations, but it was processed in the same way so that the analysis of the two binaries remains homogeneous.

For KIC\,8129189, changing from the full-light-curve to the selected-eclipse solutions produces small but measurable shifts in the red-giant parameters. In PHOEBE-2, the red-giant mass changes from $1.414$ to $1.410\,M_\odot$, and the red-giant radius changes from $4.272$ to $4.255\,R_\odot$, corresponding to shifts of 0.3\% and 0.4\%, respectively. In JKTEBOP, the red-giant mass changes from $1.417$ to $1.421\,M_\odot$, while the red-giant radius changes from $4.289$ to $4.242\,R_\odot$, corresponding to shifts of 0.3\% and 1.1\%, respectively. The companion radius and other nuisance or companion-dependent quantities show somewhat larger sensitivity to eclipse selection than the red-giant mass and radius.

For KIC\,10920813, the selected-eclipse analysis was motivated by the observed eclipse-depth variations, but these variations were not found to produce a large shift in the inferred red-giant mass or radius. The JKTEBOP full-light-curve solution has a reduced $\chi^2$ of 39.01, whereas the selected-eclipse solution has a reduced $\chi^2$ of 1.84, indicating that a single spot-free geometry cannot reproduce all eclipse morphologies simultaneously. Nevertheless, the JKTEBOP red-giant mass and radius change only from $1.371$ to $1.372\,M_\odot$ and from $5.889$ to $5.875\,R_\odot$, corresponding to shifts of less than 0.1\% and 0.2\%, respectively. The PHOEBE-2 full-light-curve and selected-eclipse solutions similarly change the red-giant mass from $1.378$ to $1.383\,M_\odot$ and the red-giant radius from $5.943$ to $5.931\,R_\odot$, corresponding to shifts of 0.4\% and 0.2\%, respectively.

The dominant effect of light-curve selection is not the same for all quantities. For KIC\,10920813, selecting less spot-affected eclipses greatly improves the goodness of fit and reduces the impact of time-variable eclipse morphology, but the red-giant mass and radius remain nearly unchanged. This suggests that the spot-related variations primarily affect the surface-brightness distribution, local eclipse depth, and nuisance or companion-dependent parameters, while the red-giant radius remains anchored by the eclipse duration, ingress and egress morphology, and radial-velocity orbit. For KIC\,8129189, where the eclipse morphology is more stable, the selected-eclipse and full-light-curve solutions still differ at the sub-percent to percent level, showing that light-curve selection contributes to the benchmark uncertainty even when obvious eclipse-depth variations are not present. The key result from this study is that although light-curve selection can affect fit quality and some nuisance or companion-dependent parameters, the benchmark red-giant mass and radius are stable at the level relevant for the scaling-relation calibration.

\subsubsection{Effect of model prescription}\label{sec:model_selection}

We also compared the JKTEBOP and PHOEBE-2 solutions to assess whether the choice of binary-star model materially changes the inferred benchmark masses and radii. This comparison is important because the two codes use different physical approximations. JKTEBOP assumes detached, nearly spherical stars and uses analytic limb-darkening prescriptions, whereas PHOEBE-2 uses a Roche-geometry mesh with atmosphere-table intensities and a more flexible treatment of surface-intensity effects.

For KIC\,8129189, the JKTEBOP and PHOEBE-2 solutions give consistent red-giant masses and radii. Comparing the two codes at fixed light-curve selection, the full-light-curve red-giant radii differ by 0.4\%, while the selected-eclipse red-giant radii differ by 0.3\%. The corresponding red-giant mass differences are 0.2\% for the full-light-curve solutions and 0.8\% for the selected-eclipse solutions.

For KIC\,10920813, the model prescription produces somewhat larger shifts in the red-giant benchmark quantities than the light-curve selection does. Comparing the two codes at fixed light-curve selection, the full-light-curve red-giant radii differ by about 0.9\%, and the selected-eclipse radii differ by about 0.9\%. The corresponding mass differences are about 0.5\% for the full-light-curve solutions and 0.8\% for the selected-eclipse solutions. The model dependence is also visible in nuisance and companion-dependent quantities, although these comparisons are not always one-to-one because the two codes parameterize some effects differently. Differences in parameterization, treatment of third light and other companion-dependent quantities should therefore be interpreted as part of the model uncertainty rather than as direct shifts in a single shared parameter.

Across both systems, neither light-curve selection nor model prescription produces a dominant, order-of-magnitude shift in the red-giant benchmark quantities. Instead, both effects contribute at the sub-percent to percent level, with the larger contribution dependent on the system and considered parameter. Light-curve selection produces the largest radius shift for KIC\,8129189, whereas the JKTEBOP--PHOEBE-2 difference is larger for KIC\,10920813. In contrast, goodness of fit, third light, companion radius, mass ratio, and other nuisance or companion-dependent quantities can be more sensitive to both the adopted light-curve selection and the model prescription.

The dominant precision floor is therefore not simply the choice between a simpler detached-binary code and a more detailed PHOEBE-2 mesh model, nor is it solely the choice between the full stitched light curve and selected eclipse sets. Instead, the relevant uncertainty reflects the combined effect of model prescription, time-variable eclipse morphology, and the formal uncertainty within each model configuration. For the scaling-relation application, this motivates adopting a conservative benchmark envelope rather than selecting the formally most precise single solution.

\subsubsection{Adopted dynamical benchmark values}\label{sec:adopted_dyn}

The previous comparisons show that the red-giant masses and radii are stable across reasonable choices of light-curve selection and binary-star model, but that the uncertainty from any single model and data configuration does not capture the full modeling uncertainty relevant for benchmark calibration. We therefore adopted final benchmark masses and radii from the envelope of the four different model and data configurations.

The individual model tables report symmetric uncertainties: JKTEBOP values are listed as nominal best-fitting values with TASK~7 one-sigma bootstrap uncertainties, while PHOEBE-2 values are listed as posterior means and standard deviations. For the adopted benchmark values, however, we used the full envelope spanned by the central 68.27\% interval of each model and data configuration, corresponding to the probability enclosed within $\pm1\sigma$ for a Gaussian distribution. The adopted central value was taken as the midpoint of this envelope, and the adopted uncertainty was taken as half of the full envelope width. This prescription means that the adopted value need not coincide exactly with the central value of any single model; instead, it summarizes the range of values supported by the complete set of dynamical solutions.

The resulting adopted red-giant parameters are $M_{\rm RG}=1.413 \pm 0.037\,M_\odot$ and $R_{\rm RG}=4.296 \pm 0.061\,R_\odot$ for KIC\,8129189, and $M_{\rm RG}=1.383 \pm 0.022\,M_\odot$ and $R_{\rm RG}=5.914 \pm 0.043\,R_\odot$ for KIC\,10920813. These correspond to mass precisions of 2.6\% and 1.6\%, and radius precisions of 1.4\% and 0.7\%, respectively. The adopted benchmark values are used in Sect.~\ref{sec:discussion} for the scaling-relation calibration.

The adopted uncertainties are larger than the uncertainties from several individual model configurations, particularly the full-light-curve JKTEBOP and PHOEBE-2 fits. This increase reflects the intended use of these systems as empirical calibrators. For calibration work, underestimating the benchmark uncertainty would artificially overweight a single modeling choice, whereas the envelope approach preserves the information from all four model and data configurations while acknowledging the sensitivity to light-curve morphology and model prescription.

\section{Seismic and stellar modeling}\label{sec:grid}

To place the red-giant components in an evolutionary context, we combined the spectroscopic constraints derived above with the global asteroseismic parameters in a grid-based stellar-modeling analysis. We used the BAyesian STellar Algorithm (BASTA; \citealt{Aguirre22}), which compares observed stellar properties with pre-computed stellar-evolution tracks in a Bayesian framework and returns posterior distributions for both the stellar properties and the model-predicted observables. The model grid used here is based on tracks computed with the Garching Stellar Evolution Code \citep{weiss2008}.

As input constraints, we adopted the atmospheric parameters of the red-giant components from the spectral analysis, specifically $T_{\rm eff}$ and $[{\rm Fe/H}]$, together with the global seismic quantities $\Delta\nu$, $\nu_{\rm max}$, and, when measured, the mixed-mode period spacing $\Delta\Pi_1$. The BASTA results presented in this work are based on the global seismic diagnostics. We also explored fits to the individual oscillation frequencies. Fits using the radial ($\ell=0$) modes converged and yielded stellar properties consistent with the solutions obtained from the global seismic constraints, whereas satisfactory solutions could not be obtained using the $\ell=1$ or $\ell=2$ modes. We therefore retained the global-seismic solutions as the BASTA results reported here.

\begin{figure}[t]
    \includegraphics[width=\columnwidth]{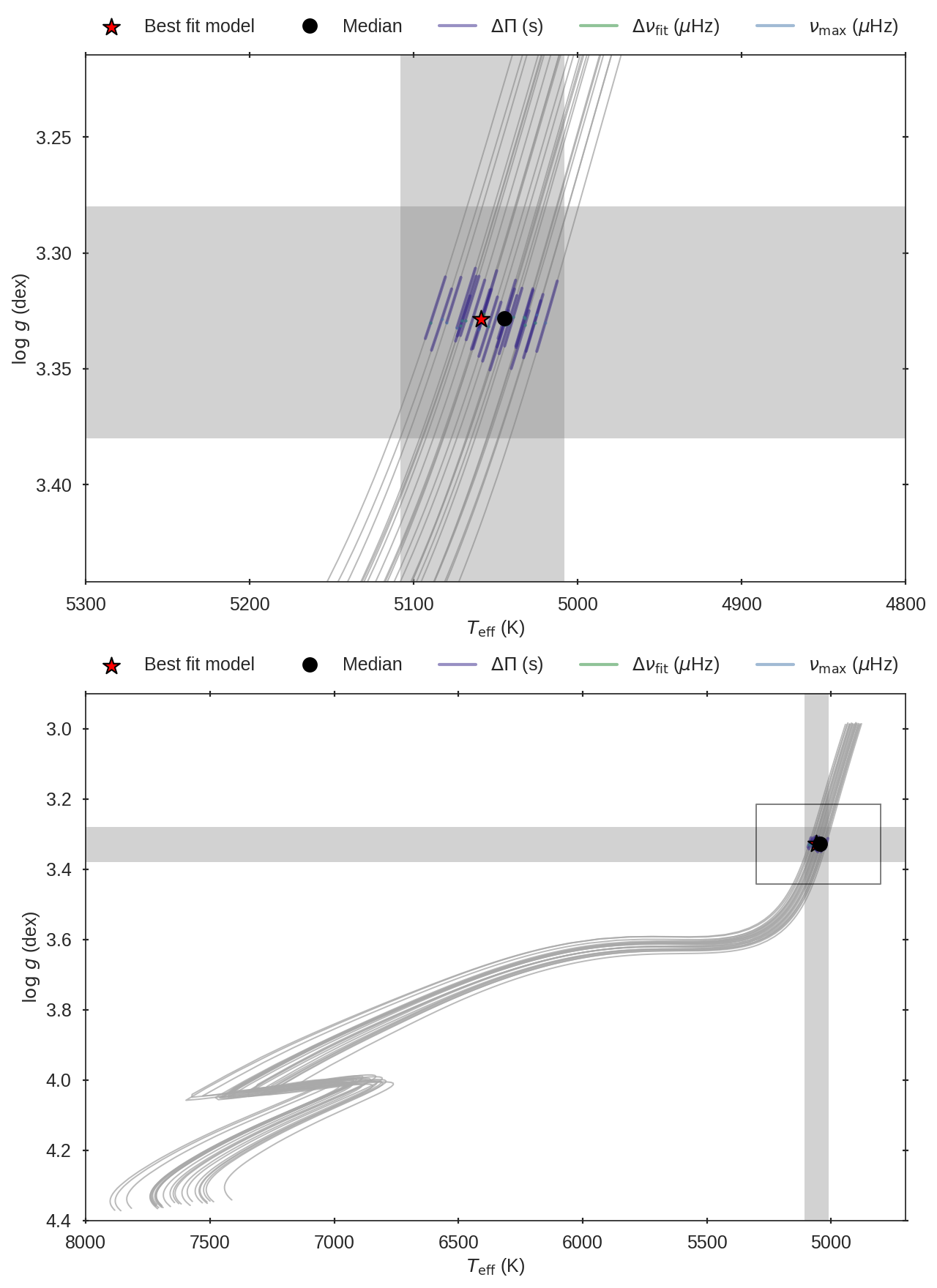}
    \caption{Kiel diagram showing the evolutionary tracks and best-fitting BASTA models for KIC\,8129189. The top panel is a zoomed view of the bottom panel in the region of interest. The individual gray lines are a subset of model tracks within $1\,\sigma$ of the final mass and metallicity solution. The black filled circle represents the median of the posteriors in  $T_{\rm eff}-\log g$ space, while the red star shows the parameters for the best-fitting model. The vertical and horizontal gray shaded rectangles denote the spectroscopic uncertainties of $T_{\rm eff}$ and $\log g$, centered at their estimated values. The small sections of colored lines for $\Delta\nu$, $\nu_{\rm max}$, and $\Delta\Pi$ highlight the parts of the grid that agree within uncertainties of the input values for those quantities. In this case, they are all mostly on top of each other.}
    \label{fig:kiel812}
\end{figure}

\begin{figure}[t]
    \includegraphics[width=\columnwidth]{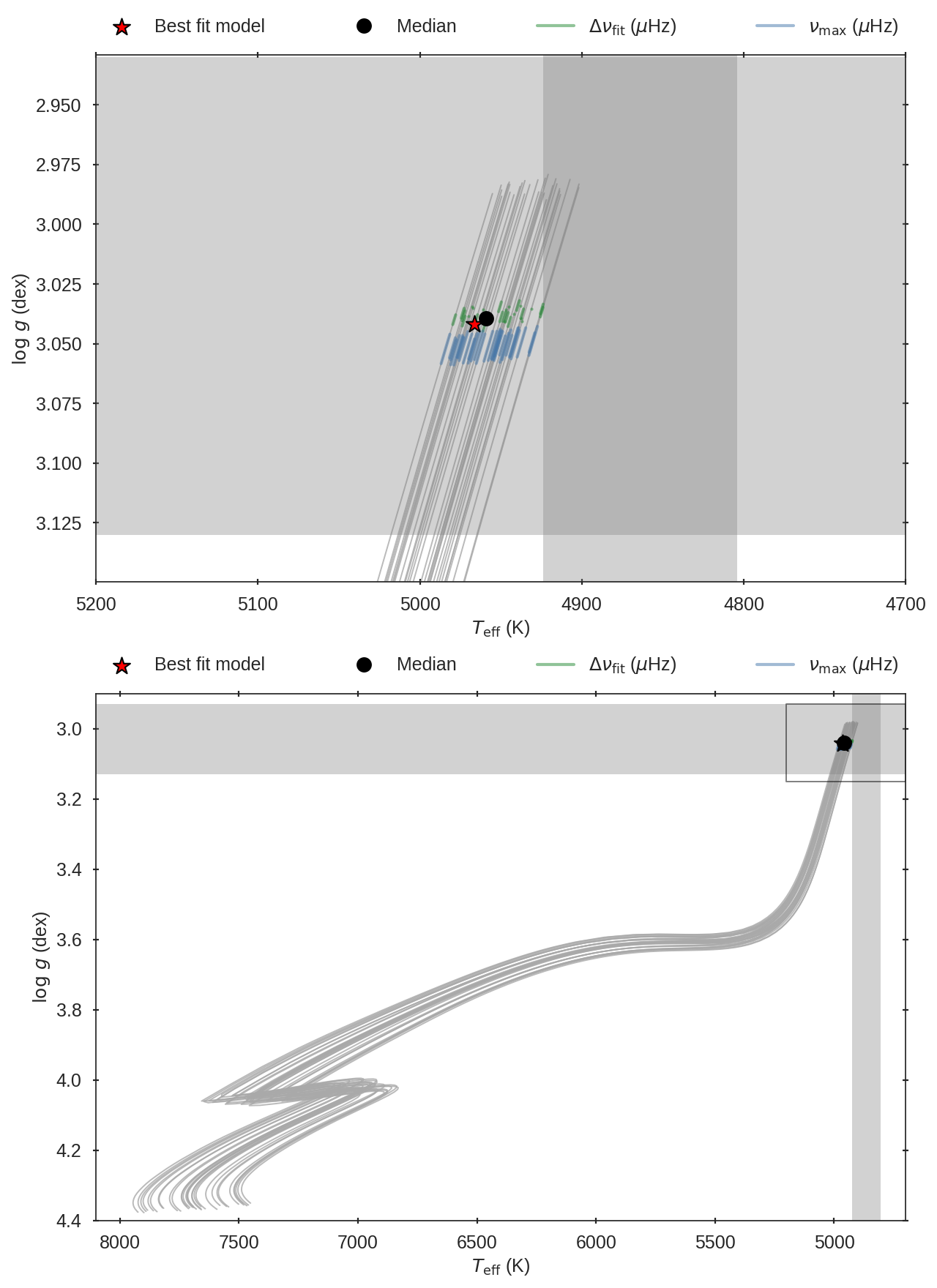}
    \caption{Same as Fig.~\ref{fig:kiel812}, but the Kiel diagram shows the evolutionary tracks and best-fitting BASTA model for KIC\,10920813. This figure omits the period spacing as a constraint in the modeling, since it was not measured.}
    \label{fig:kiel109}
\end{figure}

The resulting posterior estimates are summarized in Table~\ref{tab:atm_param}, which lists both the spectroscopic input constraints and the stellar properties and observables returned by the grid-based modeling. For KIC\,8129189, the posterior estimates give $M = 1.47 \pm 0.03\,M_\odot$, $R = 4.35 \pm 0.04\,R_\odot$, and an age of $2.47 \pm 0.16$\,Gyr. For KIC\,10920813, the corresponding values are $M = 1.31 \pm 0.08\,M_\odot$, $R = 5.84 \pm 0.14\,R_\odot$, and an age of $3.54 \pm 0.92$\,Gyr. In both cases, the models reproduce the atmospheric and global seismic constraints reasonably well, yielding slightly subsolar metallicities and model-predicted global seismic parameters that are consistent with the measured values. The corresponding evolutionary tracks and best-fitting spectroscopic and asteroseismic solutions are shown in the Kiel diagrams in Figs.~\ref{fig:kiel812} and \ref{fig:kiel109}. A direct comparison between these grid-based stellar properties, the dynamical solutions, and the scaling-relation estimates is deferred to the Discussion.

\begin{table}
\caption{Spectroscopic input constraints for the red-giant components and posterior properties from the grid-based stellar modeling with BASTA.}
\label{tab:atm_param}
\footnotesize
\centering
\begin{tabular}{lcc}
\hline
Parameter & KIC\,8129189 & KIC\,10920813\\
\hline
\multicolumn{3}{c}{Spectroscopic constraints}\\
\hline
$T_{\rm eff}$ [K] & $5168 \pm 120$ & $4864 \pm 60$\\
$[{\rm Fe/H}]$ & $-0.24 \pm 0.12$ & $-0.23 \pm 0.15$\\
$\log g$ (fixed) & $3.33$ & $3.03$\\
\hline
\multicolumn{3}{c}{Grid-based stellar modeling}\\
\hline
$T_{\rm eff}$ [K] & $5039 \pm 33$ & $4929 \pm 32$\\
$[{\rm Fe/H}]$ & $-0.12 \pm 0.05$ & $-0.21 \pm 0.04$\\
$\log g$ & $3.328 \pm 0.002$ & $3.024 \pm 0.009$\\
$\nu_{\rm max}$ [$\mu{\rm Hz}$] & $257.8 \pm 1.4$ & $129.0 \pm 2.2$\\
$\Delta\nu$ [$\mu{\rm Hz}$] & $17.88 \pm 0.05$ & $10.79 \pm 0.06$\\
$\Delta\Pi_1$ [s] & $87.9 \pm 0.4$ & $76.6 \pm 1.2$\\
$M$ [$M_\odot$] & $1.47 \pm 0.03$ & $1.31 \pm 0.08$\\
$R$ [$R_\odot$] & $4.35 \pm 0.04$ & $5.84 \pm 0.14$\\
Age [Gyr] & $2.47 \pm 0.16$ & $3.54 \pm 0.92$\\
\hline
\end{tabular}
\end{table}

\section{Discussion}\label{sec:discussion}

The need for empirical calibration of the asteroseismic scaling relations has become more important as asteroseismic masses and radii have moved from individual stellar characterization to large ensemble studies \citep[e.g.,][]{Miglio13,SilvaAguirre18}. Detached SB2 eclipsing binaries containing solar-like oscillating red giants provide direct empirical tests of these relations. Previous dynamical calibrations have shown that uncorrected solar-referenced scaling relations overestimate red-giant masses and radii on the order of 15\% and 5\%, respectively \citep{Gaulme16,Benbakoura21}. Current calibration work is therefore framed in terms of correction factors, $f_{\Delta\nu}$ and $f_{\nu_{\rm max}}$, applied to the scaling relations rather than a simple replacement of the solar reference values \citep{Valle25}. This is important because theoretical $f_{\Delta\nu}$ corrections depend on the adopted stellar-model input physics, and recent work has shown that those choices can propagate into several-percent differences in corrected masses \citep{Sharma16,Li24,Valle25}.

At the same time, recent work has suggested that the performance of the scaling relations is not uniform across red-giant parameter space \citep{Pinsonneault25,Ash24}. The present work contributes to the broader effort to calibrate and test the applicability of the scaling relations by adding two detached eclipsing binaries containing relatively small-radius red giants to the empirical benchmark sample.

\subsection{Benchmark sample and calibration inputs}

The adopted dynamical benchmark masses and radii for KIC\,8129189 and KIC\,10920813 are defined in Sect.~\ref{sec:adopted_dyn}. We used those values here to update the empirical calibration sample for the asteroseismic scaling relations. For the literature benchmark compilation, we adopted the revised KIC\,4054905 dynamical solution and scaling-relation inputs from \citet{Brogaard22}, the revised KIC\,8430105 dynamical solution and seismic parameters from \citet{Thomsen22}, and the revised KIC\,10001167 dynamical solution and scaling-relation inputs from \citet{Thomsen25}. These updates replace the older values for those systems in the calibration table. The full benchmark compilation considered in this work is listed in Table~\ref{tab:scaling_benchmark_inputs}.

\begin{table*}[t]
\caption{Benchmark red-giant inputs considered for the empirical scaling-relation calibration and diagnostic comparisons.}
\label{tab:scaling_benchmark_inputs}
\centering
\footnotesize
\setlength{\tabcolsep}{3.0pt}
\begin{tabular}{lccccccc}
\hline
System & $M_{\rm dyn}$ & $R_{\rm dyn}$ & $\nu_{\rm max}$ & $\Delta\nu$ & $T_{\rm eff}$ & State & Ref. \\
 & [$M_\odot$] & [$R_\odot$] & [$\mu{\rm Hz}$] & [$\mu{\rm Hz}$] & [K] & & \\
\hline
KIC\,4054905  & $0.954 \pm 0.009$ & $8.36 \pm 0.03$ & $48.4 \pm 0.6$ & $5.37 \pm 0.03$ & $4850 \pm 70$ & RGB & B22 \\
KIC\,4663623  & $1.41 \pm 0.08$ & $9.80 \pm 0.20$ & $54.09 \pm 0.24$ & $5.21 \pm 0.02$ & $4812 \pm 92$ & RC* & B21, G16 \\
KIC\,5640750  & $1.16 \pm 0.01$ & $13.12 \pm 0.09$ & $24.1 \pm 0.2$ & $2.969 \pm 0.006$ & $4525 \pm 75$ & RGB & T18 \\
KIC\,5786154  & $1.06 \pm 0.06$ & $11.40 \pm 0.20$ & $29.75 \pm 0.16$ & $3.523 \pm 0.014$ & $4747 \pm 100$ & RGB/RC & G16 \\
KIC\,7037405  & $1.25 \pm 0.04$ & $14.10 \pm 0.20$ & $21.75 \pm 0.14$ & $2.792 \pm 0.012$ & $4516 \pm 36$ & RGB* & G16, B18 \\
KIC\,7293054  & $1.6 \pm 0.1$ & $--$ & $42.58 \pm 0.27$ & $4.32 \pm 0.01$ & $4790 \pm 160$ & RGB/RC & B21 \\
KIC\,7377422  & $1.05 \pm 0.08$ & $9.50 \pm 0.20$ & $40.10 \pm 2.10$ & $4.643 \pm 0.052$ & $4938 \pm 110$ & RGB/RC & G16, Y26 \\
KIC\,7955301  & $1.30 \pm 0.03$ & $5.22 \pm 0.22$ & $124.9 \pm 0.4$ & $10.49 \pm 0.02$ & $4720 \pm 105$ & RGB & G22 \\
KIC\,8129189  & $1.413 \pm 0.037$ & $4.296 \pm 0.061$ & $256.0 \pm 3.7$ & $17.90 \pm 0.05$ & $5168 \pm 120$ & RGB & TW \\
KIC\,8410637  & $1.56 \pm 0.03$ & $10.74 \pm 0.11$ & $46.00 \pm 0.19$ & $4.641 \pm 0.017$ & $4699 \pm 91$ & RGB* & F13, G16, T18 \\
KIC\,8430105  & $1.254 \pm 0.014$ & $7.475 \pm 0.031$ & $76.78 \pm 0.81$ & $7.123 \pm 0.035$ & $4990 \pm 80$ & RGB & T22 \\
KIC\,9153621  & $1.1 \pm 0.2$ & $10.4 \pm 0.6$ & $38.2 \pm 0.3$ & $4.28 \pm 0.01$ & $4760 \pm 190$ & RGB/RC & B21 \\
KIC\,9246715  & $2.149 \pm 0.007$ & $8.30 \pm 0.04$ & $106.4 \pm 0.8$ & $8.31 \pm 0.02$ & $5030 \pm 45$ & RC2 & R16, G16 \\
KIC\,9540226  & $1.33 \pm 0.05$ & $12.80 \pm 0.10$ & $27.07 \pm 0.15$ & $3.216 \pm 0.013$ & $4692 \pm 65$ & RGB & G16, T18 \\
KIC\,9970396  & $1.14 \pm 0.03$ & $8.00 \pm 0.20$ & $63.70 \pm 0.16$ & $6.32 \pm 0.01$ & $4916 \pm 68$ & RGB & G16, Z20 \\
KIC\,10001167 & $0.9337 \pm 0.0077$ & $13.03 \pm 0.15$ & $19.78 \pm 0.16$ & $2.714 \pm 0.017$ & $4625 \pm 60$ & RGB* & T25 \\
KIC\,10920813 & $1.383 \pm 0.022$ & $5.914 \pm 0.043$ & $130.0 \pm 2.3$ & $10.79 \pm 0.06$ & $4864 \pm 60$ & RGB/RC & TW \\
\hline
\end{tabular}
\tablefoot{Dynamical quantities refer to the oscillating red-giant component. KIC\,7293054 and KIC\,7955301 are listed for completeness, but are not included in the joint mass-radius fit. Reference codes: F13 = \citet{Frandsen13}; R16 = \citet{Rawls16}; G16 = \citet{Gaulme16}; B18 = \citet{Brogaard18}; T18 = \citet{Themessl18}; Z20 = \citet{Zhang20}; B21 = \citet{Benbakoura21}; B22 = \citet{Brogaard22}; T22 = \citet{Thomsen22}; T25 = \citet{Thomsen25}; G22 = \citet{Gaulme22}; Y26 = \citet{Yildiz26}; TW = this work. Plain evolutionary state labels denote secure classifications using mixed-mode analysis, an asterisk denotes a favored or model-supported classification that is not secured by mixed-mode classification, and RGB/RC denotes that both interpretations remain viable. For KIC\,4663623 the dynamical quantities and evolutionary state are from B21 and the effective temperature and seismic quantities are from G16. Although mixed-mode analysis was completed for KIC\,4663623, the modes were depleted so we mark the state with an asterisk. For KIC\,7037405 the effective temperature and dynamical and seismic quantities are from G16 and the evolutionary state is from B18. For KIC\,7377422 all of the quantities are from G16, with Y26 obtaining viable solutions for both evolutionary states. For KIC\,8410637, the dynamical quantities are from F13, the effective temperature and seismic quantities are from G16, and the evolutionary state is from T18. For KIC\,9246715, the effective temperature and seismic quantities are from G16 and the dynamical quantities and evolutionary state are from R16. For KIC\,9540226 and KIC\,9970396 the effective temperature and dynamical and seismic quantities are from G16 and the evolutionary states are from T18 and Z20 respectively. For KIC\,10001167, the radius uncertainty combines the statistical uncertainty and the additional systematic term quoted by \citet{Thomsen25} in quadrature.} 
\end{table*}

The two systems analyzed here extend this benchmark compilation by adding well-characterized binaries with less-evolved red giants than most of the previously available calibrators. They therefore provide leverage in a part of red-giant parameter space that has been less densely anchored by dynamical measurements in previous calibrations \citep{Gaulme16,Benbakoura21,Gaulme22}. The practical consequence is that the updated calibration is constrained not only by more systems, but by systems that help bridge the gap between the least evolved oscillating red giants and the more luminous giants that have dominated much of the earlier benchmark sample.

The conservative mass and radius precisions are 2.6\% and 1.4\% for KIC 8129189 and 1.6\% and 0.7\% for KIC 10920813, respectively. These are comparable to or better than the typical precision of the literature benchmarks listed in Table~\ref{tab:scaling_benchmark_inputs}, while incorporating the spread among the four adopted model and data configurations. The new systems should therefore be treated as high-value calibration anchors as they have robustly derived dynamical red-giant radii and they occupy the relatively small-radius red-giant regime where the existing benchmark sample is sparse.

The grid-based stellar-modeling results from Sect.~\ref{sec:grid}, obtained from the spectroscopic and global seismic constraints, are broadly similar to the adopted dynamical benchmark values. With the conservative dynamical uncertainties adopted here, the KIC\,8129189 grid-based solution exceeds the dynamical benchmark by approximately $1.3\sigma$ in mass and $0.8\sigma$ in radius. For KIC\,10920813, the grid-based solution is smaller by about $0.8\sigma$ in mass and $0.5\sigma$ in radius. 

\subsection{Empirical scaling-relation correction}

We used the adopted calibration sample drawn from the updated benchmark compilation to derive the empirical scaling-relation correction directly. We adopted the standard solar-referenced scaling relations \citep{Brown91,Kjeldsen95}, but introduced correction factors at the level of the two underlying seismic relations, 
\begin{equation}
\frac{\Delta\nu}{\Delta\nu_\odot}=f_{\Delta\nu}\left(\frac{\rho}{\rho_\odot}\right)^{1/2},
\end{equation}
\begin{equation}
\frac{\nu_{\rm max}}{\nu_{{\rm max},\odot}}=f_{\nu_{\rm max}}\left(\frac{g}{g_\odot}\right)\left(\frac{T_{\rm eff}}{T_{{\rm eff},\odot}}\right)^{-1/2}.
\end{equation}
This parameterization calibrates the underlying $\Delta\nu$--density and $\nu_{\rm max}$--surface-gravity relations rather than fitting unrelated offsets to the final masses and radii. This form also enables direct comparison with empirical reference-value studies and model-based $f_{\Delta\nu}$ corrections \citep{Themessl18,Rodrigues17,Thomsen22,Valle25}. With this convention, the corrected mass and radius inferred from the raw solar-scaled values are
\begin{equation}
M_{\rm seis,corr}=M_{\rm seis,raw}\,f_{\Delta\nu}^{4}f_{\nu_{\rm max}}^{-3},
\end{equation}
\begin{equation}
R_{\rm seis,corr}=R_{\rm seis,raw}\,f_{\Delta\nu}^{2}f_{\nu_{\rm max}}^{-1}.
\end{equation}
Thus a single pair of correction factors must reproduce both the dynamical mass and the dynamical radius. The fit is therefore a joint fit to the coupled mass and radius residuals, not two independent calibrations. The shared factors should therefore be interpreted as effective sample-level corrections for the seismic masses and radii of the benchmark ensemble, rather than as physically exact correction factors for every individual star. The stronger dependence of mass on both $\nu_{\rm max}$ and $\Delta\nu$ means that small residual systematics in the underlying seismic relations can produce a larger fractional bias in mass than in radius. 

We used solar reference values $\nu_{{\rm max},\odot}=3090\,\mu{\rm Hz}$, $\Delta\nu_\odot=135.1\,\mu{\rm Hz}$, and $T_{{\rm eff},\odot}=5777\,{\rm K}$ \citep{Huber11,Gaulme16}. The adopted calibration sample consists of the literature detached-binary benchmarks plus the two systems from this work.

We excluded KIC\,7293054 from the joint calibration because no dynamical radius is available. We also excluded KIC\,7955301 from the calibration because it is a hierarchical triple rather than a detached red-giant eclipsing binary. In this system, the red giant is the non-eclipsing component of the outer orbit, while the observed eclipses occur in the inner main-sequence binary \citep{Gaulme22}. Consequently, the red-giant radius is more model-dependent and does not provide the same geometrically constrained red-giant radius as the detached SB2 benchmarks; it was retained only as a comparison object. The final joint calibration uses 15 red-giant benchmarks with dynamical masses, dynamical radii, $\nu_{\rm max}$, $\Delta\nu$, and $T_{\rm eff}$.

For each star \(i\), we constructed a logarithmic mass--radius residual vector \({\bf y}_i\) comparing the dynamical mass and radius to the raw seismic-scaling mass and radius as
\begin{equation}
{\bf y}_i =
\begin{pmatrix}
\ln M_{{\rm dyn},i}-\ln M_{{\rm seis,raw},i}\\
\ln R_{{\rm dyn},i}-\ln R_{{\rm seis,raw},i}
\end{pmatrix}
=
\begin{pmatrix}
4 & -3\\
2 & -1
\end{pmatrix}
\begin{pmatrix}
\ln f_{\Delta\nu}\\
\ln f_{\nu_{\rm max}}
\end{pmatrix}
+{\bf \epsilon}_i.
\end{equation}
We then solved for the shared calibration parameters from the stacked residual vectors of all calibrators using generalized least squares. The covariance matrix for ${\bf \epsilon}_i$ includes the quoted dynamical mass and radius uncertainties, together with the propagated uncertainties from $\nu_{\rm max}$, $\Delta\nu$, and $T_{\rm eff}$. Because the raw seismic mass and radius share the same three observables, the off-diagonal covariance term from these shared inputs is retained. 

The adopted fit gives
\begin{equation}
 f_{\Delta\nu}=0.9777^{+0.0045}_{-0.0048},\qquad
 f_{\nu_{\rm max}}=1.024^{+0.008}_{-0.008},
\end{equation}
where the uncertainties are the central 68.27\% intervals from 5000 bootstrap resamples of the benchmark systems in the calibration sample. These values correspond to multiplicative corrections
\begin{equation}
 C_M=f_{\Delta\nu}^{4}f_{\nu_{\rm max}}^{-3}=0.850^{+0.019}_{-0.020}
\end{equation}
for the raw seismic masses and
\begin{equation}
 C_R=f_{\Delta\nu}^{2}f_{\nu_{\rm max}}^{-1}=0.9333^{+0.0083}_{-0.0090}
\end{equation}
for the raw seismic radii. In other words, the fitted correction reduces the raw seismic masses by about 15\% and the raw seismic radii by about 7\%. This is consistent with the direction and approximate scale of the offsets found by \citet{Gaulme16} and \citet{Benbakoura21} for red giants in eclipsing binaries. It is also consistent with the model-based calibration of \citet{Li22}, who used radial-mode frequencies, global seismic parameters, and spectroscopic constraints in grid-based stellar models and found comparable mass and radius offsets.

\begin{figure}[ht!]
\centering
\includegraphics[width=0.4\textwidth]{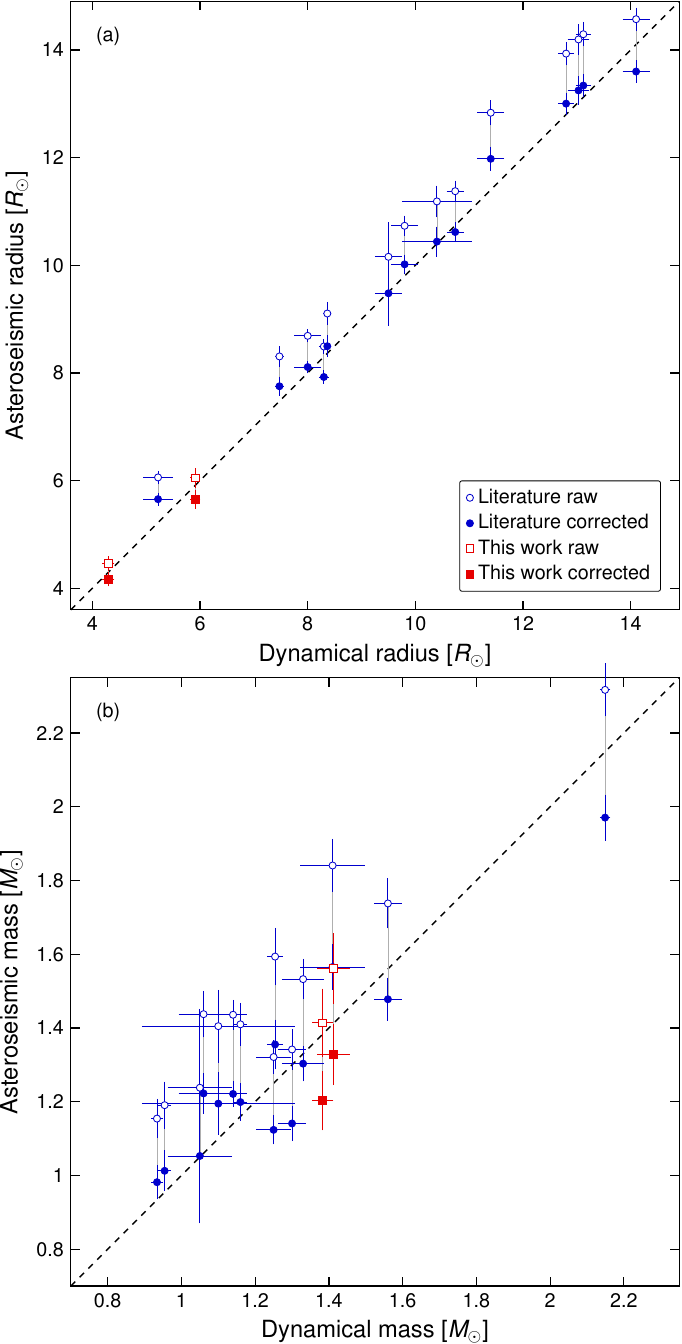}
\caption{Asteroseismic masses and radii compared with dynamical benchmark values for the adopted scaling-relation calibration sample. KIC\,7955301 is shown for comparison but was not included in the calibration. Open symbols show the raw solar-referenced scaling-relation values, while filled symbols show the values after applying the adopted correction factors $f_{\Delta\nu}=0.9777$ and $f_{\nu_{\rm max}}=1.024$. Circles indicate literature systems and squares indicate the two systems analyzed in this work. Horizontal error bars show dynamical uncertainties; vertical error bars show the propagated uncertainties from $\nu_{\rm max}$, $\Delta\nu$, and $T_{\rm eff}$, scaled by the corresponding correction factor for the filled symbols. The common uncertainty in the fitted correction factors is not included in the individual point error bars. The dashed line marks equality between seismic and dynamical values. The two panels are stacked so that the figure can be placed in a single column.}
\label{fig:scaling_calibration}
\end{figure}

Fig.~\ref{fig:scaling_calibration} shows the calibration graphically. The open symbols show the raw solar-referenced scaling-relation values, and the filled symbols show the values after applying the adopted joint correction factors. The adopted correction improves the overall agreement of the calibration sample with the one-to-one relation. Leave-one-out tests confirm that no individual benchmark materially controls the adopted correction factors (Appendix \ref{sec:append_scaling_loo}).

\subsection{Direct tests of the underlying seismic relations}
\label{sec:direct_scaling}

The joint mass--radius calibration constrains a coupled combination of the $\Delta\nu$--density and $\nu_{\rm max}$--surface-gravity relations. To examine the two underlying relations separately, we also computed the direct per-star quantities
\begin{equation}
 f_{\Delta\nu,i}=\frac{\Delta\nu_i/\Delta\nu_\odot}{(\rho_i/\rho_\odot)^{1/2}}
\end{equation}
and
\begin{equation}
 f_{\nu_{\rm max},i}=\frac{\nu_{{\rm max},i}/\nu_{{\rm max},\odot}}{(g_i/g_\odot)(T_{{\rm eff},i}/T_{{\rm eff},\odot})^{-1/2}},
\end{equation}
using the dynamical densities and surface gravities. This diagnostic separates deviations in the large-frequency-separation scaling from deviations in the frequency-of-maximum-power scaling, which is useful because departures from strict homology in $\Delta\nu$ and additional physical dependencies in the $\nu_{\rm max}$ relation have been discussed extensively in the literature \citep[e.g.,][]{Belkacem11,White11,Viani17}. The resulting per-star correction factors and adopted atmospheric parameters are listed in Table~\ref{tab:direct_scaling_factors}.

\begin{table*}[t]
\caption{Direct per-star tests of the underlying asteroseismic scaling relations.}
\label{tab:direct_scaling_factors}
\centering
\footnotesize
\setlength{\tabcolsep}{3.5pt}
\begin{tabular}{lccccc}
\hline
System & $[\mathrm{M/H}]$ & $f_{\Delta\nu,i}$ & $f_{\Delta\nu,\mathrm{Sch}}$ & $f_{\nu_{\rm max},i}$ & Ref. \\
& [dex] & & & & \\
\hline
KIC\,4054905 & $-0.35 \pm 0.08$\tablefootmark{b} & $0.984 \pm 0.009$ & $0.943 \pm 0.004$ & $1.051 \pm 0.020$ & B22 \\
KIC\,4663623 & $+0.16 \pm 0.04$\tablefootmark{c} & $0.998 \pm 0.042$ & $0.991 \pm 0.001$ & $1.092 \pm 0.077$ & B21, G16 \\
KIC\,5640750 & $-0.29 \pm 0.09$\tablefootmark{a} & $0.970 \pm 0.011$ & $0.939 \pm 0.002$ & $1.024 \pm 0.021$ & T18 \\
KIC\,5786154 & $-0.06 \pm 0.06$\tablefootmark{d} & $0.976 \pm 0.038$ & $0.948 \pm 0.003$; $0.979 \pm 0.002$ & $1.074 \pm 0.073$ & G16 \\
KIC\,7037405 & $-0.13 \pm 0.06$\tablefootmark{c} & $0.979 \pm 0.026$ & $0.940 \pm 0.002$ & $0.991 \pm 0.043$ & G16, B18 \\
KIC\,7293054 & $+0.11 \pm 0.26$\tablefootmark{d} & $--$ & $0.964 \pm 0.008$; $0.989 \pm 0.008$ & $--$ & B21 \\
KIC\,7377422 & $-0.33 \pm 0.06$\tablefootmark{d} & $0.984 \pm 0.050$ & $0.947 \pm 0.007$; $0.971 \pm 0.003$ & $1.038 \pm 0.106$ & G16, Y26 \\
KIC\,7955301 & $-0.01 \pm 0.12$\tablefootmark{a} & $0.813 \pm 0.052$ & $0.957 \pm 0.004$ & $0.768 \pm 0.067$ & G22 \\
KIC\,8129189 & $-0.24 \pm 0.12$\tablefootmark{d} & $0.993 \pm 0.025$ & $0.981 \pm 0.009$ & $1.024 \pm 0.044$ & TW \\
KIC\,8410637 & $+0.02 \pm 0.08$\tablefootmark{a} & $0.968 \pm 0.018$ & $0.957 \pm 0.003$ & $0.993 \pm 0.030$ & F13, G16, T18 \\
KIC\,8430105 & $-0.41 \pm 0.10$\tablefootmark{b} & $0.962 \pm 0.009$ & $0.958 \pm 0.005$ & $1.029 \pm 0.020$ & T22 \\
KIC\,9153621 & $-0.35 \pm 0.21$\tablefootmark{d} & $1.027 \pm 0.129$ & $0.944 \pm 0.008$; $0.973 \pm 0.006$ & $1.144 \pm 0.246$ & B21 \\
KIC\,9246715 & $+0.03 \pm 0.02$\tablefootmark{a} & $1.003 \pm 0.008$ & $0.994 \pm 0.001$ & $1.030 \pm 0.014$ & R16, G16, H19 \\
KIC\,9540226 & $-0.31 \pm 0.09$\tablefootmark{a} & $0.946 \pm 0.021$ & $0.942 \pm 0.003$ & $0.974 \pm 0.041$ & G16, T18 \\
KIC\,9970396 & $-0.18 \pm 0.07$\tablefootmark{c} & $0.992 \pm 0.039$ & $0.956 \pm 0.004$ & $1.069 \pm 0.061$ & G16, Z20 \\
KIC\,10001167 & $-0.46 \pm 0.13$\tablefootmark{b} & $0.978 \pm 0.018$ & $0.928 \pm 0.003$ & $1.042 \pm 0.028$ & T25 \\
KIC\,10920813 & $-0.23 \pm 0.15$\tablefootmark{d} & $0.977 \pm 0.014$ & $0.958 \pm 0.005$; $0.984 \pm 0.004$ & $0.977 \pm 0.028$ & TW \\
\hline
\end{tabular}
\tablefoot{The empirical correction factors and their symmetric uncertainties are the midpoint and half-width of the central 68.27\% intervals. The Schimak values are calculated from the relations in Appendix~B of \citet{Schimak26} using the seismic inputs listed in Table~\ref{tab:scaling_benchmark_inputs} and the adopted $[\mathrm{M/H}]$ values. KIC\,7293054 and KIC\,7955301 are included for comparison but are excluded from the adopted detached-binary calibration. 
\tablefoottext{a}{Directly published global metallicity for the oscillating red-giant component.}
\tablefoottext{b}{Global metallicity calculated from published $[\mathrm{Fe/H}]$ and $[\alpha/\mathrm{Fe}]$. For KIC\,4054905 the published relation uses the coefficients adopted by \citet{Brogaard22}; for KIC\,8430105 and KIC\,10001167 we used the Salaris conversion adopted by \citet{Thomsen22}.}
\tablefoottext{c}{APOGEE survey global metallicity. These values are the APOGEE measurements listed by \citet{Gaulme16}.}
\tablefoottext{d}{No direct $[\mathrm{M/H}]$, measured $[\alpha/\mathrm{Fe}]$, or suitable survey $[\mathrm{M/H}]$ was identified; $[\mathrm{M/H}]=[\mathrm{Fe/H}]$ is therefore adopted as a last-resort scaled-solar approximation.}
For systems listed as RGB/RC, the Schimak column lists the RGB and RC calculations in that order. Schimak uncertainties include propagation of the listed $\nu_{\rm max}$, $\Delta\nu$, $T_{\rm eff}$, and $[\mathrm{M/H}]$ uncertainties. The direct $[\mathrm{M/H}]$ values for KIC\,5640750, KIC\,8410637, and KIC\,9540226 are from T18; the value for KIC\,9246715 is from \citet{Helminiak19} (H19). Reference codes are otherwise the same as in Table~\ref{tab:scaling_benchmark_inputs}.}
\end{table*}

\begin{figure}[t]
\centering
\includegraphics[width=\columnwidth]{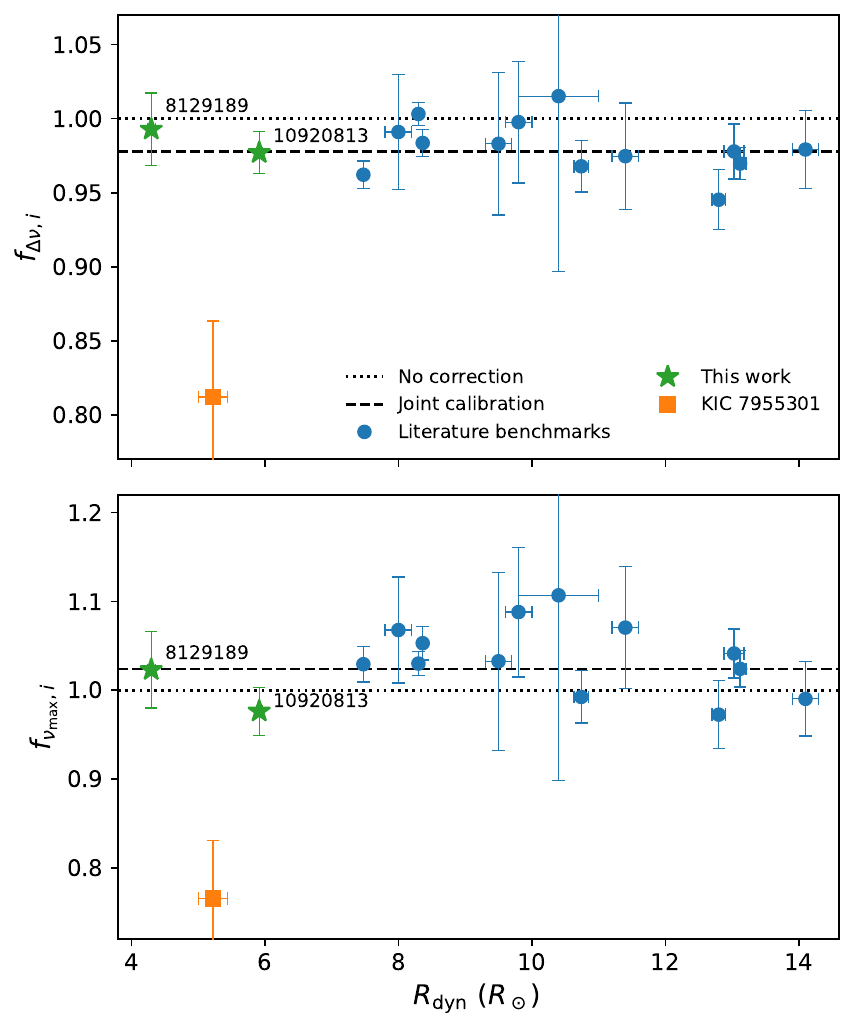}
\caption{Direct tests of the underlying asteroseismic scaling relations using dynamical benchmark densities and surface gravities. The upper panel shows the per-star correction factor for the $\Delta\nu$--density relation; the lower panel shows the corresponding correction factor for the $\nu_{\rm max}$--surface-gravity relation. The uncorrected scaling is marked by the dotted line and the adopted calibration is marked by the dashed line. Literature benchmarks are shown as blue circles and the two systems analyzed in this work are shown as green stars. KIC\,7955301 (orange square) is plotted for comparison but is excluded from the adopted calibration. }
\label{fig:direct_scaling_diagnostics}
\end{figure}

Fig.~\ref{fig:direct_scaling_diagnostics} shows that KIC\,8129189 and KIC\,10920813 fall within the benchmark distribution in both direct scaling quantities. They therefore extend the lower-radius end of the red-giant benchmark sample without appearing as outliers in either of the two underlying seismic relations. KIC\,7955301, by contrast, is a clear outlier in both panels, with $f_{\Delta\nu}\simeq0.81$ and $f_{\nu_{\rm max}}\simeq0.77$. This supports excluding it from the adopted detached-binary calibration while still showing it as a useful comparison case: although it provides a dynamical mass through its hierarchical triple configuration, it does not behave like the detached SB2 benchmark sample in the direct scaling-relation diagnostics.
 
We find no statistically significant monotonic trend of either direct correction factor with dynamical radius over the benchmark radius range. This result is unchanged whether the comparison object KIC\,7955301 is included or omitted. This supports using a constant correction as a sample-level summary, but does not imply that the physically appropriate correction factors are intrinsically constant; dependencies on stellar properties, such as effective temperature, metallicity, and evolutionary state, may remain.

\begin{figure}[t]
\centering
\includegraphics[width=\columnwidth]{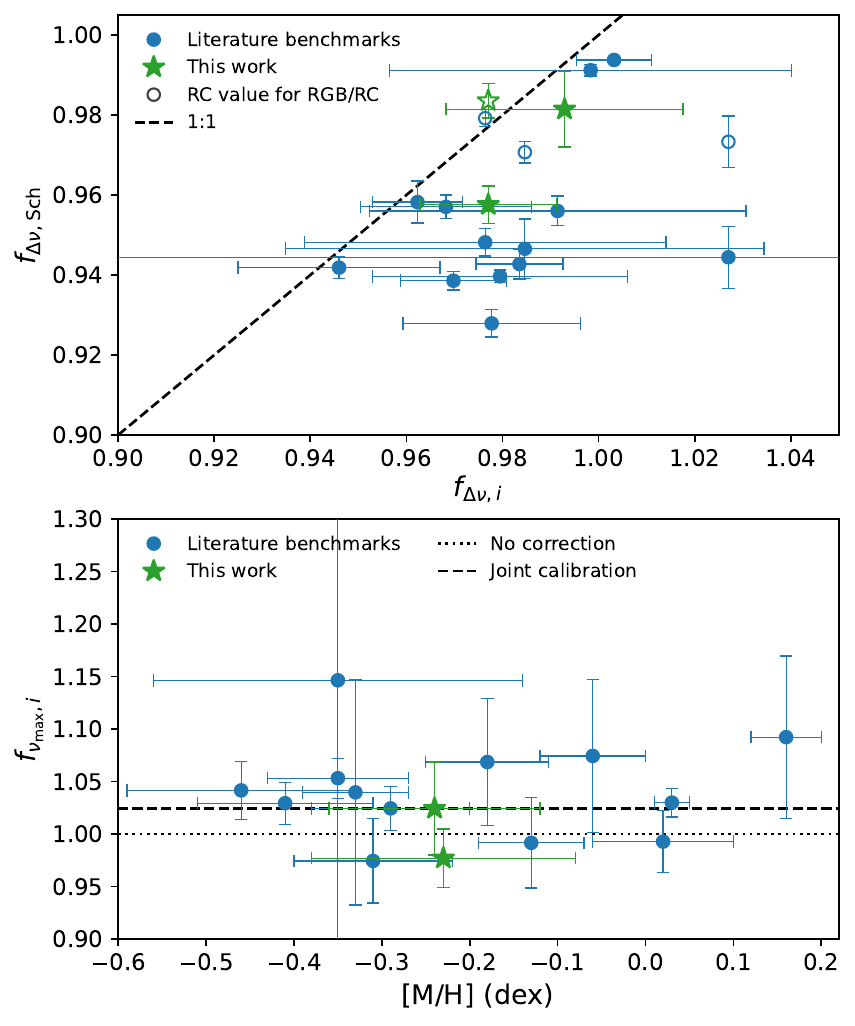}\\
\caption{Comparison of the dynamically derived correction factors with model-based predictions and as a function of metallicity. Top: dynamically derived $f_{\Delta\nu,i}$ values compared with the stellar-model-based $f_{\Delta\nu,\rm Sch}$ values calculated using Appendix~B of \citet{Schimak26}. For systems with ambiguous RGB/RC classifications, both are shown. Bottom: dynamically derived $f_{\nu_{\rm max},i}$ values as a function of $[\mathrm{M/H}]$. The uncorrected scaling relation is marked by the dotted line and the adopted calibration is marked by the dashed line. Literature benchmarks are shown as blue circles and the two systems analyzed in this work as green stars. Note: KIC\,7955301 is not shown in either plot to allow scatter of benchmark systems to be clearly seen.}
\label{fig:schimakMetallicity}
\end{figure}

To investigate these dependencies, we compared our dynamically derived $f_{\Delta\nu,i}$ values with the stellar-model-based calculations of \citet{Li23} and \citet{Schimak26}. \citet{Li23} developed a method for calculating $f_{\Delta\nu}$ for RGB stars using stellar models to account for departures from the approximate $\Delta\nu\propto\sqrt{\rho}$ relation, including the effect of near-surface modeling errors on $\Delta\nu$. \citet{Schimak26} extended this approach to red-clump stars. Based on the seismic parameters, effective temperatures, and evolutionary states listed in Table~\ref{tab:scaling_benchmark_inputs}, together with the metallicity estimates listed in Table~\ref{tab:direct_scaling_factors}, we used the formulation in Appendix~B of \citet{Schimak26} to calculate model-based $f_{\Delta\nu,\rm Sch}$ values for comparison with our dynamically derived $f_{\Delta\nu,i}$ values. We find that the model-based $f_{\Delta\nu,\rm Sch}$ values are systematically lower than the dynamically derived $f_{\Delta\nu,i}$ values across the calibration sample, as shown in Fig.~\ref{fig:schimakMetallicity}. The only exceptions occur for two systems with ambiguous RGB/RC classifications, for which adopting the alternative RC prescription places the model prediction slightly above the one-to-one relation.

We next considered metallicity as a possible source of variation in $f_{\nu_{\rm max},i}$. \citet{Viani17} showed that the usual $\nu_{\rm max}\propto gT_{\rm eff}^{-1/2}$ relation neglects dependencies on mean molecular weight and $\Gamma_1$, which can introduce a composition dependence. More recent detailed modeling has also suggested a metallicity dependence in $f_{\nu_{\rm max}}$, although its inferred strength can depend on assumptions in the stellar models \citep{Li24}. Studies of very metal-poor stars have further found increasing departures from the standard $\nu_{\rm max}$ scaling toward lower metallicity \citep{Huber24,Lindsay25,Lundkvist25}. Motivated by these results, we examined our dynamically derived $f_{\nu_{\rm max},i}$ values as a function of $[\mathrm{M/H}]$ in the lower panel of Fig.~\ref{fig:schimakMetallicity}. Over the metallicity range spanned by our calibration sample, $-0.46\lesssim[\mathrm{M/H}]\lesssim+0.16$, we find no statistically significant monotonic trend. Our dynamical benchmark sample therefore provides no evidence for a metallicity dependence in $f_{\nu_{\rm max}}$ over this relatively limited metallicity range, but does not probe the very metal-poor regime in which stronger departures have been reported.

\subsection{Implications for benchmark calibration and seismic applications}

The adopted correction improves the median seismic-to-dynamical ratios from 1.216 to 1.034 in mass and from 1.086 to 1.014 in radius. The corrected median ratios therefore lie close to unity, although the reduced $\chi^2$ of the adopted fit remains 1.59, indicating that a constant pair of correction factors does not fully describe the star-to-star scatter. As shown in the preceding subsection, however, neither derived correction factor exhibits a statistically significant monotonic relationship with dynamical radius, and $f_{\nu_{\rm max}}$ shows no statistically significant monotonic relationship with metallicity over the range sampled by the benchmark stars. The model-based $f_{\Delta\nu}$ predictions likewise do not account for the residual scatter: they are systematically lower than the dynamically derived values and would therefore not bring the individual benchmarks into closer agreement with the empirical calibration. A constant pair of correction factors is thus a useful empirical summary of the present benchmark sample, but the current data do not identify a simple dependence on radius or metallicity, nor do the tested model-based corrections explain the remaining discrepancies. This does not imply that the physically appropriate correction factors are universal. Studies have suggested dependencies on evolutionary state, luminosity, metallicity, and stellar-model assumptions \citep{Rodrigues17,Pinsonneault25,Ash24,Valle25}. Broader benchmark coverage will be required to determine whether such structure becomes apparent over a wider range of red-giant parameter space.

The practical impact of the correction is largest for applications that rely on red-giant masses. The adopted calibration corresponds to a mass correction of $C_M=0.850$ and a radius correction of $C_R=0.9333$ relative to the raw solar-referenced scaling relations. This is similar in direction and scale to the offsets reported by \citet{Gaulme16} and \citet{Benbakoura21}, who found that uncorrected scaling relations overestimated red-giant masses and radii by about 15\% and 5\%, respectively. The present fit gives a slightly stronger median correction, especially in radius, but remains broadly consistent with the empirical eclipsing-binary picture: uncorrected seismic scaling relations tend to overestimate red-giant masses and radii.

This has direct consequences for ensemble asteroseismology and Galactic archaeology. At fixed composition and evolutionary state, red-giant age estimates are strongly tied to mass, because the main-sequence lifetime decreases steeply with increasing stellar mass. A systematic overestimate of seismic mass therefore leads to a systematic underestimate of stellar age. The adopted correction reduces raw seismic masses by about 15\%, implying that ages inferred from uncorrected scaling-relation masses would be biased young relative to ages inferred from the corrected mass scale. The exact age shift depends on the stellar-model grid, metallicity, evolutionary state, and mass-loss assumptions, so we do not translate the correction into a single universal age offset. However, for stars comparable to those represented by the benchmark sample, the sign of the effect is robust: applying the correction moves the inferred ages toward older values than would be obtained from the raw scaling relations. The two systems presented here occupy a useful place in that effort by extending the calibration to smaller radii. Their conservative dynamical uncertainties keep them from dominating the fit by formal weight, while their locations at $R_{\rm RG}\simeq 4.3\,R_\odot$ and $5.9\,R_\odot$ make them valuable anchors for the less-evolved red-giant regime.

\section{Conclusions}

We presented a spectroscopic, asteroseismic, and dynamical analysis of KIC\,8129189 and KIC\,10920813, two detached SB2 eclipsing binaries containing solar-like oscillating red giants with main-sequence companions. We add these systems to the empirical benchmark sample used to calibrate the asteroseismic scaling relations, extending the sample toward less-evolved red giants and providing new dynamical anchors in a part of parameter space that has been sparsely sampled by previous calibrations.

We measured radial velocities for both components, extracted global seismic parameters from eclipse-masked \textit{Kepler} light curves, and modeled the binary orbits with both JKTEBOP and PHOEBE-2. For each system, we compared full-light-curve solutions with selected-eclipse solutions in order to assess the sensitivity of the dynamical parameters to time-variable eclipse morphology and model prescription. The red-giant masses and radii remain stable across the tested model and data configurations, while some nuisance and companion-dependent quantities are more sensitive to light-curve selection and binary-model assumptions.

The light-curve comparisons also provide guidance for detached red-giant eclipsing binaries with photometric variability. Although KIC\,10920813 shows clear time-variable photometric structure, the red-giant masses and radii inferred from the full light curve and from selected eclipses remain consistent. For systems with variability at this level, spot-related changes in eclipse morphology do not appear to dominate the uncertainty budget for the fundamental red-giant parameters. The full light curve should remain the baseline data set because it uses all available eclipses and surrounding out-of-eclipse data, providing the strongest global constraints and generally yielding more precise formal stellar parameters. Selected-eclipse solutions provide a targeted check on the sensitivity of the derived stellar parameters to time-variable eclipse morphology and can inform the adopted systematic uncertainty.

The model comparison also clarifies the role of PHOEBE-2 for well-detached red-giant eclipsing binaries. Although PHOEBE-2 provides a more physically complete treatment of stellar surfaces, passband intensities, and binary geometry, its substantially higher computational cost did not translate into materially different red-giant benchmark masses or radii for the two detached systems analyzed here. For similar systems, JKTEBOP appears sufficient for obtaining robust dynamical masses and radii, provided that the sensitivity to light-curve selection and time-variable eclipse morphology is first tested. PHOEBE-2 remains the preferred tool for systems where the assumptions of a spherical detached-binary model are less secure, including systems with stronger tidal distortion, substantial irradiation or reflection effects, and partial detachment or near Roche-lobe filling.

To define benchmark values suitable for scaling-relation calibration, we adopted red-giant masses and radii from the envelope of the tested JKTEBOP and PHOEBE-2 solutions. We obtain $M_{\rm RG}=1.413 \pm 0.037\,M_\odot$ and
$R_{\rm RG}=4.296 \pm 0.061\,R_\odot$ for KIC\,8129189, and $M_{\rm RG}=1.383 \pm 0.022\,M_\odot$ and $R_{\rm RG}=5.914 \pm 0.043\,R_\odot$ for KIC\,10920813. These correspond to red-giant mass precisions of 2.6\% and 1.6\%, and radius precisions of 1.4\% and 0.7\%, respectively. The adopted uncertainties are larger than the formal uncertainties from individual fits because they include the variation due to the effects of model prescription and light-curve selection.

Using these new benchmarks together with the updated literature benchmark sample, we fitted empirical correction factors to the underlying $\Delta\nu$--density and $\nu_{\rm max}$--surface-gravity relations. The adopted 15-star detached-binary calibration sample gives $f_{\Delta\nu}=0.9777^{+0.0045}_{-0.0048}$ and $f_{\nu_{\rm max}}=1.024^{+0.008}_{-0.008}$, corresponding to multiplicative corrections of $C_M=0.850^{+0.019}_{-0.020}$ for raw seismic masses and $C_R=0.9333^{+0.0083}_{-0.0090}$ for raw seismic radii. The correction reduces the median seismic-to-dynamical ratios from 1.216 to 1.034 in mass and from 1.086 to 1.014 in radius, consistent with previous evidence that uncorrected solar-referenced scaling relations overestimate red-giant masses and radii.

The two new systems extend the calibration into the lower-radius red-giant regime. Direct tests of the $\Delta\nu$--density and $\nu_{\rm max}$--surface-gravity relations show that both systems fall within the benchmark distribution rather than appearing as outliers. Their value is therefore not that they redefine the empirical correction, but that they strengthen its coverage at smaller red-giant radii. The direct tests reveal no significant dependence of either correction factor on radius, no significant dependence of $f_{\nu_{\rm max}}$ on metallicity, and show that model-based $f_{\Delta\nu}$ predictions are systematically lower than the dynamically derived values. Taken together, these results support a constant pair of correction factors as a useful sample-level description over the parameter range currently sampled, without implying that the corrections are universal.

The calibration has direct implications for applications that rely on seismic red-giant masses, including ensemble asteroseismology and Galactic archaeology. Uncorrected scaling-relation masses that are systematically too large imply ages that are systematically too young, with the exact age offset dependent on metallicity, evolutionary state, and mass-loss assumptions. Empirical dynamical benchmarks therefore remain essential for anchoring seismic mass and radius scales and providing the necessary corrections for application to large seismic samples.

Continued expansion of the benchmark sample is needed to evaluate the applicability of the scaling relations and to determine which dimensions of red-giant star parameter space require additional corrections. Future benchmarks are especially needed at higher luminosities, including red-clump and upper-RGB stars. Such systems will be critical for mapping any dependence of the scaling-relation correction on evolutionary state, luminosity, metallicity, and stellar-model physics.

\begin{acknowledgements}
The authors dedicate this work to our colleague and dear friend, Patrick Gaulme, who sadly passed away much too young while this manuscript was being completed. He pioneered the research topic presented here, and his legacy will certainly live on as more oscillating stars in binary systems are discovered.  La perte de Patrick est incommensurable; TA a non seulement perdu un coll\`egue inspirant mais aussi un ami et un fr\`ere d'\^ame.
This research was supported by a Los Alamos National Laboratory Center for Space and Earth Sciences grant XX8P ASF2. This paper includes data collected by the Kepler mission and obtained from the MAST data archive at the Space Telescope Science Institute (STScI). Funding for the Kepler mission is provided by the NASA Science Mission Directorate. STScI is operated by the Association of Universities for Research in Astronomy, Inc., under NASA contract NAS 5-26555. This research would not be possible without the observations obtained with the Apache Point Observatory 3.5~m telescope, which is owned and operated by the Astrophysical Research Consortium. 
This research has made use of the SIMBAD database, operated at CDS, Strasbourg, France. This research used PyRAF, a product of the Space Telescope Science Institute which is operated by AURA for NASA, and Astropy, a community-developed core Python package for Astronomy. This research made use of the JKTEBOP code (developed by John Southworth and based on the EBOP code originally written by Paul Etzel) and PHOEBE (PHysics Of Eclipsing BinariEs, open-source modeling code developed with the support of the National Science Foundation). ChatGPT (GPT-5.5 and GPT-5.6, OpenAI) was used only for editorial assistance, including language editing, organization, and clarity checks. The authors reviewed and revised all AI-assisted text and take full responsibility for the content of the manuscript.
\end{acknowledgements}

\bibliographystyle{aa}
\bibliography{aa61292-26}

\begin{appendix}
\nolinenumbers

\onecolumn
\section{Oscillation analysis tables}\label{sec:appendosc}

\footnotesize
\centering
\begin{longtable}{l l l l l l l l l }
\caption{Result of oscillation analysis of KIC\,8129189.}
\label{tab:oscillation_frequencies_kic812} \\
\hline\hline
$l$ & $\nu$ & $\sigma_\nu$ & $H$ & $\sigma_{H,+}$& $\sigma_{H,-}$ & $\Gamma$ & $\sigma_{\Gamma,+}$ & $\sigma_{\Gamma,-}$ \\
     & $\mu$Hz & $\mu$Hz & ppm$^2\,\mu{\rm Hz}^{-1}$ & ppm$^2\,\mu{\rm Hz}^{-1}$ & ppm$^2\,\mu{\rm Hz}^{-1}$ & $\mu$Hz & $\mu$Hz & $\mu$Hz\\
\hline
   0 &   203.604 &     0.007 &     355. &     155. &    108. &     0.027 &     0.012 &     0.008 \\ 
   0 &   221.400 &     0.037 &     109. &     29. &     23. &     0.111 &     0.031 &     0.024 \\ 
   0 &   239.278 &     0.012 &     330. &     77. &     62. &     0.144 &     0.025 &     0.021 \\ 
   0 &   257.041 &     0.013 &     459. &     81. &     69. &     0.089 &     0.012 &     0.011 \\ 
   0 &   274.997 &     0.065 &     88. &     11. &     10. &     0.507 &     0.059 &     0.053 \\ 
   1 &   197.509 &     0.005&     234. &     175. &     102. &     0.016 &     0.011 &    0.007 \\ 
   1 &   212.730 &     0.048 &     71. &     23. &     18. &     0.442 &     0.157 &     0.116 \\ 
   1 &   230.026 &     0.014 &    256. &     74. &     57. &     0.081 &     0.021 &     0.016 \\ 
   1 &   232.934 &     0.008 &     271. &     67. &     54. &     0.024 &     0.004 &    0.004 \\ 
   1 &   241.391 &     0.003&     275. &     89. &     67. &     0.007 &    0.001 &    0.001 \\ 
   1 &   246.604 &     0.006 &     272. &     70. &     55. &     0.018 &     0.003 &    0.003 \\ 
   1 &   248.655 &     0.010 &     672. &     123. &     104. &     0.070 &     0.010 &    0.009 \\ 
   1 &   252.657 &     0.007&    1176. &    562. &    380. &     0.019 &    0.006 &    0.005 \\ 
   1 &   257.062 &     0.016 &    794. &     410. &     270. &     0.036 &     0.014 &    0.010 \\ 
   1 &   263.371 &     0.005 &     1047. &     403. &     291. &     0.015 &     0.004 &    0.003 \\ 
   1 &   266.308 &     0.016 &     347. &     59. &     50. &     0.157 &     0.023 &     0.020 \\ 
   1 &   281.548$^\dagger$ &     0.155 &     38. &     8. &     7. &     1.032 &     0.243 &     0.197 \\ 
   2 &   201.635 &     0.019 &     355.&     155. &   108 &    0.027 &    0.012 &     0.008 \\
   2 &   219.259 &     0.027 &     109. &     29. &     23. &     0.111 &     0.031 &     0.024 \\
   2 &   237.275 &     0.017 &     330. &     77. &     62. &     0.144 &     0.025 &     0.021 \\ 
   2 &   255.353 &     0.018 &     459. &     81. &     69. &     0.089 &     0.012 &     0.011 \\ 
   2 &   273.123 &     0.119 &     88. &     11. &     10. &     0.507 &     0.059 &     0.053 \\  
\hline
\end{longtable}
\tablefoot{Oscillation frequencies $\nu$ are expressed in $\mu$Hz, heights $H$ in ppm$^2\,\mu{\rm Hz}^{-1}$, and linewidths [$\mu{\rm Hz}$] with their respective errors $\sigma$.  For the $l=2$ modes, the heights and linewidths are those of the $l=0$ modes.$\dagger$~Aliased frequency of 284.86 $\mu$Hz.}

\footnotesize
\centering
\begin{longtable}{l l l l l l l l l }
\caption{Result of oscillation analysis of KIC\,10920813.}
\label{tab:oscillation_frequences_kic109} \\
\hline\hline
$l$ & $\nu$ & $\sigma_\nu$ & $H$ & $\sigma_{H,+}$& $\sigma_{H,-}$ & $\Gamma$ & $\sigma_{\Gamma,+}$ & $\sigma_{\Gamma,-}$ \\
     & $\mu$Hz & $\mu$Hz & ppm$^2\,\mu{\rm Hz}^{-1}$ & ppm$^2\,\mu{\rm Hz}^{-1}$ & ppm$^2\,\mu{\rm Hz}^{-1}$ & $\mu$Hz & $\mu$Hz & $\mu$Hz\\
\hline
   0 &   110.321 &     0.033 &   319. &    63. &    53. &     0.174 &     0.032 &     0.027 \\ 
   0 &   121.135 &     0.033 &   302. &    50. &    43. &     0.165 &     0.024 &     0.021 \\ 
   0 &   131.885 &     0.036 &   285. &    39. &    34. &     0.241 &     0.030 &     0.026 \\ 
   0 &   142.596 &     0.064 &   167. &    22. &    20. &     0.388 &     0.050 &     0.044 \\ 
   0 &   152.815 &     0.242 &    40. &     7. &     6. &     0.819 &     0.162 &     0.135 \\ 
   0 &   164.789 &     0.032 &   190. &    97. &    64. &     0.088 &     0.044 &     0.030 \\ 
   1 &   115.580 &     0.006 &   870. &    576. &    346. &     0.012 &     0.008 &     0.005 \\ 
   1 &   126.636 &     0.059 &   144. &    22. &    19. &     0.184 &     0.020 &     0.018 \\ 
   1 &   137.412 &     0.062 &   142.  &    34. &    28. &     0.255 &     0.063 &     0.051 \\ 
   1 &   148.018 &     0.120 &   81.  &    28. &    21. &     0.492 &     0.185 &     0.134 \\ 
   2 &   109.058 &     0.050 &   320. &    63. &    53. &     0.174 &     0.032 &     0.027 \\ 
   2 &   119.687 &     0.056 &   302. &    50. &    43. &     0.165 &     0.023 &     0.021 \\ 
   2 &   130.366 &     0.063 &   285. &    39. &    34. &     0.241 &     0.030 &     0.026 \\ 
   2 &   141.215 &     0.114 &   167. &    22. &    20. &     0.388 &     0.050 &     0.044 \\ 
   2 &   152.424 &     0.469 &   40.  &    7.  &     6. &     0.819 &     0.162 &     0.135 \\
   2 &   164.289 &     0.031 &   190. &    97. &    64. &     0.088 &     0.045 &     0.030 \\
\hline
\end{longtable}
\tablefoot{Oscillation frequencies $\nu$ are expressed in $\mu$Hz, heights $H$ in ppm$^2\,\mu{\rm Hz}^{-1}$, and linewidths [$\mu{\rm Hz}$] with their respective errors $\sigma$.  For the $l=2$ modes, the heights and linewidths are those of the $l=0$ modes.}

\clearpage

\section{Dynamical model comparisons and fit diagnostics}\label{sec:append_dyn}\label{sec:appendjktebop}

\raggedright 
This appendix collects the diagnostics used to define the conservative dynamical benchmark envelope. Appendix~\ref{sec:append_quarter} shows the quarter-to-quarter eclipse-depth variation that motivated the selected-eclipse tests. Appendices~\ref{sec:append_812} and \ref{sec:append_109} then group the JKTEBOP and PHOEBE-2 full-light-curve versus selected-eclipse comparison tables and selected-eclipse forward-model fits by system. Appendix~\ref{sec:append_offsets} shows the offsets of the accepted red-giant model solutions from the adopted benchmark values.

\subsection{Quarter-to-quarter eclipse-depth variation}\label{sec:append_quarter}
\begin{figure}[ht!]
\centering
\includegraphics[width=0.98\textwidth]{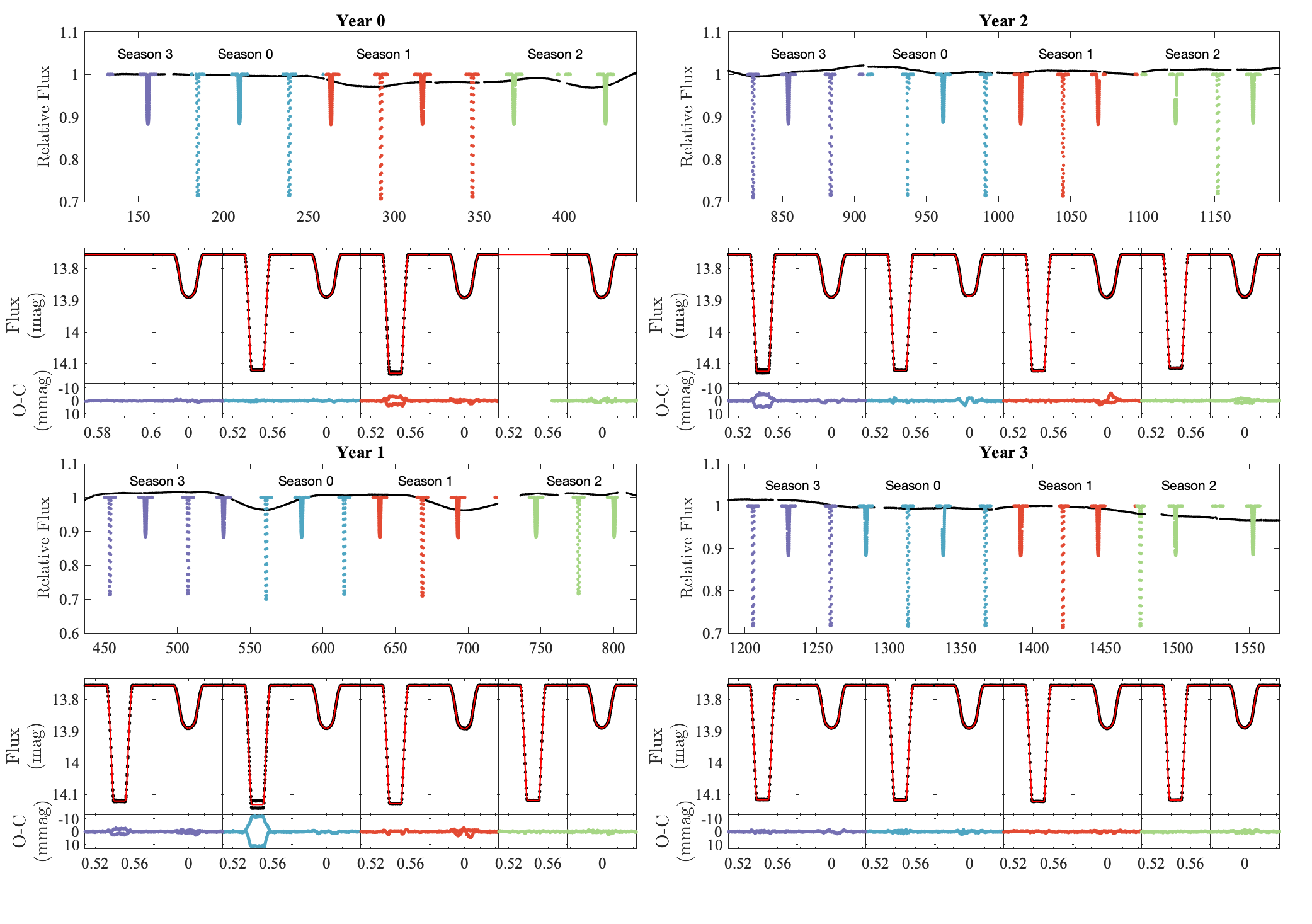}
\caption{Quarter-by-quarter variation in eclipse depth for KIC\,10920813. The changing depth motivates the selected-eclipse analysis, because a single static, spot-free binary model cannot represent all observed eclipse morphologies equally well when the visible surface-brightness distribution evolves with time.}
\label{fig:var_depth}
\end{figure}

\FloatBarrier
\subsection{KIC\,8129189 full light-curve and selected eclipse diagnostics}\label{sec:append_812}

\begin{table}[ht!]
\caption{Comparison of KIC\,8129189 JKTEBOP parameters for the full stitched light-curve and selected-eclipse solutions.}
\label{tab:jktebop_812_compare_app}
\centering
\footnotesize
\begin{tabular}{lcc}
\hline
Parameter & Full LC & Selected eclipse\\
\hline
\multicolumn{3}{c}{Fitted parameters}\\
\hline
Surface-brightness ratio $J$ & $2.550 \pm 0.094$ & $2.21 \pm 0.36$\\
$r_{\rm A}+r_{\rm B}$ & $0.06939 \pm 0.00014$ & $0.0691 \pm 0.0028$\\
$k=r_{\rm B}/r_{\rm A}$ & $0.3400 \pm 0.0018$ & $0.351 \pm 0.043$\\
Limb darkening $A_{1}$ & $0.495 \pm 0.045$ & $0.70 \pm 0.25$\\
Limb darkening $B_{1}$ & $0.361 \pm 0.028$ & $0.37 \pm 0.50$\\
Inclination $i$ [deg] & $87.191 \pm 0.017$ & $87.25 \pm 0.11$\\
$e\cos\omega$ & $-0.20070 \pm 0.00013$ & $-0.20056 \pm 0.00073$\\
$e\sin\omega$ & $-0.1963 \pm 0.0031$ & $-0.200 \pm 0.018$\\
$P$ [days] & $53.6470701 \pm 0.0000041$ & $53.64654 \pm 0.00054$\\
Ephemeris timebase $t_{0}$ [days] & $133.09234 \pm 0.00023$ & $133.0997 \pm 0.0073$\\
$K_{\rm A}$ [km\,s$^{-1}$] & $37.783 \pm 0.078$ & $37.85 \pm 0.48$\\
$K_{\rm B}$ [km\,s$^{-1}$] & $43.492 \pm 0.088$ & $43.57 \pm 0.53$\\
$\gamma$ [km\,s$^{-1}$] & $-11.20568 \pm 0.00081$ & $-11.180 \pm 0.021$\\
Third light $L_{3}$ & 0.001 fixed & 0.001 fixed\\
Reduced $\chi^2$ & $2.97$ & $2.97$\\
\hline
\multicolumn{3}{c}{Derived parameters}\\
\hline
Orbital eccentricity $e$ & $0.28075$ & $0.28341$\\
Periastron longitude $\omega$ [deg] & $224.37$ & $224.96$\\
$M_{\rm A}$ [$M_\odot$] & $1.4175^{+0.007}_{-0.0047}$ & $1.416^{+0.033}_{-0.04}$\\
$M_{\rm B}$ [$M_\odot$] & $1.2314^{+0.0061}_{-0.0042}$ & $1.230^{+0.029}_{-0.036}$\\
$R_{\rm A}$ [$R_\odot$] & $4.2894^{+0.012}_{-0.0075}$ & $4.313^{+0.043}_{-0.075}$\\
$R_{\rm B}$ [$R_\odot$] & $1.4592^{+0.011}_{-0.0051}$ & $1.7^{+0.2}_{-0.2}$\\
$\log g_{\rm A}$ [cgs] & $3.3248$ & $3.3355$\\
$\log g_{\rm B}$ [cgs] & $4.2008$ & $4.1842$\\
$\rho_{\rm A}$ [$\rho_\odot$] & $0.01797$ & $0.01862$\\
$\rho_{\rm B}$ [$\rho_\odot$] & $0.39710$ & $0.37457$\\
\hline
\end{tabular}
\tablefoot{Star~A denotes the red-giant component. Fitted parameters are listed as the nominal best-fitting value with the TASK~7 one-sigma bootstrap uncertainty. Masses and radii are listed as the bootstrap median with the central 68.27\% interval, which is used in constructing the conservative benchmark envelope.}
\end{table}

\begin{table}[ht!]
\caption{Comparison of KIC\,8129189 PHOEBE-2 parameters for the full stitched light-curve and selected-eclipse posterior samples.}
\label{tab:phoebe_812_compare_app}
\footnotesize
\centering

\begin{tabular}{lcc}
\hline
Parameter & Full LC & Selected eclipse\\
\hline
\multicolumn{3}{c}{Sampled parameters}\\
\hline
$r_{\rm A}+r_{\rm B}$ & $0.069262 \pm 0.000088$ & $0.06897 \pm 0.00013$\\
$k=r_{\rm B}/r_{\rm A}$ & $0.3422 \pm 0.0016$ & $0.340 \pm 0.003$\\
Inclination $i$ [deg] & $87.194 \pm 0.004$ & $87.2272 \pm 0.0039$\\
Eccentricity $e$ & $0.28047 \pm 0.00044$ & $0.2842 \pm 0.0031$\\
Argument of periastron $\omega$ [deg] & $224.3 \pm 0.1$ & $225.08 \pm 0.68$\\
Mass ratio $q=M_{\rm B}/M_{\rm A}$ & $0.8703 \pm 0.0026$ & $0.8701 \pm 0.0082$\\
$a\sin i$ [$R_\odot$] & $82.681 \pm 0.034$ & $82.59 \pm 0.36$\\
$P$ [days] & $53.647162 \pm 0.000021$ & $53.6471 \pm 0.0004$\\
$t_{0,\mathrm{supconj}}$ [days] & $133.092652 \pm 0.000001$ & $133.0957 \pm 0.0054$\\
$T_{\mathrm{eff,B}}/T_{\mathrm{eff,A}}$ & $1.2680 \pm 0.0023$ & $1.2679 \pm 0.0067$\\
$\gamma$ [km\,s$^{-1}$] & $-11.31 \pm 0.01$ & $-11.190 \pm 0.093$\\
Third-light fraction $l_{3,\mathrm{frac}}$ & $0.00237 \pm 0.00062$ & $0.00521 \pm 0.00073$\\
$\sigma_{\ln f}$ & $-7.815 \pm 0.019$ & $-8.060 \pm 0.051$\\
\hline
\multicolumn{3}{c}{Propagated parameters}\\
\hline
$M_{\rm A}+M_{\rm B}$ [$M_\odot$] & $2.6446 \pm 0.0032$ & $2.636 \pm 0.035$\\
$M_{\rm A}$ [$M_\odot$] & $1.41400 \pm 0.00027$ & $1.410 \pm 0.023$\\
$M_{\rm B}$ [$M_\odot$] & $1.2306 \pm 0.0035$ & $1.226 \pm 0.014$\\
$R_{\rm A}$ [$R_\odot$] & $4.2717 \pm 0.0021$ & $4.25 \pm 0.02$\\
$R_{\rm B}$ [$R_\odot$] & $1.4619 \pm 0.0062$ & $1.448 \pm 0.012$\\
$a$ [$R_\odot$] & $82.780 \pm 0.034$ & $82.69 \pm 0.36$\\
$r_{\rm A}$ & $0.051603 \pm 0.000017$ & $0.051457 \pm 0.000032$\\
$r_{\rm B}$ & $0.017660 \pm 0.000082$ & $0.01751 \pm 0.00015$\\
\hline
\end{tabular}
\tablefoot{Values are posterior means and standard deviations. The selected-eclipse model uses the same eclipse set as the selected-eclipse JKTEBOP model. Star~A denotes the red-giant component.}
\end{table}
\clearpage

\begin{figure*}[ht!]
\centering
\begin{minipage}[t]{0.485\textwidth}
\centering
\includegraphics[width=\linewidth]{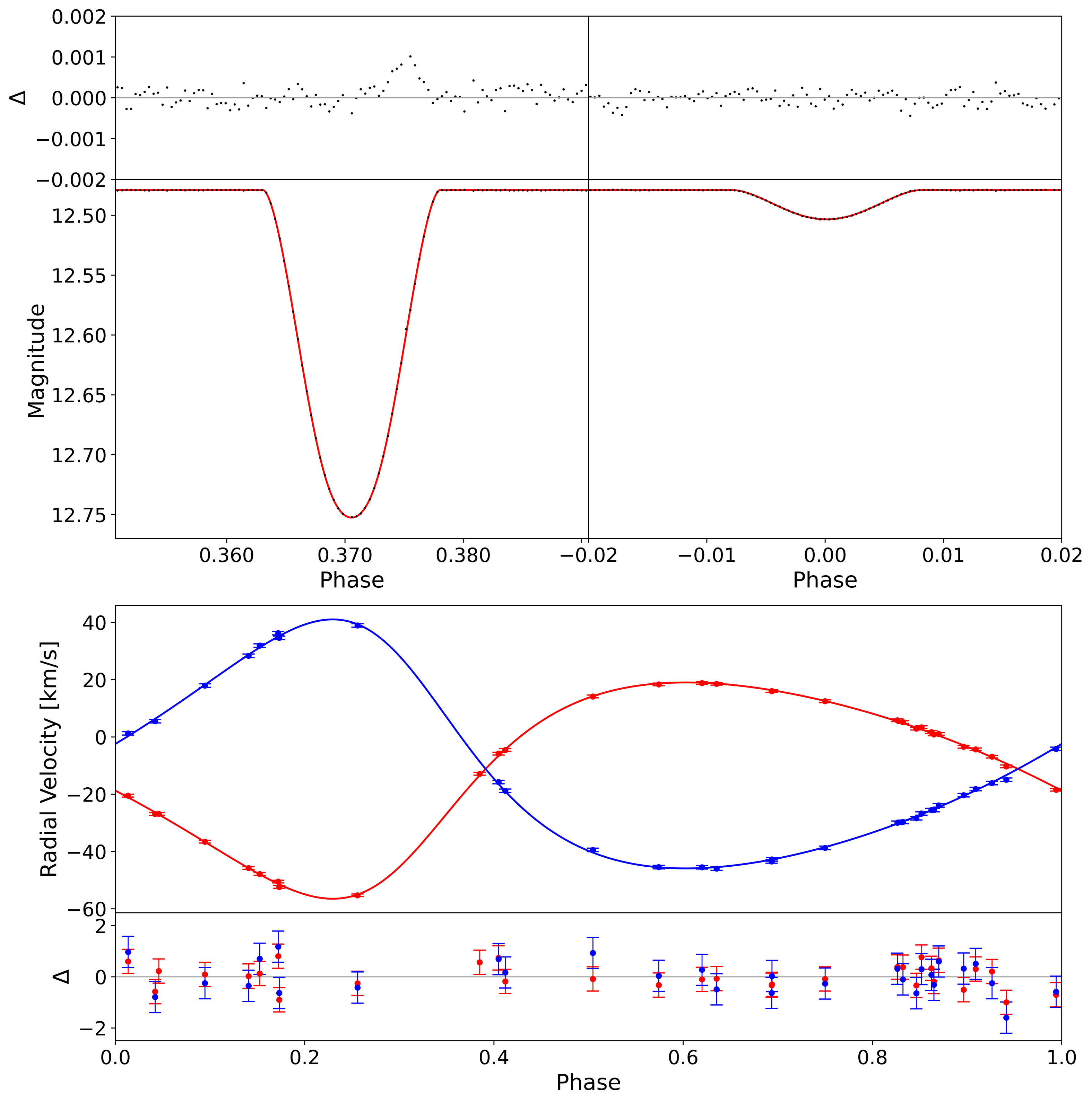}
\caption{Selected-eclipse JKTEBOP light-curve and radial-velocity fit for KIC\,8129189.}
\label{fig:jktebop_selected_812_app}
\end{minipage}
\hfill
\begin{minipage}[t]{0.485\textwidth}
    \centering
    \includegraphics[width=\linewidth]{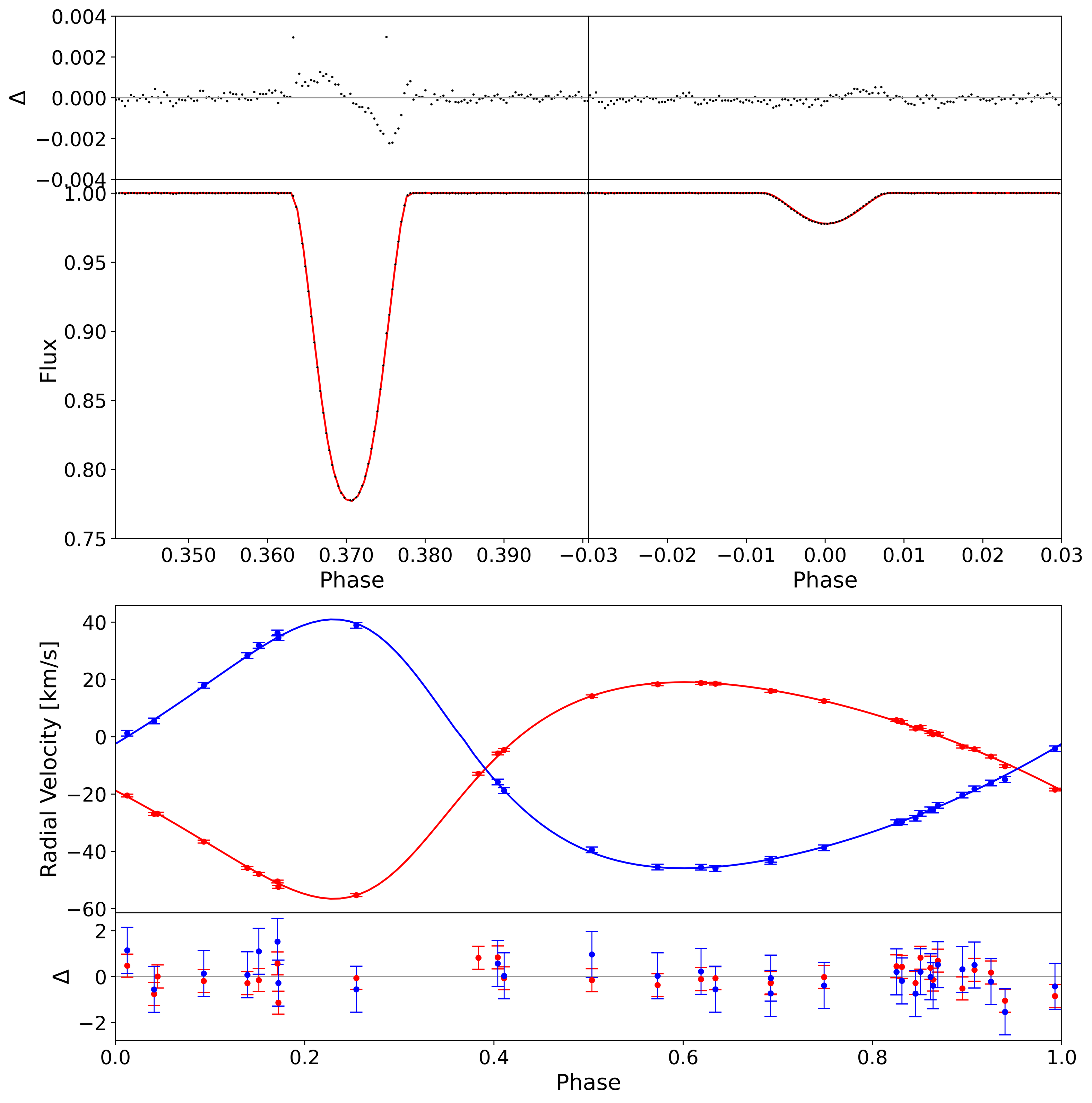}
\caption{Selected-eclipse PHOEBE-2 light-curve and radial-velocity fit for KIC\,8129189. This model uses the same selected eclipse set as the JKTEBOP fit in Fig.~\ref{fig:jktebop_selected_812_app}.}
\label{fig:phoebe_selected_812_app}
\end{minipage}
\end{figure*}

\subsection{KIC\,10920813 full light-curve and selected eclipse diagnostics}\label{sec:append_109}

\begin{table}[ht!]
\caption{Comparison of KIC\,10920813 JKTEBOP parameters for the full stitched light-curve and selected-eclipse solutions. }
\label{tab:jktebop_109_compare_app}
\footnotesize
\centering

\begin{tabular}{lcc}
\hline
Parameter & Full LC & Selected eclipse\\
\hline
\multicolumn{3}{c}{Fitted parameters}\\
\hline
Surface-brightness ratio $J$ & $2.986 \pm 0.067$ & $3.206 \pm 0.036$\\
$r_{\rm A}+r_{\rm B}$ & $0.09687 \pm 0.00028$ & $0.09709 \pm 0.00012$\\
$k=r_{\rm B}/r_{\rm A}$ & $0.3709 \pm 0.0013$ & $0.37759 \pm 0.00086$\\
Limb darkening $A_{1}$ & $0.500 \pm 0.014$ & $0.461 \pm 0.012$\\
Limb darkening $B_{1}$ & $0.419 \pm 0.055$ & $0.554 \pm 0.025$\\
Inclination $i$ [deg] & $89.99 \pm 0.21$ & $89.88 \pm 0.14$\\
$e\cos\omega$ & $0.0690166 \pm 0.0000093$ & $0.069027 \pm 0.000016$\\
$e\sin\omega$ & $-0.18113 \pm 0.00051$ & $-0.17845 \pm 0.00085$\\
$P$ [days] & $53.740819 \pm 0.000014$ & $53.74099 \pm 0.00010$\\
Ephemeris timebase $t_{0}$ [days] & $155.55173 \pm 0.00026$ & $155.55210 \pm 0.00040$\\
$K_{\rm A}$ [km\,s$^{-1}$] & $39.2111 \pm 0.0022$ & $39.1952 \pm 0.0043$\\
$K_{\rm B}$ [km\,s$^{-1}$] & $40.7728 \pm 0.0011$ & $40.7615 \pm 0.0036$\\
$\gamma$ [km\,s$^{-1}$] & $-20.7705 \pm 0.0014$ & $-20.7767 \pm 0.0025$\\
Third light $L_{3}$ & 0.020 fixed & 0.020 fixed\\
Reduced $\chi^2$ & $39.01$ & $1.84$\\
\hline
\multicolumn{3}{c}{Derived parameters}\\
\hline
Orbital eccentricity $e$ & $0.19384$ & $0.19133$\\
Periastron longitude $\omega$ [deg] & $290.86$ & $291.15$\\
$M_{\rm A}$ [$M_\odot$] & $1.37142^{+0.00031}_{-0.00026}$ & $1.37218^{+0.00054}_{-0.00048}$\\
$M_{\rm B}$ [$M_\odot$] & $1.31889^{+0.00026}_{-0.00021}$ & $1.31945^{+0.00046}_{-0.00042}$\\
$R_{\rm A}$ [$R_\odot$] & $5.8923^{+0.021}_{-0.0037}$ & $5.8761^{+0.0087}_{-0.0052}$\\
$R_{\rm B}$ [$R_\odot$] & $2.1864^{+0.02}_{-0.0024}$ & $2.2188^{+0.0081}_{-0.003}$\\
$\log g_{\rm A}$ [cgs] & $3.0351$ & $3.0374$\\
$\log g_{\rm B}$ [cgs] & $3.8796$ & $3.8663$\\
$\rho_{\rm A}$ [$\rho_\odot$] & $0.00671$ & $0.00677$\\
$\rho_{\rm B}$ [$\rho_\odot$] & $0.12653$ & $0.12084$\\
\hline
\end{tabular}
\tablefoot{Star~A denotes the red-giant component. Fitted parameters are listed as the nominal best-fitting value with the TASK~7 one-sigma bootstrap uncertainty. Masses and radii are listed as the bootstrap median with the central 68.27\% interval, which is used in constructing the conservative benchmark envelope.}
\end{table}
\FloatBarrier

\begin{table}[ht!]
\caption{Comparison of KIC\,10920813 PHOEBE-2 parameters for the full stitched light-curve and selected-eclipse posterior samples.}
\label{tab:phoebe_109_compare_app}
\footnotesize
\centering

\begin{tabular}{lcc}
\hline
Parameter & Full LC & Selected eclipse\\
\hline
\multicolumn{3}{c}{Sampled parameters}\\
\hline
$r_{\rm A}+r_{\rm B}$ & $0.096981 \pm 0.000037$ & $0.097156 \pm 0.000029$\\
$k=r_{\rm B}/r_{\rm A}$ & $0.365305 \pm 0.000095$ & $0.36749 \pm 0.00012$\\
Inclination $i$ [deg] & $89.932 \pm 0.038$ & $89.978 \pm 0.016$\\
Eccentricity $e$ & $0.19929 \pm 0.00026$ & $0.19949 \pm 0.00029$\\
Argument of periastron $\omega$ [deg] & $290.239 \pm 0.028$ & $290.211 \pm 0.032$\\
Mass ratio $q=M_{\rm B}/M_{\rm A}$ & $0.975 \pm 0.021$ & $0.9545 \pm 0.0078$\\
$a\sin i$ [$R_\odot$] & $83.670 \pm 0.044$ & $83.48 \pm 0.34$\\
$P$ [days] & $53.7408177 \pm 0.0000001$ & $53.74088 \pm 0.00028$\\
$t_{0,\mathrm{supconj}}$ [days] & $155.551747 \pm 0.000003$ & $155.55215 \pm 0.00064$\\
$T_{\mathrm{eff,B}}/T_{\mathrm{eff,A}}$ & $1.28874 \pm 0.00019$ & $1.29265 \pm 0.00022$\\
$\gamma$ [km\,s$^{-1}$] & $-20.23 \pm 0.26$ & $-20.800 \pm 0.095$\\
Third-light fraction $l_{3,\mathrm{frac}}$ & $0.012250 \pm 0.000049$ & $0.005285 \pm 0.000068$\\
$\sigma_{\ln f}$ & $-6.614 \pm 0.007$ & $-7.635 \pm 0.032$\\
\hline
\multicolumn{3}{c}{Propagated parameters}\\
\hline
$M_{\rm A}+M_{\rm B}$ [$M_\odot$] & $2.7213 \pm 0.0043$ & $2.703 \pm 0.033$\\
$M_{\rm A}$ [$M_\odot$] & $1.378 \pm 0.013$ & $1.383 \pm 0.021$\\
$M_{\rm B}$ [$M_\odot$] & $1.343 \pm 0.016$ & $1.320 \pm 0.014$\\
$R_{\rm A}$ [$R_\odot$] & $5.9433 \pm 0.0028$ & $5.931 \pm 0.024$\\
$R_{\rm B}$ [$R_\odot$] & $2.171 \pm 0.001$ & $2.1797 \pm 0.0089$\\
$a$ [$R_\odot$] & $83.670 \pm 0.044$ & $83.48 \pm 0.34$\\
$r_{\rm A}$ & $0.071032 \pm 0.000027$ & $0.071047 \pm 0.000022$\\
$r_{\rm B}$ & $0.025948 \pm 0.000011$ & $0.02611 \pm 0.00001$\\
\hline
\end{tabular}
\tablefoot{Values are posterior means and standard deviations. The selected-eclipse model uses the same eclipse set as the selected-eclipse JKTEBOP model. Star~A denotes the red-giant component.}
\end{table}

\begin{figure*}[ht!]
\centering
\begin{minipage}[t]{0.485\textwidth}
\centering
\includegraphics[width=\linewidth]{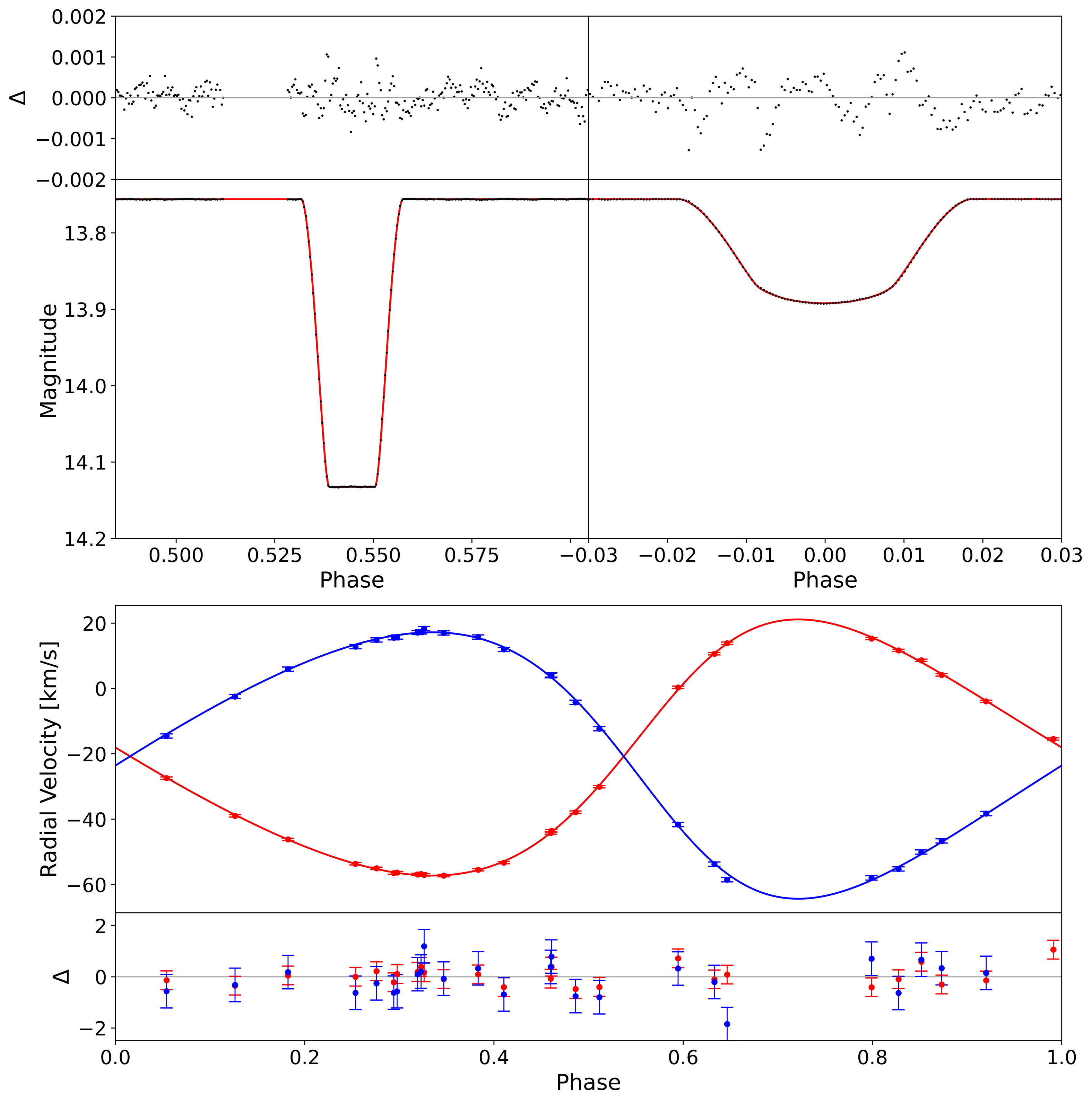}
\caption{Selected-eclipse JKTEBOP light-curve and radial-velocity fit for KIC\,10920813.}
\label{fig:jktebop_selected_109_app}    
\end{minipage}
\begin{minipage}[t]{0.485\textwidth}
    \centering
\includegraphics[width=\linewidth]{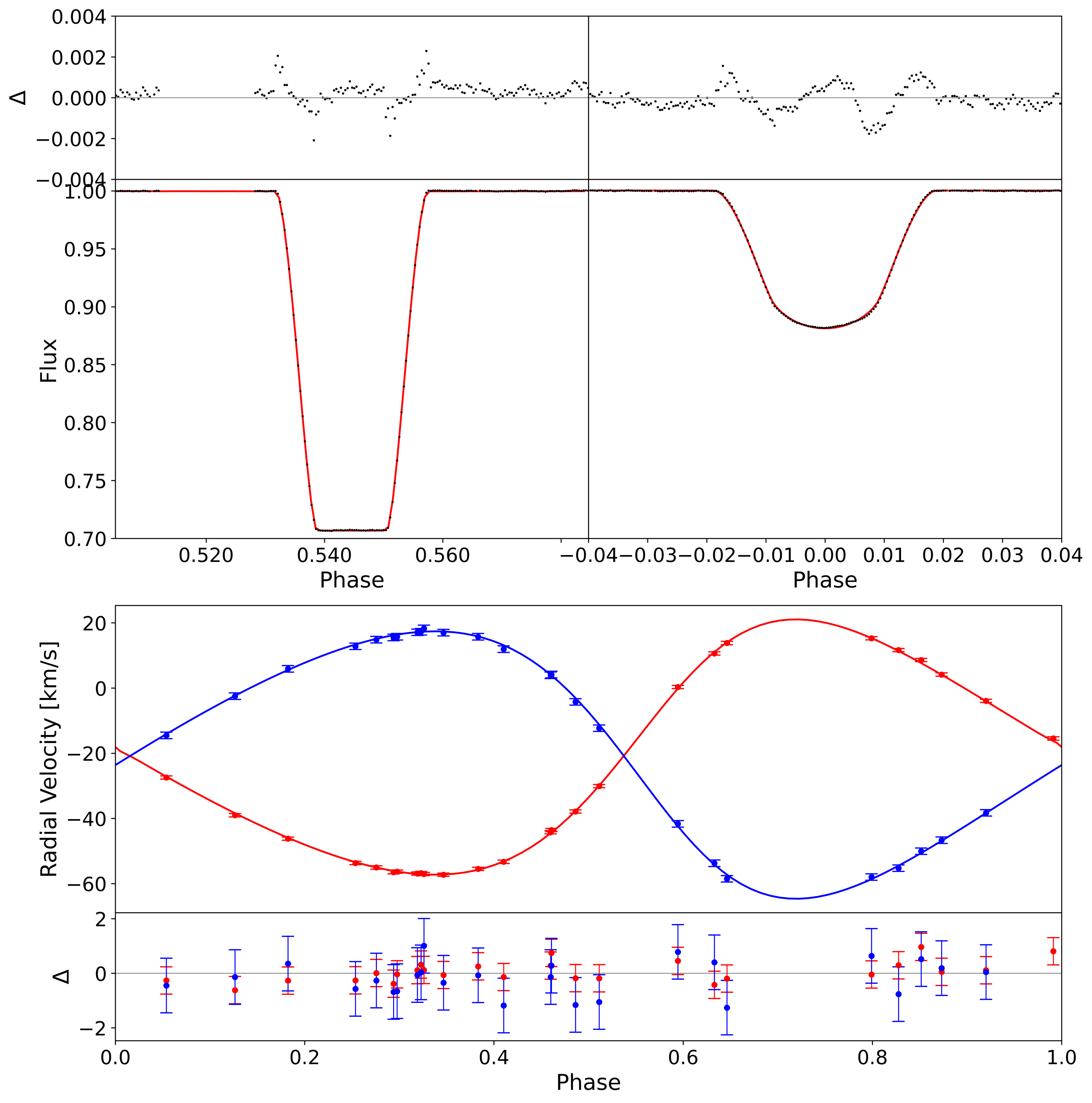}
\caption{Selected-eclipse PHOEBE-2 light-curve and radial-velocity fit for KIC\,10920813. This model uses the same selected eclipse set as the JKTEBOP fit in Fig.~\ref{fig:jktebop_selected_109_app}.}
\label{fig:phoebe_selected_109_app}
\end{minipage}
\end{figure*}

\clearpage
\FloatBarrier
\subsection{Model offsets from the adopted benchmark values}\label{sec:append_offsets}

\begin{figure}[ht!]
\centering
\includegraphics[width=0.75\textwidth]{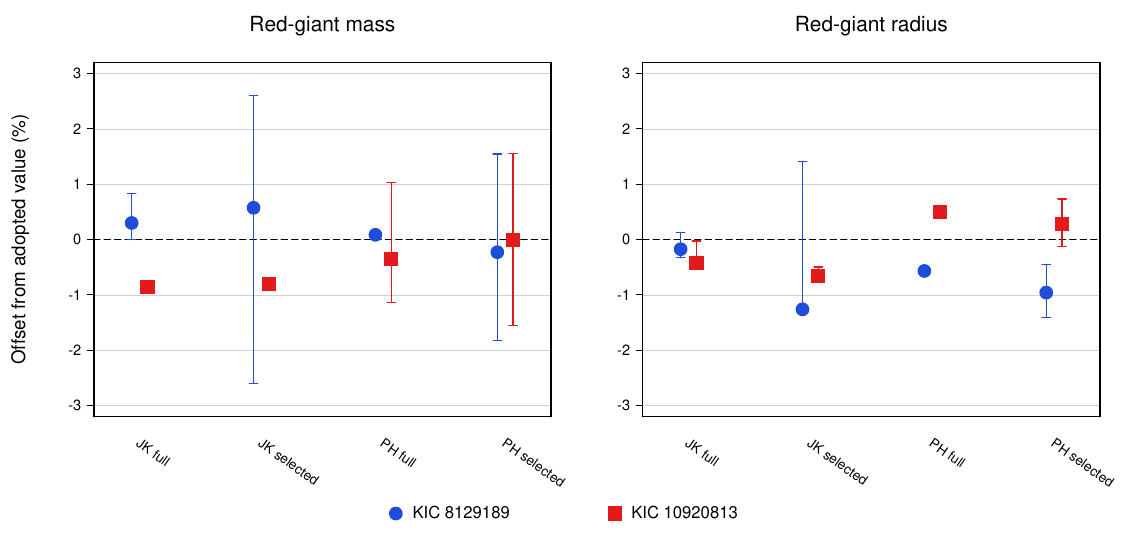}
\caption{Offsets in the red-giant mass and radius from the four accepted dynamical model and data configurations, relative to the adopted conservative benchmark value for each system. JK and PH denote JKTEBOP and PHOEBE-2, respectively; full and selected denote the full-light-curve and selected-eclipse fits. Error bars show the central 68.27\% interval for each individual model or posterior, expressed relative to the adopted value. The dashed line marks the adopted conservative value. The clustering of the central values shows that the red-giant masses and radii are stable, while the error bars illustrate why the benchmark precision is set by the full model and data envelope rather than by the formally most precise single fit.}
\label{fig:dynamical_model_comparison}
\end{figure}

\FloatBarrier
\section{Leave-one-out scaling-relation calibration}\label{sec:append_scaling_loo}

\begin{table*}[ht!]
\caption{Leave-one-out tests for the adopted empirical scaling-relation calibration.}
\label{tab:scaling_loo}
\footnotesize
\centering

\begin{tabular}{lrrrrrrr}
\hline
Omitted system & $f_{\Delta\nu}$ & $f_{\nu_{\rm max}}$ & $C_M$ & $C_R$ & $\Delta C_M$ & $\Delta C_R$ & $\chi^2_\nu$ \\
\hline
KIC\,4054905 & 0.9765 & 1.0211 & 0.8541 & 0.9339 & 0.0036 & 0.0006 & 1.661 \\
KIC\,4663623 & 0.9777 & 1.0227 & 0.8542 & 0.9347 & 0.0038 & 0.0014 & 1.615 \\
KIC\,5640750 & 0.9786 & 1.0243 & 0.8534 & 0.9350 & 0.0030 & 0.0017 & 1.674 \\
KIC\,5786154 & 0.9787 & 1.0234 & 0.8561 & 0.9360 & 0.0056 & 0.0027 & 1.468 \\
KIC\,7037405 & 0.9773 & 1.0271 & 0.8420 & 0.9299 & -0.0085 & -0.0033 & 1.454 \\
KIC\,7377422 & 0.9776 & 1.0241 & 0.8505 & 0.9333 & 0.0000 & -0.0000 & 1.715 \\
KIC\,8129189 & 0.9768 & 1.0238 & 0.8484 & 0.9320 & -0.0021 & -0.0013 & 1.657 \\
KIC\,8410637 & 0.9785 & 1.0271 & 0.8461 & 0.9322 & -0.0044 & -0.0011 & 1.654 \\
KIC\,8430105 & 0.9796 & 1.0251 & 0.8547 & 0.9360 & 0.0043 & 0.0027 & 1.570 \\
KIC\,9153621 & 0.9776 & 1.0240 & 0.8508 & 0.9334 & 0.0003 & 0.0001 & 1.708 \\
KIC\,9246715 & 0.9741 & 1.0243 & 0.8378 & 0.9264 & -0.0126 & -0.0069 & 1.242 \\
KIC\,9540226 & 0.9800 & 1.0274 & 0.8506 & 0.9348 & 0.0001 & 0.0015 & 1.655 \\
KIC\,9970396 & 0.9767 & 1.0200 & 0.8577 & 0.9353 & 0.0073 & 0.0021 & 1.587 \\
KIC\,10001167 & 0.9771 & 1.0218 & 0.8544 & 0.9344 & 0.0039 & 0.0011 & 1.664 \\
KIC\,10920813 & 0.9780 & 1.0266 & 0.8456 & 0.9317 & -0.0049 & -0.0015 & 1.531 \\
\hline
\end{tabular}
\tablefoot{Each row gives the correction factors obtained by repeating the joint mass--radius calibration after omitting one included benchmark system. The quantities $\Delta C_M$ and $\Delta C_R$ are computed relative to the adopted calibration using the 15-star detached-binary benchmark sample. The small changes in the correction factors show that the empirical calibration is not dominated by any individual included system. The adopted calibration, excluding KIC\,7955301, gives $f_{\Delta\nu}=0.9777$, $f_{\nu_{\rm max}}=1.0242$, $C_M=0.8505$, and $C_R=0.9333$. The maximum absolute shifts are $|\Delta C_M|=0.0126$ and $|\Delta C_R|=0.0069$, corresponding to relative changes of 1.5\% and 0.7\%, respectively.}
\end{table*}

\end{appendix}
\end{document}